\documentclass[prd,amsmath,amssymb,superscriptaddress,preprintnumbers,twocolumn,nofootinbib,10pt]{revtex4-1}

\pdfoutput=1
\usepackage{graphicx}
\usepackage{dcolumn}
\usepackage{bm}
\usepackage{amssymb}
\usepackage{latexsym}
\usepackage{booktabs}
\usepackage{amsmath}
\usepackage{multirow}
\usepackage{url}
\usepackage{footnote}
\usepackage{float}
\usepackage{threeparttable}
\usepackage[colorlinks=true, linkcolor=blue, citecolor=blue]{hyperref}
\usepackage[bottom]{footmisc}

\usepackage[normalem]{ulem}
\usepackage{color}
\usepackage{array}
\usepackage{enumerate}
\usepackage{adjustbox}
\usepackage{multirow}

\usepackage{hyperref}

\usepackage{makecell}
\usepackage{diagbox}
\usepackage{epstopdf}
\usepackage{epsfig}
\usepackage{longtable}
\usepackage{supertabular}
\usepackage{algorithm}
\usepackage{pifont}
\usepackage{algorithmic}
\usepackage{changepage}
\usepackage{setspace}

\usepackage{xcolor}  
\usepackage{enumitem} 

\begin{document}

\title{Model-independent cosmological inference after the DESI DR2 data with improved inverse distance ladder }

\author{Jia-Le Ling}
\affiliation{Liaoning Key Laboratory of Cosmology and Astrophysics, College of Sciences, Northeastern University, Shenyang 110819, China}

\author{Guo-Hong Du}
\affiliation{Liaoning Key Laboratory of Cosmology and Astrophysics, College of Sciences, Northeastern University, Shenyang 110819, China}

\author{Tian-Nuo Li}
\affiliation{Liaoning Key Laboratory of Cosmology and Astrophysics, College of Sciences, Northeastern University, Shenyang 110819, China}

\author{Jing-Fei Zhang}
\affiliation{Liaoning Key Laboratory of Cosmology and Astrophysics, College of Sciences, Northeastern University, Shenyang 110819, China}

\author{Shao-Jiang Wang}\thanks{Corresponding author}\email{schwang@itp.ac.cn}
\affiliation{Institute of Theoretical Physics, Chinese Academy of Sciences, Beijing 100190, China}
\affiliation{Asia Pacific Center for Theoretical Physics, Pohang 37673, Korea}

\author{Xin Zhang}\thanks{Corresponding author}\email{zhangxin@neu.edu.cn}
\affiliation{Liaoning Key Laboratory of Cosmology and Astrophysics, College of Sciences, Northeastern University, Shenyang 110819, China}
\affiliation{MOE Key Laboratory of Data Analytics and Optimization for Smart Industry, Northeastern University, Shenyang 110819, China}
\affiliation{National Frontiers Science Center for Industrial Intelligence and Systems Optimization, Northeastern University, Shenyang 110819, China}

\begin{abstract}
Recently, the baryon acoustic oscillations (BAO) measurements from the Dark Energy Spectroscopic Instrument (DESI) Collaboration survey have suggested hints of dynamical dark energy, challenging the standard $\Lambda $ cold dark matter ($\Lambda $CDM) model.
 In this work, we adopt an improved inverse distance ladder approach based on the latest cosmological data to provide a model-independent perspective, employing a global parametrization based on cosmic age (PAge). Our analysis incorporates DESI DR2 BAO measurements, cosmic
chronometer (CC) data, and type Ia supernovae observations from either the DESY5 or PantheonPlus datasets. For the DESY5+DESI DR2+CC datasets, we obtain $H_0 = 67.91 \pm 2.33~\mathrm{km~s^{-1}~Mpc^{-1}}$. This value is consistent with the Planck 2018 result, while showing $2.0 \sigma$ tension with the SH0ES measurement. Furthermore, by mapping specific cosmological models into PAge approximation parameter space $(p_{\mathrm{age}}, \eta)$, our model-independent analysis reveals a notable deviation from the $\Lambda \mathrm{CDM}$ model, as indicated by the DESY5 and DESI DR2 datasets. Finally, DESY5+DESI DR2+CC datasets provide nearly decisive evidence favoring the PAge model over the standard $\Lambda \mathrm{CDM}$ model. These findings highlight the need for further investigation into the expansion history to better understand the deviations from the $\Lambda \mathrm{CDM}$ model.

\end{abstract}

\maketitle
\section{INTRODUCTION}

The simplest $\Lambda$ cold dark matter ($\Lambda\mathrm{CDM}$) model has been remarkably consistent with most astronomical observations across different epochs and scales of the Universe \cite{SupernovaSearchTeam:1998fmf,SupernovaCosmologyProject:1998vns,Brout:2022vxf,KiDS:2020suj,DES:2017qwj,ACT:2020gnv}. However, the persistent discrepancy in the value of the Hubble constant $H_0$ inferred from the early and late Universe, known as the Hubble tension, is challenging the $\Lambda\mathrm{CDM}$ model \cite{DiValentino:2020zio,Perivolaropoulos:2021jda,Shah:2021onj, Verde:2019ivm}. Among various discrepancies, the most significant is the over $5 \sigma$ divergence between the Planck 2018 estimate of $H_0 = 67.4\pm0.5\,\mathrm{km\,s^{-1}\,Mpc^{-1}}$, indirectly inferred from the cosmic microwave background (CMB) data assuming the $\Lambda \mathrm{CDM}$ model \cite{Planck:2018vyg}, and the SH0ES result of $H_0=73.04\pm1.04 \,\mathrm{km\ s^{-1}\ Mpc^{-1}}$ directly measured from the type Ia supernovae (SNe) calibrated using the Cepheid variables \cite{Riess:2021jrx}. The tension cannot be merely explained by a statistical fluctuation, even after the reanalysis of systematic uncertainties \cite{Planck:2019evm, Zhang:2017aqn}. Therefore, the irreconcilable tension may indicate underlying systematics or new physics beyond $\Lambda\mathrm{CDM}$ \cite{DiValentino:2021izs,Bernal:2016gxb,Abdalla:2022yfr,CosmoVerseNetwork:2025alb}. 

However, no consensus has yet been reached on resolving the tension based on current data. Given this discrepancy between the early and late Universe, proposed scenarios can broadly be classified into early-time or late-time solutions.

In early-time solutions, new physics is introduced to reduce the sound horizon, $r_\mathrm{d}$, thereby increasing $H_0$ while preserving the baryon acoustic oscillation (BAO) measurement of $r_\mathrm{d} H_0$. One realization involves modifying expansion history by introducing additional energy components around the epoch of matter-radiation equality, such as early dark energy~\cite{Poulin:2018cxd, Karwal:2016vyq, Mortsell:2018mfj, Kamionkowski:2014zda, Berghaus:2019cls} or extra neutrino species~\cite{Zhang:2014dxk,Zhao:2016ecj,Zhao:2017urm,Feng:2019jqa,Kreisch:2019yzn, Wyman:2013lza, Sakstein:2019fmf, Zhang:2014nta}.
Alternatively, early-time solutions may also involve modifying the recombination history~\cite{Sekiguchi:2020teg, Chiang:2018xpn, Mirpoorian:2024fka} or introducing primordial magnetic fields~\cite{Jedamzik:2020krr, Jedamzik:2025cax, Jedamzik:2013gua}. However, these early-time solutions are in tension with other cosmological observables, such as the fluctuation amplitude $\sigma_8$\footnote{However, using $\sigma_8$ may bias the comparison between models with high $H_0$ and $\Lambda \mathrm{CDM}$, since the reference scale $R=8h^{-1}~\mathrm{Mpc}$ depends on $h$, which changes the interpretation of $\sigma_8$ \cite{Forconi:2025cwp, Sanchez:2020vvb}.} and the integrated Sachs-Wolfe effect, as discussed in the no-go arguments of Refs.~\cite{Hill:2020osr, Krishnan:2020obg, Jedamzik:2020zmd, Vagnozzi:2023nrq, Dutta:2019pio}.
Late-time solutions modify the Hubble parameter $H(z)$ in the late universe, so that it can accommodate the SH0ES $H_0$ result locally while leaving the CMB anisotropy spectrum unchanged, such as the deformation of the dark-energy equation of state \cite{Wang:2016lxa,Agrawal:2019lmo, Li:2019yem,Li:2004rb,Shapiro:2000dz, Zhang:2012uu, Li:2012spm, Zhang:2015rha, Feng:2016djj}, modified gravity \cite{Aluri:2022hzs,Capozziello:2011et,DeFelice:2010aj}, and parameter transitions in the late Universe \cite{Mortonson:2009qq,Kazantzidis:2019nuh}.
Analogous to early-time solutions, various no-go arguments have also been raised against the late-time scenarios. The requirement of consistency between low redshift cosmological distance measurements from BAO and SNe imposes strong constraints on the late-time solutions \cite{Benevento:2020fev,Yang:2021flj}.
Moreover, these solutions cannot reconcile the $H_0$ tension and $S_8$ tension\footnote{The $S_8$ tension is alleviated as indicated by the updated Kilo-Degree Survey cosmic shear constraints with improved redshift distribution estimation and calibration, which found consistency ($0.73\sigma $) with Planck \cite{Wright:2025xka}.}  simultaneously, as Refs.~\cite{Guo:2018ans,Gao:2021xnk,Heisenberg:2022gqk,Dutta:2019pio,Alestas:2021xes} pointed out.

In light of the plethora of proposed cosmological models and corresponding no-go arguments, it is compelling to explore new and independent measurements of $H_0$. For instance, techniques such as strong gravitational lensing time delay \cite{H0LiCOW:2018tyj,DES:2019fny} and gravitational wave standard sirens \cite{Guidorzi:2017ogy,Song:2025ddm,Jin:2025dvf} provide an independent perspective on this tension. We can also employ other independent calibrators to replace Cepheid variables in the distance ladder, such as megamasers \cite{Pesce:2020xfe}, the tip of the red giant branch \cite{Freedman:2021ahq}, surface brightness fluctuations \cite{Blakeslee:2021rqi}, and Mira variables \cite{Huang:2019yhh}.

The inverse distance ladder (IDL) calibrates the SNe via anchored BAO instead of Cepheid variables, providing a model-insensitive method to infer $H_0$ \cite{BOSS:2014hhw,Cuesta:2014asa,Verde:2016ccp,Lemos:2018smw,Yu:2017iju, Gomez-Valent:2018hwc,Haridasu:2018gqm,Gomez-Valent:2023uof}. In the traditional IDL, the BAO data are anchored using an $r_\mathrm{d}$ prior derived from the CMB or big bang nucleosynthesis. Then the data are used to calibrate SNe to give an absolute distance. Finally, the distance is extrapolated to $z=0$ to infer $H_0$ under a fiducial $H(z)$ model \cite{SDSS:2009ocz,Heavens:2014rja,DES:2017txv,eBOSS:2020yzd,DES:2024ywx,DESI:2024mwx,DES:2018rjw}.

It should be noted that the $r_\mathrm{d}$ prior introduces dependence on early-Universe physics \cite{Aylor:2018drw}. The effect of this $r_\mathrm{d}$ prior has been systematically evaluated in Refs.~\cite{Barua:2024gei,Luongo:2024fww}. To mitigate such dependence, the $r_\mathrm{d}$ prior can be replaced with other independent measurements. The cosmic chronometer (CC) measures $H(z)$ directly, rather than through integrals of $H(z)$, such as those involved in lensing distances from strong lensing systems~\cite{Wojtak:2019tkz,Taubenberger:2019qna,Wang:2021kxc,Gong:2024yne}, thus offering higher sensitivity to cosmological parameters~\cite{Moresco:2024wmr,Vagnozzi:2021tjv,Jimenez:2019onw}. CC data have been widely applied as an effective replacement for $r_\mathrm{d}$ in constructing IDL \cite{Guo:2024pkx,Luongo:2024fww,Favale:2023lnp,Cai:2021weh}.

Furthermore, adopting a fiducial $H(z)$ derived from a specific cosmological model in IDL will inevitably introduce model dependence. To avoid this, the traditional approach is to parametrize $H(z)$ via a Taylor expansion in terms of $z$ or $y\equiv1-a=z/(1+z)$, known as the cosmography method \cite{DES:2024ywx,Pourojaghi:2024bxa,Luongo:2024fww,Lemos:2018smw,Cattoen:2007sk,Zhang:2016urt,Chiba:1998tc,Capozziello:2011tj,Aviles:2012ay}. Alternatively, purely data-driven techniques such as the Gaussian process (GP) have been employed to reconstruct $H(z)$ \cite{Guo:2024pkx,Jiang:2024xnu,He:2021rzc,Mukherjee:2024ryz,Yang:2022jkf}. However, both approaches possess intrinsic limitations. The Taylor expansion is problematic at $z \sim 1$ \cite{Banerjee:2020bjq,Cai:2021weh,Cai:2022dkh,Zhang:2016urt}. GP methods are sensitive to the choice of kernel functions and tend to yield
smaller errors at the cost of strongly correlated results \cite{OColgain:2021pyh,Kjerrgren:2021zuo}. To address these limitations, we adopt a global parametrization based on the cosmic age (PAge) \cite{Huang:2020mub,Luo:2020ufj} and its extension, a more accurate parametrization based on cosmic age (MAPAge) \cite{Cai:2022dkh,Huang:2021aku}, to accurately reproduce late-time cosmological behaviors.

The PAge model can be interpreted as an effective description of homogeneous late-time scenarios. The studies \cite{Huang:2020evj, Huang:2021tvo, Li:2022inq} have employed the PAge model to analyze gamma-ray bursts, redshift-space distortions, and quasars. References~\cite{Cai:2021weh, Cai:2022dkh, Huang:2024erq} have derived no-go theorems against late-time solutions through the comparison between the PAge (denoted as late-time scenarios) and $\Lambda \mathrm{CDM}$ models.

Recently, the Dark Energy Spectroscopic Instrument (DESI) Collaboration reported a $3.9\sigma$ hint of dynamical dark energy in the first data release (DR1), which was further increased to $4.2\sigma$ in the second data release (DR2) \cite{DESI:2024mwx,DESI:2025zgx}. This finding has sparked intense discussions about possible new physics and systematic uncertainties \cite{Malekjani:2024bgi,Colgain:2024mtg,Jia:2024wix,Wang:2024pui,Huang:2025som,Cortes:2024lgw,Efstathiou:2024xcq,Abreu:2025zng,Pang:2024qyh,Fikri:2024klc,Feng:2025mlo,Du:2024pai,Li:2024qus,Li:2024qso,Li:2025owk,Du:2025iow,Pang:2025lvh,Wang:2025ljj,Wang:2024dka,DESI:2025fii,Pan:2025qwy,You:2025uon,Pan:2025psn,Silva:2025hxw,Qiang:2025cxp,Cai:2025mas,Cline:2025sbt,Rodrigues:2025tfg,Li:2025ops,Bhattacharjee:2025xeb,Gialamas:2025pwv,Wang:2025dtk,Cline:2025sbt,Li:2025eqh,Li:2025ula,Li:2025dwz,Li:2025htp,Gonzalez-Fuentes:2025lei,Gomez-Valent:2024ejh, Wang:2025znm,Du:2025xes,Feng:2025wbz}. This behavior, which deviates from $\Lambda\mathrm{CDM}$ can be effectively described by the PAge model. This motivates us to propose a PAge-improved IDL to yield fully model-independent cosmological constraints in this work. Our results will provide model-independent perspectives on the $H_0$ tension and the hint of dynamical dark energy. 

The remainder of this paper is organized as follows. In Sec.~\ref{sec2}, we introduce the PAge model and cosmological datasets used in our analysis. We present the cosmological constraints and related discussion in Sec.~\ref{sec3}. The paper concludes with a summary in Sec.~\ref{sec4}.

\section{METHODOLOGY}\label{sec2}

\subsection{The PAge and MAPAge parametrization}\label{sec2.1}

Within the IDL, the PAge (MAPAge) parametrization is used to extrapolate $H(z)$ to $z=0$, thereby determining $H_0$ in a cosmological-model-independent manner. The PAge (MAPAge) rests on two assumptions:

\begin{enumerate}[label=(\roman*)]  
    \item At $z\gg 1$, the Universe is matter dominated, and the brief radiation-dominated era is neglected.\label{assumption1}
    \item In PAge, the product of the Hubble parameter $H$ and the cosmic time $t$, can be approximated by a quadratic function of $t$. In MAPAge, this product can be approximated by a cubic function of $t$. \label{assumption2}
\end{enumerate}

\begin{figure*}[htbp]
	\resizebox{\textwidth}{!}{
		\centering
		\includegraphics[width=\linewidth]{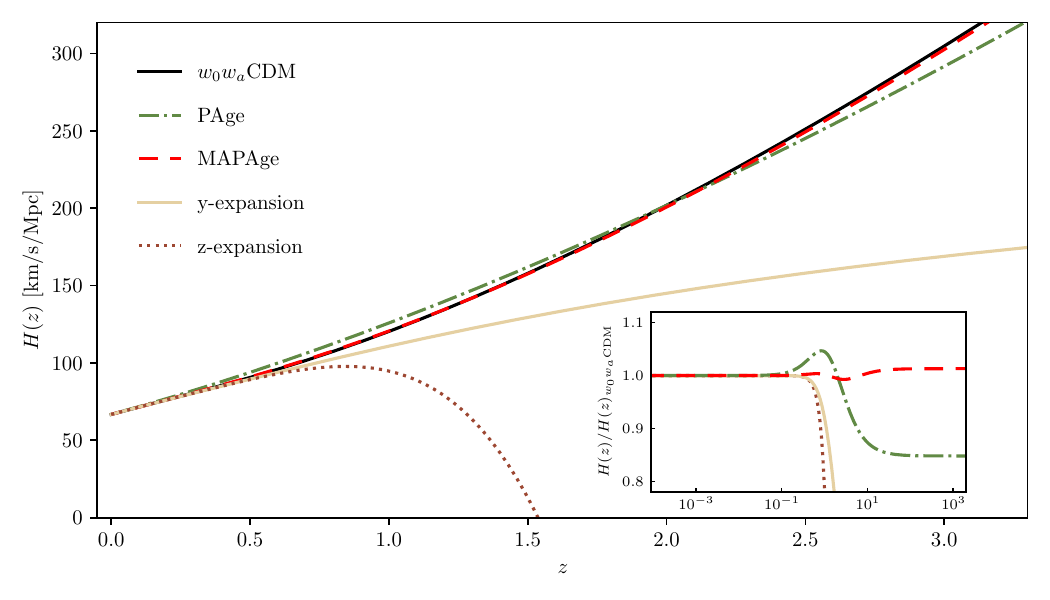}
	}
	\centering
	\caption{Comparison of approximation accuracy for PAge (green dash-dotted), MAPAge (red dashed), Taylor expansion in $z$ (brown dotted), and $y$ (light beige), targeting the fiducial $w_0 w_a \mathrm{CDM}$ (black) model. The lower-right inset figure shows normalized Hubble parameter $H(z)/H(z)_{w_0w_a \mathrm{CDM}}$ up to $z\sim10^3$. We fix the parameters of the $w_0w_a\rm CDM$ model to DESI DR2 results: $\Omega_\mathrm{m}= 0.32, ~H_0=66.74 ~\mathrm{km~s^{-1}~Mpc^{-1}} ,~w_0=-0.75,~ w_a=-0.86$ \cite{DESI:2025zgx}.}
	\label{fig1}
\end{figure*}

These assumptions are independent of any specific cosmological model, such as dark energy models and modified gravity theories \cite{Huang:2020mub,Cai:2021weh,Cai:2022dkh,Huang:2024erq,Wang:2024nsi}. Rather than parametrizing the dark energy equation of state, PAge directly parametrizes the cosmic expansion history, providing a more straightforward probe of that history than equation of state based approaches such as $w_0w_a\mathrm{CDM}$. Compared with traditional parametrizations, such as expansions in $z$ or $y$ \cite{Capozziello:2011tj,Zhang:2016urt,Visser:2003vq}, PAge (MAPAge) can parametrize $H(z)$ over a wider redshift range with higher accuracy. As shown in Fig.~\ref{fig1}, PAge (MAPAge) can approximate the target $w_0w_a\mathrm{CDM}$ model up to $z \sim 10^3$, whereas $z$ and $y$ expansion develop deviation at $z>1$ \cite{Hu:2022udt,Cai:2021weh}. Moreover, for target models where $H(z)$ evolves rapidly, a Taylor expansion in terms of $z$ or $y$ requires high-order terms to accurately capture the behavior. However, since $t$ is defined through the integral of $H(z)$, it evolves more smoothly. As a result, even pronounced changes in $H(z)$ can typically be captured by a low-order expansion in $t$. Thus, the PAge (MAPAge) model serves as a global, model-independent parametrization that can be effectively adopted in IDL.

In PAge, we first rewrite $H(t)t$ as
\begin{equation}
    H(t)t=H(t)(t-t_0)+H(t)t_0,
    \label{1}
\end{equation}
where $t_0$ denotes the current cosmic age. Then we perform a Taylor expansion of $H(t)$ around $t_0$,
\begin{equation}
        H(t)=H_0 +\left. \dot{H} \, \right|_{t = t_0}(t - t_0)   + \ldots.
        \label{2-2}
    \end{equation}
Substituting Eq.~(\ref{2-2}) into Eq.~(\ref{1}) yields
\begin{align}
H(t)t=&H_0t_0+H_0(t-t_0)(1-H_0t_0-H_0t_{0}q_0 )\nonumber\\
&-H^2_0(t-t_0)^2(1+q_0-\frac{H_0t_0}{2}(2+3q_0+j_0))+\dots.
\label{eqeq1}
\end{align}
The deceleration and jerk parameters, $q_0$ and $j_0$, arise from the Taylor expansion of $H(t)$ in Eq.~(\ref{2-2}). As required by assumption~\ref{assumption2}, we keep terms up to quadratic order in Eq.~(\ref{eqeq1}). To meet the assumption~\ref{assumption1}, $ H(t)t = 2/3$ in the limit $t\to0$, we introduce the following modification of our expansion,
\begin{equation}
     \lim\limits_{t\to0}H(t)t=H_0t_0+H_0(0-t_0)(1-H_0t_0-H_0t_0q_0)-AH^2_0t^2_0,
      \label{new eq2}
\end{equation}
where $A$ is an undetermined parameter. The limit $ \lim\limits_{t\to0} H(t)t = 2/3$ is met only if $A=1+q_0-\frac{2}{3H_0^2t^2_0}$. Therefore, the final expansion becomes
\begin{align}
       H(t)t=&H_0t_0+H_0(t-t_0)(1-H_0t_0-H_0t_{0}q_0 )\nonumber\\
       &-H^2_0(t-t_0)^2(1+q_0-\frac{2}{3H_0^2t^2_0}). 
\end{align}
After setting 
\begin{equation}
    p_\mathrm{age}=H_0t_0,~ \eta=1-\frac{3}{2}p^2_\mathrm{age}(1+q_0),
    \label{4}
\end{equation}the final expansion is written as
\begin{equation}
        \frac{H(t)}{H_0}=1+\frac{2}{3}(1-\eta \frac{H_0t}{p_{\mathrm{age}}})(\frac{1}{H_0t}-\frac{1}{p_\mathrm{age}}),
         \label{3}
\end{equation}where $\eta$ quantifies the extent to which the real universe deviates from the pure matter universe with $\eta=0$ in Eq.~(\ref{3}). To obtain $z(t)$, we can replace $H(t)$ in Eq.~(\ref{3}) with its definition $H(t)=-\mathrm{d}z/\mathrm{d}t/(1+z)$, and integrate both sides,
\begin{equation}
   1+z=\left(\frac{p_{\mathrm{age}}}{H_0 t}\right)^{2/3}
\mathrm{e}^{\frac{1}{3}\left(1-\frac{H_0 t}{p_{\mathrm{age}}}\right)
\left(3p_{\mathrm{age}}+\eta \frac{H_0 t}{p_{\mathrm{age}}}-\eta-2\right)}.
\label{5}
\end{equation}

Again, in Eq.~(\ref{5}), $\eta$ quantify the deviation from the matter-dominated Universe $1+z=(t_0/{t})^{\frac{2}{3}}$. 

Similarly, we can parametrize $H(t)t$ with cubic function of $t$, which is referred to as MAPAge in Ref.~\cite{Huang:2021aku}. It takes the following forms:
\begin{align}\label{eq:MAPAge}
\frac{H}{H_0}=1+\frac23&\left[1-(\eta_1+\eta_2)\frac{H_0t}{p_\mathrm{age}}+\eta_2\left(\frac{H_0t}{p_\mathrm{age}}\right)^2\right]\nonumber\\
&\times\left(\frac{1}{H_0t}-\frac{1}{p_\mathrm{age}}\right),
\end{align}
where $\eta_1$ equals to $\eta$ in PAge. $\eta_2$ is introduced to further parametrize $H(t)t$ and is given by
\begin{equation}
  \eta_2=1-\frac{3}{4}p^2_{\mathrm{age}}(2+j_0+3q_0), 
\label{5-5}
\end{equation}

The PAge model can map specific dark energy models into a two-parameter space $(p_\mathrm{age},~\eta)$. Thus, the deviation from the $\Lambda \mathrm{CDM}$ model is clearly illustrated in the $(p_\mathrm{age},~\eta)$ parameter space. For simplicity, the mapping procedure is carried out by equating the deceleration parameter $q_0$ between the PAge model and the specific cosmological model, followed by computing $p_{\mathrm{age}} = H_0 t_0$ for the reference model. Subsequently, $\eta$ can be derived from Eq.~(\ref{4}). More details about model mapping are provided in Ref.~\cite{Luo:2020ufj}. 

In the following, we explicitly demonstrate how to map the specific dark energy models discussed in this paper into $(p_\mathrm{age},~\eta)$, including $\Lambda \mathrm{CDM}$, $w \mathrm{CDM}$, and $w_0w_a \mathrm{CDM}$.

For $\Lambda \mathrm{CDM}$ with $H(a)=H_0\sqrt{\Omega_{\mathrm{m}}a^{-3}+(1-\Omega_{\mathrm{m}})}$ when mapped to the PAge model, the deceleration parameter $q_0$ is
\begin{equation}
    q_0=-1-\frac{\dot{H}}{H^2_0}=-1+\frac{3}{2}\Omega_{\mathrm{m}}
    \label{6},
\end{equation}
and $ p_{\mathrm{age}}$ reads
\begin{equation}
    p_{\mathrm{age}}=\int^1_0 \frac{\mathrm{d}a}{a\sqrt{\Omega_{\mathrm{m}}a^{-3}+(1-\Omega_{\mathrm{m}})}}.
    \label{7}
\end{equation}

The $w_0w_a \mathrm{CDM}$ model can be mapped in the same manner as
\begin{equation}
    q_0=-1+\frac{3}{2}\Omega_{\mathrm{m}}+\frac{3(1+w_0)}{2}(1-\Omega_{\mathrm{m}}),
    \label{8}
\end{equation}
and the corresponding $p_{\mathrm{age}}$ takes

\begin{equation}
p_{\mathrm{age}}=
\int_0^1 \frac{\mathrm{d}a}{a\,\sqrt{
\Omega_{\mathrm{m}}\,a^{-3} + (1 - \Omega_{\mathrm{m}})
a^{-3(1+w_0+w_a)}
e^{3w_a(1 - a)}
}}.
\label{9}
\end{equation}

Setting $w_a=0$ and $w_0=w$, the results are applicable to the $w\mathrm{CDM}$ model.

\subsection{Cosmological datasets}\label{sec2.3}
In this section, we present an overview of various combinations of cosmological data used to construct the IDL, including SNe data from PantheonPlus and DESY5, BAO measurements from DESI DR2 and Sloan Digital Sky Survey (SDSS), and CC data.

\subsubsection{SNe}
    SNe are known as the standard candles due to their uniform intrinsic luminosity. The SNe data records the apparent magnitude $m_{\mathrm{B}}(z)$. The luminosity distance $D_{\mathrm{L}}(z)$ can be derived from the magnitude--redshift relation
    \begin{equation}
        m_{\mathrm{B}}(z)= 5~\mathrm{lg}\frac{D_{\mathrm{L}}(z)}{1 0~\mathrm{pc}}+M_{\mathrm{B}}=5~\mathrm{lg}\frac{D_{\mathrm{L}}(z)H_0}{c}-5a_{\mathrm{B}},
        \label{10}
    \end{equation}
    where $M_{\mathrm{B}}$ is the absolute magnitude of SNe. $a_{\mathrm{B}}\equiv-\frac{1}{5}[5~\mathrm{lg}(c/10~\mathrm{pc}/ H_0)+M_{\mathrm{B}}]$ is the intercept of the magnitude--redshift relation Eq.~(\ref{10}). It is evident that $H_0$ and $M_{\mathrm{B}}$ are degenerate in $a_{\mathrm{B}}$, making it impossible to constrain both $H_0$ and $M_{\mathrm{B}}$ with SNe data alone. In IDL, we incorporate BAO and CC (described below) to break this degeneracy, avoiding the dependence on external prior.

        The chi-squared statistic for SNe is defined as follows
    \begin{equation}
        \chi^2_{\mathrm{SNe}}=\Delta \mathbf{D}^{\mathrm{T}}\mathrm{C}^{-1}_{\mathrm{stat+sys}}\Delta \mathbf{D},
    \end{equation}
 where $\Delta {D}  _i=\mu_i-\mu_{\mathrm{model}}(z_i)$ is the distance moduli residual vector. Distance modulus is defined as $\mu\equiv m_{\mathrm{B}}-M_{\mathrm{B}}=5\mathrm{lg}(D_{\mathrm{L}}/1 0~\mathrm{pc})$. $\mathrm{C}^{-1}_{\mathrm{stat+sys}}$ denotes the inverse combined statistical and systematic covariance matrix of the SNe sample.
 
We use two different SNe datasets, including PantheonPlus\footnote{PantheonPlus data available at \url{https://github.com/PantheonPlusSH0ES/DataRelease}} and DESY5\footnote{DESY5 data available at \url{https://github.com/des-science/DES-SN5YR}}. 

The PantheonPlus SNe compilation consists of 1701 light curves of 1550 spectroscopically confirmed SNe Ia in the redshift range $0.001<z<2.26$ \cite{Scolnic:2021amr}. To reduce the effect of strong peculiar velocity dependence at low redshift \cite{Brout:2022vxf}, only 1550 data points in the range $0.01<z<2.26$ are used.  

The DESY5 SNe sample is the largest and deepest single sample survey to date consisting of 1635 SNe Ia \cite{DES:2024jxu, DES:2024hip, DES:2024upw} distributed in $0.10<z<1.3$ complemented by 194 low-redshift SNe Ia spanning $0.025<z<0.1$ \cite{Hicken:2009df, Hicken:2012zr, Krisciunas:2017yoe,Foley:2017zdq}. 
The DESY5 data provide the Hubble diagram redshift, corrected apparent magnitudes $m^\mathrm{corr.}_{\mathrm{B}}(z)$, distance moduli $\mu(z)$, and the corresponding distance moduli errors $\sigma_{ij}$ and related quantities. Since the distance moduli contain no information about $M_{\mathrm{B}}$, we follow Refs.~\cite{Huang:2025som, DES:2024ywx} to use the corrected apparent magnitudes to construct the distance moduli residual as $\Delta \mathbf{D}= m_{\mathrm{B}}^{\mathrm{corr.}}-M_{\mathrm{B}}-5\mathrm{lg}({D_{\mathrm{L}}(z)}/{1 0~\mathrm{pc}})$ to constrain $M_{\mathrm{B}}$.

\subsubsection{BAO}
The BAO records the imprint of early-Universe plasma fluctuations in the late-time large-scale structure. BAO data provide three types of distances with respect to a fiducial cosmology, measured along different directions. Along the line-of-sight direction, BAO measures the Hubble distance,
\begin{equation}
    \frac{D_\mathrm{H}(z)}{r_\mathrm{d}}=\frac{c}{r_\mathrm{d}H(z)},
\end{equation}
and along the transverse direction, BAO measures the (comoving) angular diameter distance,
\begin{equation}
    \frac{D_\mathrm{M}(z)}{r_\mathrm{d}}=  \frac{(1+z)D_{\mathrm{A}}(z)}{r_\mathrm{d}}=\frac{D_{\mathrm{L}}(z)}{(1+z)r_\mathrm{d}}.
\end{equation}
Taking the spherical average of both directions, BAO measures
\begin{equation}
    \frac{D_\mathrm{V}(z)}{r_\mathrm{d}}
    =\frac{[zD_\mathrm{M}(z)^2D_\mathrm{H}(z)]^{1/3}}{r_\mathrm{d}}
    ,
\end{equation}
where $r_\mathrm{d}$ is treated as a free parameter in our analysis to eliminate the model dependence introduced by the $r_\mathrm{d}$ prior in traditional IDL.

The chi-squared statistic for BAO takes
\begin{equation}
         \chi^2_{\mathrm{BAO}}= \mathbf{\Delta}^{\mathrm{T}}\mathrm{C}^{-1}_{\mathrm{BAO}} \mathbf{\Delta},   
\end{equation}
where $\mathbf{\Delta}$ denotes the residual vector between the BAO data and the corresponding model values. $\mathrm{C}^{-1}_{\mathrm{BAO}}$ is the inverse covariance matrix of the BAO data.

Two BAO datasets from DESI\footnote{DESI DR2 data available at \url{https://github.com/CobayaSampler/bao_data/tree/master/desi_bao_dr2}.} and SDSS\footnote{SDSS data available at \url{https://svn.sdss.org/public/data/eboss/DR16cosmo/tags/v1_0_0/likelihoods/BAO-plus/}} are used to delve into the role that BAO plays in the cosmological constraint, and these datasets are listed in Table \ref{table:Combined}. 

We adopt the DESI DR2 BAO measurements which include $D_{\mathrm{M}}(z)/ r_{\mathrm{d}},~D_{\mathrm{H}}(z)/r_{\mathrm{d}},~D_\mathrm{V}(z)/r_{\mathrm{d}} $ extracted from large scale structure traces across multiple distinct redshift bins \cite{DESI:2025zgx}. 

The SDSS BAO data are employed in this paper to cross-check our findings and connect with the existing literature. It includes two low-redshift data points 6dFGS \cite{Beutler:2011hx}, SDSS DR7 MGS \cite{Ross:2014qpa}, and SDSS DR16 data \cite{eBOSS:2020yzd}.

\begin{table*}[htbp]
\setlength\tabcolsep{15pt}
\renewcommand{\arraystretch}{1.0}
\centering
\caption{The BAO samples utilized in this work including DESI DR2 data and SDSS data.}
\label{table:Combined}
\begin{tabular}{lcccc}
\toprule[0.8pt]

Tracer & \({z_{\mathrm{eff}}}\) & \({D_{\mathrm{M}}/r_{\mathrm{d}}}\) & \({D_{\mathrm{H}}/r_{\mathrm{d}}}\) & \({D_\mathrm{V}/r_{\mathrm{d}}} 
 \text{ or } {r_{\mathrm{d}}/D_\mathrm{V}} \)\\

\midrule[0.4pt]
\multicolumn{5}{l}{\textbf{DESI DR2}}\\
BGS             & 0.30  & $\cdots$              & $\cdots$                      & 7.94 \(\pm\) 0.08 \\
LRG1             & 0.51  & 13.59 \(\pm\) 0.17     & 21.86 \(\pm\) 0.43     & $\cdots$  \\
LRG2             & 0.71  & 17.35 \(\pm\) 0.18     & 19.46 \(\pm\) 0.33     & $\cdots$  \\
LRG3 + ELG1       & 0.93  & 21.58 \(\pm\) 0.15     & 17.64 \(\pm\) 0.19     & $\cdots$  \\
ELG2             & 1.32  & 27.60 \(\pm\) 0.32     & 14.18 \(\pm\) 0.22     & $\cdots$  \\
QSO             & 1.48  & 30.51  \(\pm\) 0.76                    & 12.82  \(\pm\) 0.52             & $\cdots$   \\
Ly\(\alpha\) & 2.33  & 38.99 \(\pm\) 0.94     & 8.63 \(\pm\) 0.10      & $\cdots$   \\
\midrule[0.4pt]
\multicolumn{5}{l}{\textbf{SDSS}}\\
6dFGS           & 0.106 & $\cdots$            & $\cdots$                        & 0.336 \(\pm\) 0.015 \\
MGS             & 0.15  & $\cdots$            & $\cdots$               & 4.47 \(\pm\) 0.17 \\
LRG             & 0.70  & 17.86 \(\pm\) 0.33     & 19.33 \(\pm\) 0.53     & $\cdots$   \\
ELG             & 0.85  & $\cdots$              & $\cdots$                & \(18.33 ^{+0.57}_{-0.62}\) \\
QS0             & 1.48  & 30.69 \(\pm\) 0.80     & 13.26 \(\pm\) 0.55     & $\cdots$   \\
Ly\(\alpha\)    & 2.33  & 37.6 \(\pm\) 1.9       & 8.93 \(\pm\) 0.28      & $\cdots$   \\
Ly\(\alpha\) QSO& 2.33  & 37.3 \(\pm\) 1.7       & 9.08 \(\pm\) 0.34      & $\cdots$   \\

\bottomrule[0.8pt]

\end{tabular}
\end{table*}

\begin{table}[htbp]
\setlength\tabcolsep{5pt}
\renewcommand{\arraystretch}{1.0}
\centering
\caption{32 CC data points with $1\sigma$ error.}
\label{tab:3-line}
\begin{tabular}{lcc}
\hline
\toprule[0.8pt]

${z}$ & ${H(z)~\mathrm{[km~s^{-1}~Mpc^{-1}}]} $ & Ref.\\
\midrule[0.4pt]
0.09    & 69 $\pm$ 12      & \cite{Jimenez:2003iv} \\
0.17    & 83 $\pm$ 8       & \cite{Simon:2004tf} \\
0.27    & 77 $\pm$ 14      & \cite{Simon:2004tf} \\
0.4     & 95 $\pm$ 17      & \cite{Simon:2004tf} \\
0.9     & 117 $\pm$ 23     & \cite{Simon:2004tf} \\
1.3     & 168 $\pm$ 17     & \cite{Simon:2004tf} \\
1.43    & 177 $\pm$ 18     & \cite{Simon:2004tf} \\
1.53    & 140 $\pm$ 14     & \cite{Simon:2004tf} \\
1.75    & 202 $\pm$ 40     & \cite{Simon:2004tf} \\
0.48    & 97 $\pm$ 62      & \cite{Stern:2009ep} \\
0.88    & 90 $\pm$ 40      &\cite{Stern:2009ep} \\
0.1791  & 75 $\pm$ 4       & \cite{Moresco:2012jh} \\
0.1993  & 75 $\pm$ 5       & \cite{Moresco:2012jh} \\
0.3519  & 83 $\pm$ 14      & \cite{Moresco:2012jh} \\
0.5929  & 104 $\pm$ 13     & \cite{Moresco:2012jh} \\
0.6797  & 92 $\pm$ 8       & \cite{Moresco:2012jh} \\
0.7812  & 105 $\pm$ 12     & \cite{Moresco:2012jh} \\
0.8754  & 125 $\pm$ 17     & \cite{Moresco:2012jh} \\
1.037   & 154 $\pm$ 20     & \cite{Moresco:2012jh} \\
0.07    & 69 $\pm$ 19.6    & \cite{Zhang:2012mp}\\
0.12    & 68.6 $\pm$ 26.2  & \cite{Zhang:2012mp} \\
0.2     & 72.9 $\pm$ 29.6  & \cite{Zhang:2012mp} \\
0.28    & 88.8 $\pm$ 36.6  & \cite{Zhang:2012mp}\\
1.363   & 160 $\pm$ 36.6   & \cite{Moresco:2015cya} \\
1.965   & 186.5 $\pm$ 50.4 & \cite{Moresco:2015cya}\\
0.3802  & 83 $\pm$ 13.5    & \cite{Moresco:2016mzx} \\
0.4004  & 77 $\pm$ 10.2    & \cite{Moresco:2016mzx} \\
0.4247  & 87.1 $\pm$ 11.2  & \cite{Moresco:2016mzx} \\
0.4497  & 92.8 $\pm$ 12.9  & \cite{Moresco:2016mzx} \\
0.4783  & 80.9 $\pm$ 9     & \cite{Moresco:2016mzx} \\
0.47    & 89 $\pm$ 49.6    & \cite{Ratsimbazafy:2017vga} \\
0.80    & 113.1 $\pm$ 28.5 & \cite{Jiao:2022aep} \\
\bottomrule[0.8pt]
\end{tabular}
\label{table:CC}
\end{table}

\subsubsection{CC}
CC measurements are incorporated to break the $r_\mathrm{d}$ and $H_0$ degeneracy in BAO data, serving a role analogous to the $r_\mathrm{d}$ prior in traditional IDL. It directly measures the Hubble parameter through massive passively evolving galaxies with old stellar populations and very low star formation rates \cite{Moresco:2023zys}. By comparing two galaxies formed at the same time but separated by a small redshift interval, the Hubble parameter $H(z)$ can be calculated using the differential age method \cite{Jimenez:2001gg}, without assuming any cosmological model,
\begin{equation}
    H(z)=-\frac{1}{1+z}\frac{\mathrm{d}z}{\mathrm{d}t},
\end{equation}
where ${\mathrm{d}z}/{\mathrm{d}t}$ is the derivative of redshift with respect to the look-back time. Its nature as a direct measurement of $H_0$ makes it highly sensitive to cosmological parameters.

As CC data are not entirely independent, we incorporate correlations through the use of a covariance matrix when available. We adopt 32 CC data points, listed in Table \ref{table:CC}, including 15 correlated measurements with the corresponding covariance matrix compiled in Ref.~\cite{Moresco:2020fbm} and 17 uncorrelated ones.   For the correlated measurements, we adopt the covariance matrix as described in Refs.~\cite{Moresco:2020fbm, Moresco:2022phi},
\begin{equation}
    \mathrm{C}_{ij}=\mathrm{C}^{\mathrm{stat}}_{ij}+\mathrm{C}^{\mathrm{sys}}_{ij},
\end{equation}
where $\mathrm{C}^{\mathrm{stat}}_{ij}$ represents the diagonal covariance matrix containing statistical uncertainties, $\mathrm{C}^{\mathrm{sys}}_{ij}$ accounts for systematic effects, including those from stellar population models, metallicity assumptions, and potential contamination from residual young stellar components.

In this work, we adopt the recommended covariance matrix combination from \cite{mmorescoCCcovariance},
\begin{equation}
\mathrm{C}^{\mathrm{tot}}_{ij}=\mathrm{C}^{\mathrm{stat}}_{ij}+\mathrm{C}_{ij}^{\mathrm{met}}+\mathrm{C}_{ij}^{\mathrm{sps,ooo}}+\mathrm{C}_{ij}^{\mathrm{IMF}},
\end{equation}
where $\mathrm{C}_{ij}^{\mathrm{met}}$ denotes the contribution from uncertainties in stellar metallicity estimates. $\mathrm{C}_{ij}^{\mathrm{sps,ooo}}$ accounts for uncertainties arising from the odd one out stellar population synthesis (SPS) model. $\mathrm{C}_{ij}^{\mathrm{IMF}}$ reflects uncertainties due to the adopted initial mass function (IMF).
For a detailed discussion of CC data, we refer the reader to Ref.~\cite{Moresco:2020fbm}. 

For data points not discussed in Ref.~\cite{Moresco:2020fbm}, only the diagonal components are considered. The final $\chi^2$ is
\begin{equation}
\chi_{\mathrm{CC}}^{2}=\sum_{i=1}^{17}\left[\frac{H_{\mathrm{obs}}\left(z_{i}\right)-H_{\mathrm{model}}\left(z_{i}\right)}{\sigma_{H,_{i}}}\right]^{2}+\Delta \mathbf{H}^{\rm T} \mathrm{C}_{\mathrm{tot}}^{-1} \Delta \mathbf{H},
\end{equation}
where $\sigma_{\mathrm{H}}$ are the diagonal components mentioned above, and $\Delta \mathbf{H} $
is the residual vector between the CC data and the model values.

We consider four datasets, namely DESY5 + DESI DR2 + CC, DESY5 + SDSS + CC, PantheonPlus + DESI DR2 + CC, and PantheonPlus + SDSS + CC. The final combined likelihood, $-2\mathrm{ln}\mathcal{L}=\chi^2_{\mathrm{SNe}}+\chi^2_{\mathrm{BAO}}+\chi^2_{\mathrm{CC}}$, is used to constrain the PAge model with five free parameters $\{ \eta,~p_{\mathrm{age}},~M_{\mathrm{B}},~H_0,~r_\mathrm{d}\}$. Then we employ the Markov Chain Monte Carlo (MCMC) sampler \texttt{emcee}\footnote{\url{https://github.com/dfm/emcee}} \cite{Foreman-Mackey:2012any} to explore the parameter space and obtain best-fit parameters through $\chi^2$ minimization.
For comparison, we also consider the $\Lambda \mathrm{CDM}$ model, with four free parameters $\{ \Omega_\mathrm{m},~M_{\mathrm{B}},~H_0,~r_\mathrm{d} \}$.
Uniform priors are adopted for the following parameter ranges $ \eta\in[-2,2],~p_{\mathrm{age}}\in[0.15,2],~M_{\mathrm{B}}\in[-21,-19],~H_0\in[60,80]~\mathrm{km~s^{-1}~Mpc^{-1}},~r_\mathrm{d}\in [120,180]~\mathrm{Mpc},~\Omega_{\mathrm{m}}\in [0.2,0.4]$.
Posterior samples are analyzed and visualized using \texttt{GetDist}\footnote{\url{https://github.com/cmbant/getdist}}  \cite{Lewis:2019xzd}. For model comparison, we use the \texttt{MCEvidence}\footnote{\url{https://github.com/yabebalFantaye/MCEvidence}}
 algorithm as described in Ref.~\cite{Heavens:2017afc} to compute the Bayesian evidence directly from the MCMC chain.

\begin{table*}[htbp]
\setlength\tabcolsep{3pt}
\renewcommand{\arraystretch}{1.5}
\centering
\caption{Cosmological constraints on the PAge and $\Lambda\mathrm{CDM}$ models based on the DESY5+DESI DR2+CC, DESY5+SDSS+CC, PantheonPlus+DESI DR2+CC, and PantheonPlus+SDSS+CC datasets. Here, $H_0$ and $r_\mathrm{d}$ are in units of $\mathrm{km~s^{-1}~Mpc^{-1}}$ and $\mathrm{Mpc}$.}
\label{tab:PAgeLCDM}
\begin{tabular}{lcccccc}
\toprule[0.8pt]
 Dataset & \(\eta\) & \(p_{\mathrm{age}}\) & \(\Omega_{\mathrm{m}}\) & \(M_{\mathrm{B}}\) & \(H_0\) & \(r_{\mathrm{d}}\) \\
 \midrule[0.4pt]
\multicolumn{6}{l}{\textbf{PAge}}\\
DESY5+DESI DR2+CC           & 0.23 \(\pm\) 0.04 & 0.94 \(\pm\) 0.01 & $\cdots$            &  \(-19.36\pm\) 0.07 & 67.91 \(\pm\) 2.33 &$ 145.0 ^{+4.5}_{-5.2}$ \\
DESY5+SDSS+CC        & 0.16 \(\pm\) 0.06 & 0.95 \(\pm\) 0.01 & $\cdots$             &  \(-19.35\pm\) 0.07 & 67.74 \(\pm\) 2.36 & $143.5 ^{+4.6}_{-5.3}$ \\
PantheonPlus+DESI DR2+CC    & 0.31 \(\pm\) 0.06 & 0.96 \(\pm\) 0.01 & $\cdots$              &  \(-19.38\pm\) 0.07 & 68.97 \(\pm\) 2.42 & $145.0 ^{+4.5}_{-5.2}$ \\
PantheonPlus+SDSS+CC & 0.25 \(\pm\) 0.07 & 0.96 \(\pm\) 0.02 & $\cdots$              &  \(-19.39\pm\) 0.08 & 68.58 \(\pm\) 2.49 & $143.4^{+4.6}_{-5.3} $ \\
\midrule[0.4pt]
\multicolumn{6}{l}{$\bm {\Lambda \mathrm{CDM}}$}\\
DESY5+DESI DR2+CC          & $\cdots$               & $\cdots$                & 0.32 \(\pm\) 0.01 &  \(-19.35\pm\) 0.07 & 68.99 \(\pm\) 2.35 & $145.2 ^{+4.5}_{-5.1}$ \\
DESY5+SDSS+CC       & $\cdots$           & $\cdots$               & 0.35 \(\pm\) 0.01 &  \(-19.38\pm\) 0.07 & 67.46 \(\pm\) 2.37 & $143.4^{+4.6}_{-5.3}$   \\
PantheonPlus+DESI DR2+CC    & $\cdots$               & $\cdots$               & 0.30 \(\pm\) 0.01 &  \(-19.37\pm\) 0.07 & 69.75 \(\pm\) 2.39 & $145.0^{+4.5}_{-5.2}$ \\
PantheonPlus+SDSS+CC & $\cdots$              & $\cdots$               & 0.32 \(\pm\) 0.02 &  \(-19.39\pm\) 0.08 & 69.08 \(\pm\) 2.51 & $143.2^{+4.6}_{-5.3}$ \\
\bottomrule[0.8pt]
\end{tabular}
\end{table*}

\section{RESULTS AND DISCUSSION}\label{sec3}

\begin{figure*}[htbp]
	\resizebox{\textwidth}{!}{
		\centering
		\includegraphics[width=\linewidth]{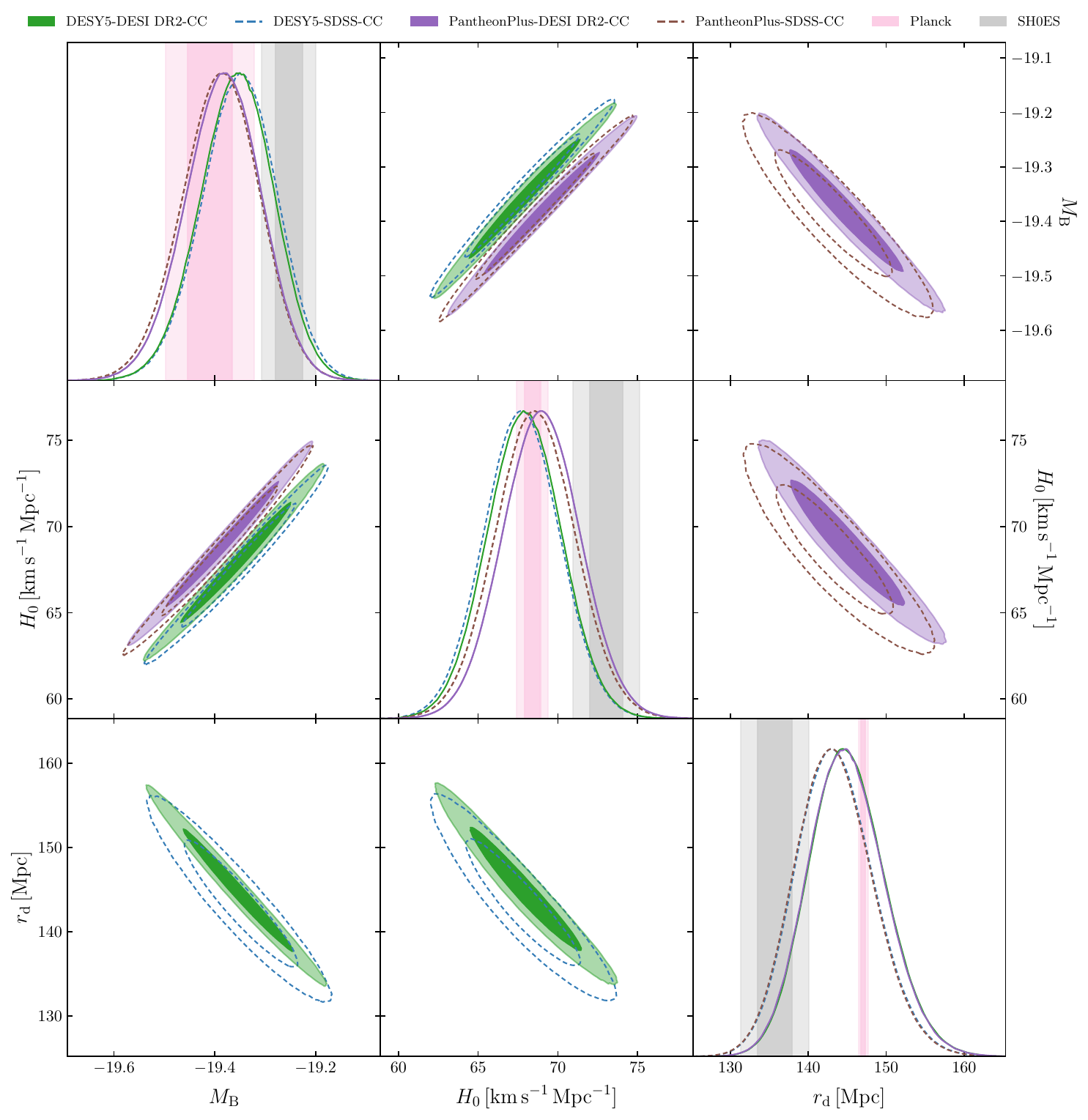}
	}
	\centering
	\caption{Constraints on the PAge model from DESY5+DESI DR2+CC (green), DESY5+SDSS+CC (blue, dotted), PantheonPlus+DESI DR2+CC (purple), and PantheonPlus+SDSS+CC (brown, dotted). Planck 2018 (pink) and SH0ES (grey) benchmarks are included to illustrate the $H_0$ tension both from the direct $H_0$ constraints and from those inferred via $M_\mathrm{B}$ and $r_\mathrm{d}$. The SH0ES-equivalent $r_\mathrm{d}$ and Planck-equivalent $M_\mathrm{B}$ are obtained by equating $H_0r_\mathrm{d}$
    and $a_\mathrm{B}$, respectively. Four datasets are presented in the $M_{\mathrm{B}}$–$H_0$ plane to illustrate the shift in the intercept of the magnitude-redshift relation vividly.}
	\label{Fig: Main-corner}
\end{figure*}

Our primary goal is to derive a model-independent estimate of $H_0$, as well as other cosmological parameters $M_{\mathrm{B}}$ and $r_{\mathrm{d}}$, using the PAge-improved IDL. The summary of the full cosmological results is presented in Table~\ref{tab:PAgeLCDM}, along with the corresponding $\Lambda \mathrm{CDM}$ result for comparison. The one- and two-dimensional posterior distributions of the cosmological parameters in the PAge model are presented in Fig.~\ref{Fig: Main-corner}. For the DESY5+DESI DR2+CC datasets, the PAge model yields \begin{equation}
    H_0 = 67.91 \pm 2.33~\mathrm{km~s^{-1}~Mpc^{-1}},
\end{equation}
at the $1\sigma$ confidence level. Our estimate agrees with the Planck 2018 result of $H_0=67.4\pm0.5~\mathrm{km~s^{-1}~Mpc^{-1}}$ \cite{Planck:2018vyg}. However, a mild tension persists at the $2.0\sigma$ level in comparison with the local SH0ES measurement of $H_0=73.04 \pm 1.04~\mathrm{km~s^{-1}~Mpc^{-1}}$ \cite{Riess:2021jrx}. Other datasets exhibit the same agreement with the Planck result and a similar tension level with the SH0ES measurements. Specifically, DESY5+SDSS+CC datasets show a \(2.1 \sigma\) discrepancy, while PantheonPlus+DESI DR2+CC and PantheonPlus+SDSS+CC datasets exhibit \(1.6\sigma\) and \(1.7 \sigma\) discrepancies, respectively. Our results show larger uncertainty than other traditional IDLs \cite{DES:2024ywx,BOSS:2014hhw,DES:2018rjw} due to the substitution of the Gaussian $r_{\mathrm{d}}$ prior with the more uncertain CC measurements. 

From DESY5+DESI DR2+CC, we obtain $M_{\mathrm{B}}=-19.36\pm0.07$, consistent with the Gaussian processes estimate $M_\mathrm{B}=-19.31^{+0.09}_{-0.11}$ \cite{Favale:2023lnp}, with $M_\mathrm{B}=-19.303$ derived from tip of the red giant branch with full sample in Hubble Space Telescope or James Webb Space Telescope \cite{Li:2025lfp}, and $1.43 \sigma$ differing from the SH0ES result $M_{\mathrm{B}}=-19.253\pm0.027$ \cite{Riess:2021jrx}. For sound horizon, we find $r_{\mathrm{d}}=145.0^{+4.5}_{-5.2}~\mathrm{Mpc}$, in agreement with Planck result $r_\mathrm{d}=147.05\pm0.30~\rm Mpc$ and $ r_\mathrm{d}=146.0^{+4.2}_{-5.1},~141.9^{+5.6}_{-4.9}~\mathrm{Mpc}$ reported in Refs.~\cite{Gomez-Valent:2021hda, Favale:2023lnp}. As shown in Fig.~\ref{Fig: Main-corner}, despite the constraints on $M_\mathrm{B}$ and $r_\mathrm{d}$ being broad and remain $\sim 1\sigma$ consistent with Planck and SH0ES, the value of $H_0$ differs from SH0ES by $2.0\sigma$. This higher discrepancy stems from $a_\mathrm{B}$ offset between DESY5 and PantheonPlus and difference in $H_0r_\mathrm{d}$ inferred from DESI and SDSS. As reported in Ref.~\cite{Efstathiou:2024xcq}, DESY5 and PantheonPlus  exhibit an offset of $\sim0.04$ magnitude in their overlap. Moreover, Ref.~\cite{Huang:2025som} finds that $-5a_\mathrm{B}$ for DESY5 is higher than that for PantheonPlus. From Eq.~(\ref{10}), at fixed $M_\mathrm{B}$, a larger value of $-5a_{\mathrm{B}}$ implies smaller $H_0$, thereby increasing the tension with SH0ES. As shown in the $H_0-M_\mathrm{B}$ plane of Fig.~\ref{Fig: Main-corner}, holding $M_\mathrm{B}$ fixed, DESY5 favors a lower value of $H_0$, consistent with Refs.~\cite{Colgain:2024mtg,Notari:2024rti,DESI:2024mwx}. Note that the value of $H_0r_\mathrm{d}$ inferred from DESI is higher than those from SDSS~\cite{Mukherjee:2024ryz, DESI:2024mwx}, which would, for fixed $r_\mathrm{d}$, raise the inferred $H_0$. Nevertheless, this effect is smaller than the effect of SNe as shown in $H_0-M_\mathrm{B}$ and $H_0-r_\mathrm{d}$ planes across four datasets in Fig.~\ref{Fig: Main-corner}. Consequently, although \(M_\mathrm{B}\) is within \(1.4\sigma\) of SH0ES, the inferred \(H_0\) differs from the SH0ES value by \(2.0\sigma\).

\begin{figure*}[!t]
	\resizebox{\textwidth}{!}{
		\centering
		\includegraphics[width=\linewidth]{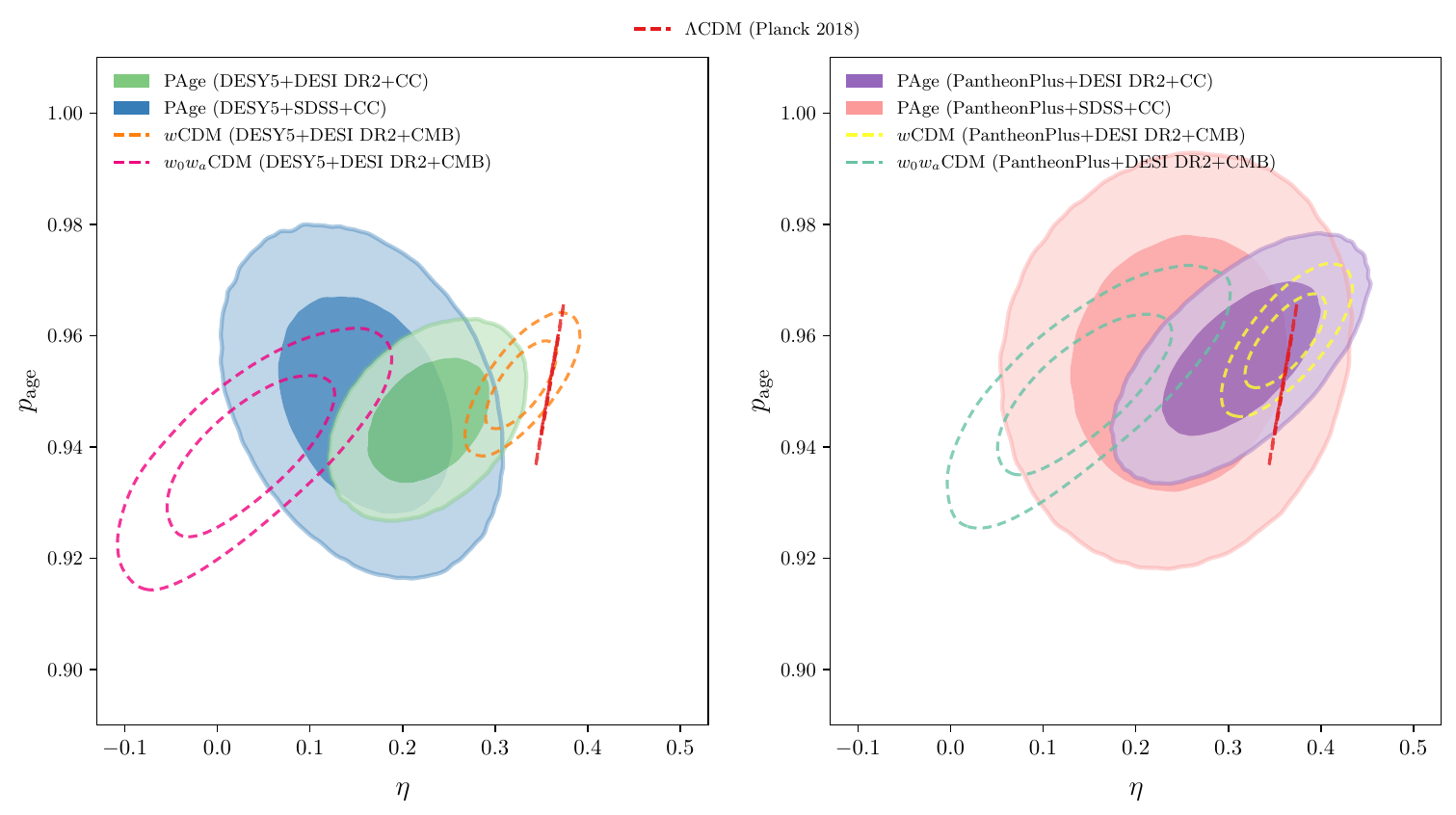}
	}
	\centering
	\caption{The constraints on $\eta$ and $p_\mathrm{age}$ in the PAge model obtained from DESY5+DESI DR2+CC (green), DESY5+SDSS+CC (blue), PantheonPlus+DESI DR2+CC (purple), and PantheonPlus+SDSS+CC (pink).
The DESI DR2 and Planck 2018 results are introduced for comparison.
Since $\Lambda \mathrm{CDM}$ lacks one parameter compared to the PAge model, it is mapped as a straight line in the parameter space.
    }
	\label{Fig: sub-corner1}
\end{figure*}

 \begin{figure*}[!t]
	\resizebox{\textwidth}{!}{
		\centering
		\includegraphics[width=1\textwidth]{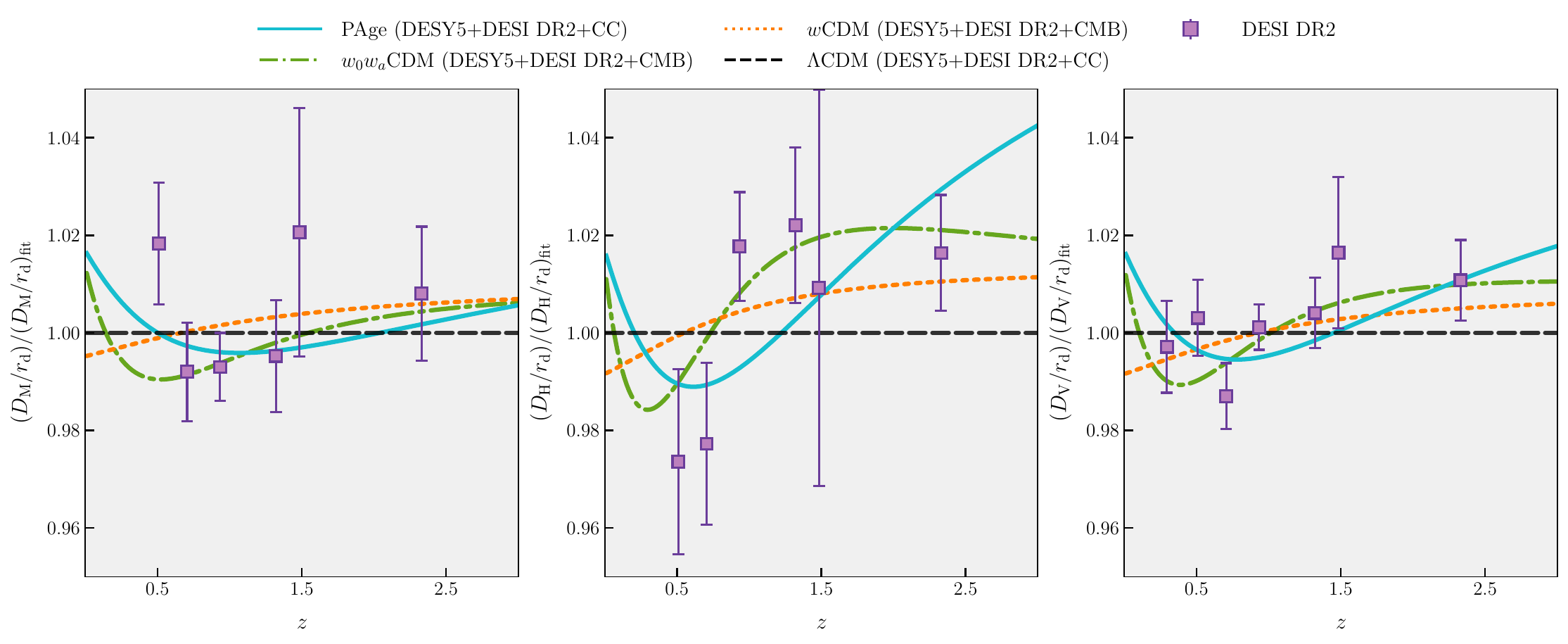}
	}
	\centering
	\caption{The BAO distance for the PAge (blue), $w_0w_a\mathrm{CDM}$ (green), and $w\mathrm{CDM}$ (orange) models in DESY5 + DESI DR2 + CC.  These distances are normalized by the best-fit $\Lambda \mathrm{CDM}$ model (black) in DESY5+DESI+CC. The data of DESI (purple, filled) are drawn with error bars.}
	\label{Fig: reconstruct}
\end{figure*}

 \begin{figure*}[htbp]
	\resizebox{\textwidth}{!}{
		\centering
		\includegraphics[width=1\textwidth]{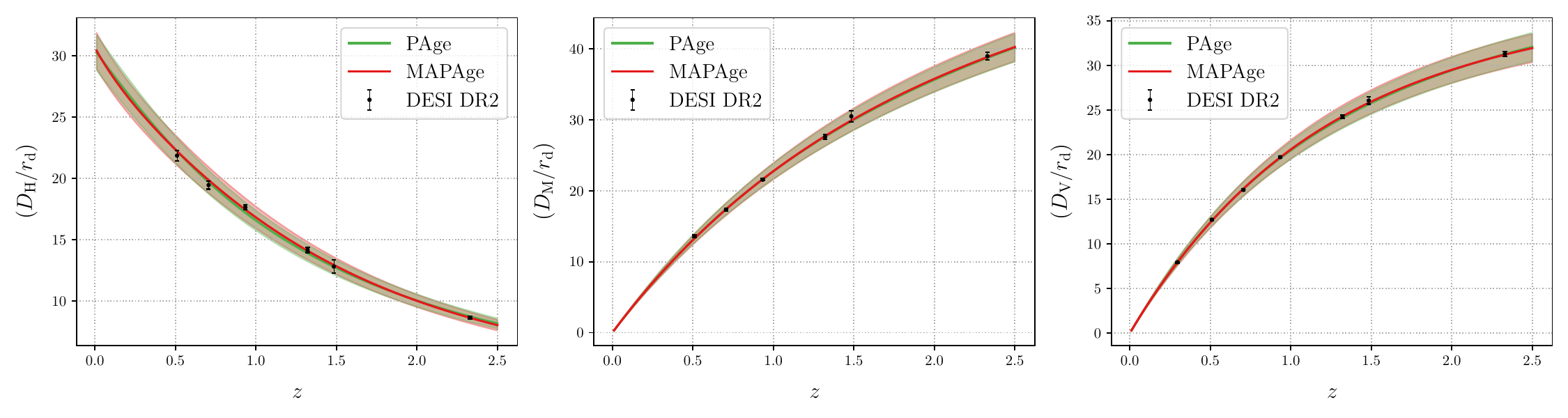}
	}
	\centering
	\caption{The reconstructed BAO distance for the PAge (green), MAPAge (red), incorporating DESI DR2 data. The shaded bands denote the corresponding $1\sigma$ uncertainties.}
	\label{Fig: trend}
\end{figure*}

\begin{figure*}[htbp]
  \centering
  \begin{minipage}[t]{0.48\textwidth}
    \centering
    \includegraphics[width=\linewidth]{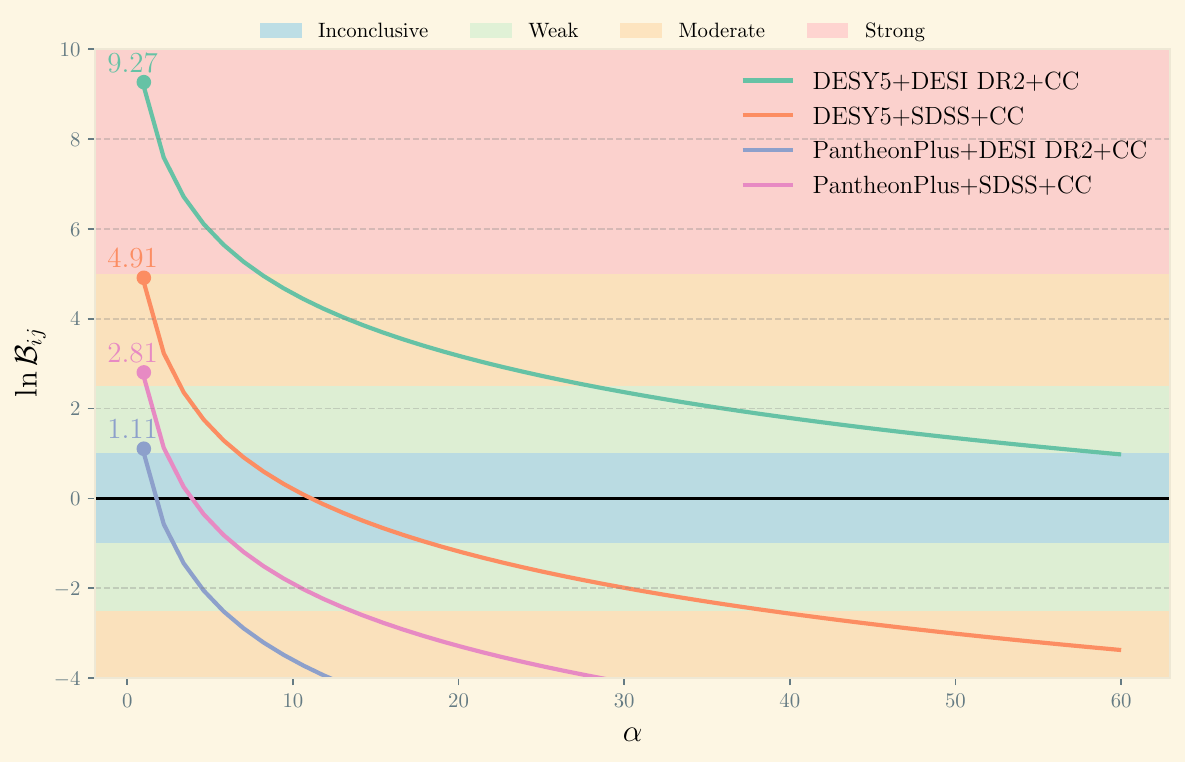}
  \end{minipage}\hfill
  \begin{minipage}[t]{0.48\textwidth}
    \centering
    \includegraphics[width=\linewidth]{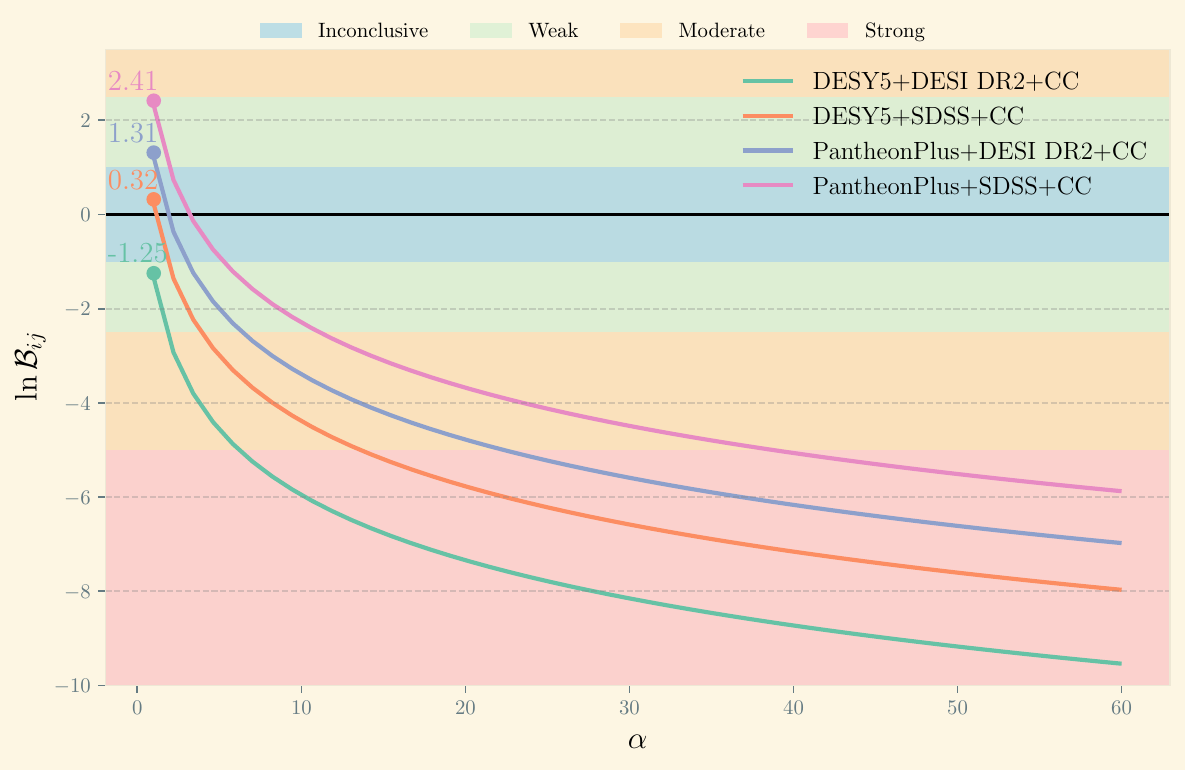}
  \end{minipage}
  \caption{Bayesian model comparison of PAge with $\Lambda\rm CDM$ (left panel) and $w_0w_a\mathrm{CDM}$ model (right panel). Positive values indicate a preference for the PAge model in each dataset. The impact of prior widths for $\eta $ and $p_{\mathrm{age}}$ is assessed by broadening the original prior width $d$ to $d^*=\alpha d$, with $\alpha$ denoting the broadening factor.}
  \label{fig:both}
\end{figure*}

After examining how the choice of SNe and BAO datasets influences the cosmological parameters, we now focus on the approximation parameters of the PAge model itself, namely $\eta$ and $p_{\mathrm{age}}$. These two parameters play a central role in offering insights into potential late-time modifications to cosmic expansion.
The posterior distributions in the \((p_{\mathrm{age}}, \eta)\) plane for the four datasets are shown in Fig.~\ref{Fig: sub-corner1}. For comparison, previous results from $\Lambda \mathrm{CDM}$ \cite{Planck:2018vyg} and DESI DR2 \cite{DESI:2025zgx} are mapped into the PAge model. The mapping procedure is detailed in Eqs.~(\ref{4}), (\ref{6}), (\ref{7}), (\ref{8}), and (\ref{9}). Note that the PAge model contains one additional parameter compared to $\Lambda \mathrm{CDM}$, which causes the mapped region of $\Lambda \mathrm{CDM}$ in the $(p_{\mathrm{age}}, \eta)$ space to appear as a line. We adopt the DESI DR2 results for $w \mathrm{CDM}$ and $w_0w_a \mathrm{CDM}$ models~\cite{DESI:2025zgx}. As shown in Fig.~\ref{Fig: sub-corner1}, it is evident that our results deviate from the $\Lambda \mathrm{CDM}$ model, showing a lower value of $\eta$ for the DESY5 dataset. In contrast, for the PantheonPlus samples, different BAO combinations remain consistent with $\Lambda \mathrm{CDM}$.

To better understand these results, we reconstruct the BAO distance predicted by the PAge, $w \mathrm{CDM}$ and $w_0w_a \mathrm{CDM}$ in Fig.~\ref{Fig: reconstruct}, based on our best-fit  $\Lambda \mathrm{CDM}$ model in the DESY5+DESI DR2+CC datasets. The PAge model exhibits a distinctive trend driven by the DESI DR2 data. In comparison to the $\Lambda \mathrm{CDM}$ model, it predicts slightly enhanced BAO distances at very low redshifts, suppressed distances at intermediate redshifts, and elevated values again at high redshifts across all configurations. While the $w \mathrm{CDM}$ model attempts to capture this behavior, it fails to reproduce the variation across the relevant redshift range. In contrast, the $w_0w_a \mathrm{CDM}$ model, which incorporates dynamically evolving equation of state, is more capable of reproducing this characteristic trend.

These findings suggest that the DESI DR2 data may not be fully compatible with the $\Lambda\mathrm{CDM}$ model, which lack the flexibility to capture such rapid variations in the expansion history, thereby hinting at a preference for dynamical dark energy. PAge provides a model-independent description of the underlying behaviors of the late-time Universe. However, it should be noted that the PAge model rests on specific assumptions. Thus, PAge may not capture inhomogeneous cosmologies, nor scenarios that violate the assumptions~\ref{assumption1} and \ref{assumption2}.

Then, we fit the MAPAge model to DESY5+DESI DR2+CC instead of PAge, to assess the impact of the chosen truncated order in Eq.~(\ref{1}). Our final constraints are: $\eta_1=-0.01 \pm0.09,~\eta_2=0.65\pm0.22,~p_{\mathrm{age}}=0.93\pm0.01,~H_0=67.57\pm2.33~\mathrm{km~s^{-1}~Mpc^{-1}},~r_{\mathrm{d}}=144.7^{+4.5}_{-5.2}~\mathrm{Mpc}$. The constrains of $p_{\mathrm{age}},~H_0,~M_{\mathrm{B}},~r_{\mathrm{d}}$ agree with PAge, except for $\eta~(\eta_1)$, which shifts toward more negative values and has a larger uncertainty in MAPAge, caused by the added cubic term. Furthermore, the reconstructed BAO distances $D_{\mathrm{H}}/r_\mathrm{d},~D_{\mathrm{M}}/r_\mathrm{d},~D_{\mathrm{v}}/r_\mathrm{d}$ are shown in Fig.~\ref{Fig: trend}. The results exhibit no significant difference from PAge in either central values or uncertainties. The enlarged uncertainty in $\eta~(\eta_1)$ minimally affects $D_{\mathrm{H}}/r_\mathrm{d},~D_{\mathrm{M}}/r_\mathrm{d},~D_{\mathrm{v}}/r_\mathrm{d}$, demonstrating stability with respect to the chosen truncated order in Eq.~(\ref{1}).

To quantitatively evaluate the performance of the PAge model relative to the $\Lambda\mathrm{CDM}$ model,  we perform Bayesian inference across different datasets. The MAPAge does not show significant different with PAge, as demonstrated by Bayesian inference. Therefore, for simplicity , we just focus on PAge model. Specifically, the Bayes factors $\mathrm{ln}\mathcal{B}_{ij}$ are calculated in logarithmic space,
\begin{equation}
    \mathrm{ln}\mathcal{B}_{ij}=\mathrm{ln}Z_i-\mathrm{ln}Z_j,
    \label{23}
\end{equation}
where $Z_i$ and $Z_j$ are Bayesian evidence of PAge and $\Lambda\mathrm{CDM}$ models, respectively. The Bayesian evidence is defined as 
\begin{equation}
    Z\equiv \int P(D \mid \theta, M) \, P(\theta \mid M) \, P(M) \, \mathrm{d}\theta,
\end{equation}
where $P(D \mid \theta, M)$ is the likelihood of the data $D$ given the parameters $\theta$ and model $M$, $P(\theta \mid M)$ is the prior probability of $\theta$ under $M$, and $P(M)$ is the prior of $M$. The extent of model preference is assessed using the Jeffreys scale, as described in Ref.~\cite{Trotta:2008qt}.
\begin{itemize}
    \item $|\mathrm{ln} \mathcal{B}_{ij}|<1$, inconclusive evidence.
    \item $1\leq |\mathrm{ln} \mathcal{B}_{ij}|<2.5$, weak evidence.
    \item $2.5\leq |\mathrm{ln} \mathcal{B}_{ij}|<5$, moderate evidence.
    \item $5\leq |\mathrm{ln} \mathcal{B}_{ij}|<10$, strong evidence.
    \item $10\leq |\mathrm{ln} \mathcal{B}_{ij}|$, decisive evidence.
\end{itemize} 

Given the impact of prior width on our results, we follow Refs.~\cite{Patel:2024odo, SolaPeracaula:2018wwm} and broaden the original prior width $d$ by a factor $\alpha$, so that the new prior width is $d^* = \alpha d$. In the case of uniform priors \cite{Patel:2024odo, SolaPeracaula:2018wwm}, the new Bayes factors $\ln \mathcal{B}^*_{ij}$ are modified into 
\begin{equation}
    \ln \mathcal{B}_{ij}^{*}  = \ln \left[ \frac{(\alpha_{\nu_1} \alpha_{\nu_2} \alpha_{\nu_3} \dots )_j}{(\alpha_{\nu_1} \alpha_{\nu_2} \alpha_{\nu_3} \dots )_i} \right] + \ln \mathcal{B}_{ij},
    \label{25}
\end{equation} where $(\alpha_{\nu_1} \alpha_{\nu_2} \alpha_{\nu_3} \dots )_i$ denotes the product of prior-width broadening factors $\alpha _{\nu _k}$ for parameters $\nu_k $ in PAge and $(\alpha_{\nu_1} \alpha_{\nu_2} \alpha_{\nu_3} \dots )_j$ denotes the corresponding product for $\Lambda \mathrm{CDM}$. We consider only the impact of the prior widths of $\eta,~p_{\mathrm{age}}$ in PAge. For simplicity, we set $\alpha_\eta =~\alpha_{p_{\mathrm{age}}}=\alpha$. So the Eq.~(\ref{25}) becomes 
\begin{equation}
        \ln \mathcal{B}_{ij}^{*}  =  \ln \mathcal{B}_{ij}-2\ln\alpha.
        \label{26}
\end{equation}
In the left panel of Fig.~\ref{fig:both}, we present the final model comparison results, with the PAge model assessed relative to the fiducial $\Lambda \mathrm{CDM}$ across four datasets. Positive values indicate a preference for the PAge model, while negative values indicate a preference for the $\Lambda \mathrm{CDM}$ model. Specifically, when $ \alpha=1$ we find $  \mathrm{ln} \mathcal{B}_{ij} = 9.17,~4.81,~ 1.01,~ \text{and}~ 2.71$ for DESY5+DESI DR2+CC, DESY5+SDSS+CC, PantheonPlus+DESI DR2+CC, and PantheonPlus+SDSS+CC, respectively. It is clear that the PAge model consistently provides a better fit than the standard $\Lambda \mathrm{CDM}$ across all datasets. In particular, for the DESY5+DESI DR2+CC datasets, the Bayes factor is 9.17, indicating nearly decisive evidence in favor of the PAge model over the $\Lambda\mathrm{CDM}$ model. Varying $\alpha$ can switch the preferred model. When $\alpha \ge 12,~2,~4$ for DESY5+SDSS+CC, PantheonPlus+DESI DR2+CC, and PantheonPlus+SDSS+CC, respectively, the Bayes factor favors $\Lambda\mathrm{CDM}$. By contrast, for DESY5+DESI DR2+CC, the preference flips for $\alpha \ge 98$. 

Moreover, to assess the performance of PAge compared to $w_0w_a \mathrm{CDM}$, we perform Bayesian inference in the same manner. The prior ranges used for $w_0w_a\mathrm{CDM}$ in this work are $ w_0\in[-3,1], ~w_a\in[-3,2],~M_{\mathrm{B}}\in[-21,-19],~H_0\in[60,80]~\mathrm{km~s^{-1}~Mpc^{-1}},~r_\mathrm{d}\in [120,180]~\mathrm{Mpc},~\Omega_{\mathrm{m}}\in [0.2,0.4]$. We also explore the impact of prior width by broadening the prior widths on $\eta$ and $p_{\mathrm{age}}$. The Bayes factors take the same form as Eq.~(\ref{26}), where $i$ denotes PAge and $j$ denotes $w_0w_a\mathrm{CDM}$. The final results are shown in the right panel of Fig.~\ref{fig:both}. For $\alpha=1$, we find $  \ln \mathcal{B}_{ij} = -1.25,~0.32,~ 1.31,~ \text{and}~ 2.41$ for DESY5+DESI DR2+CC, DESY5+SDSS+CC, PantheonPlus+DESI DR2+CC, and PantheonPlus+SDSS+CC, respectively. These results indicate at most weak evidence in favor of $w_0w_a \mathrm{CDM}$ model over PAge model. For DESY5+SDSS+CC, PantheonPlus+DESI DR2+CC, and PantheonPlus+SDSS+CC, the evidence shifts in favor of $w_{0}w_{a}\mathrm{CDM}$ when $\alpha\ge 2,~2,~4$, respectively. By contrast, for DESY5+DESI DR2+CC, PAge is favored only when $\alpha\le0.5$. These conclusions also depend on the chosen prior width for $w_0,~w_a$, which are held fixed here for simplicity. Widening those prior widths on $w_{0}$ and $w_{a}$ tends to strengthen the evidence in favor of the PAge model.

\section{CONCLUSION}\label{sec4}
The tension between the early-Universe and late-Universe measurements of $H_0$ has become one of the most significant challenges in modern cosmology. Recently, DESI DR1 \cite{DESI:2024mwx} and DR2 \cite{DESI:2025zgx} have further sharpened this tension with addition hints of dynamical dark energy in $3.9\sigma$ and $4.2\sigma$ level. In the context of the $H_0$ tension, numerous theoretical models have been proposed, although many models face corresponding no-go arguments. Similarly, the hint of dynamical dark energy has been critically examined through the reanalyzes of the DESY5 SNe data \cite{Huang:2025som,Efstathiou:2024xcq,DES:2025tir,Gialamas:2024lyw} and the DESI BAO measurements \cite{Zheng:2024qzi,Mukherjee:2025fkf,Liu:2024gfy}.

To provide a model-independent perspective, we employ the PAge-improved IDL in this work to constrain cosmological parameters. Our work is based on four datasets, namely DESY5+DESI DR2+CC, DESY5+SDSS+CC, 
PantheonPlus+DESI DR2+CC,
and PantheonPlus+SDSS+CC, allowing us to investigate the respective roles of SNe and BAO data in constraining cosmological parameters. For DESY5+DESI DR2+CC, we find $H_0=67.91\pm2.33~ \mathrm{km~s^{-1}~Mpc^{-1}}$, which agrees with the Planck 2018 result, but exhibits a mild $2.0\sigma$ tension with the SH0ES measurement. The other datasets also remain consistent with Planck 2018. Compared to SH0ES, the resulting tensions are $2.1\sigma$, $1.6\sigma,$ and $1.7\sigma$ for DESY5+SDSS+CC, PantheonPlus+DESI DR2+CC, and PantheonPlus+SDSS+CC, respectively. In our results, the DESY5 data tend to favor a lower value of \( H_0 \) and a higher value of \( M_\mathrm{B} \), which correspond to a shift in the intercept of the magnitude--redshift relation within the \( H_0 \)--\( M_\mathrm{B} \) parameter space.
Additionally, the DESI DR2 data indicate a preference for a higher value of $r_\mathrm{d}$ compared to SDSS data, with a difference of approximately 2 Mpc.

Mapping specific cosmological models into the PAge parameter space $(p_{\mathrm{age}},\eta)$, our results align with the DESI DR2 hints of dynamical dark energy and deviate from $\Lambda \mathrm{CDM}$. In particular, the reconstructed BAO distance of the PAge model reveals a special trend compared to best-fit $\Lambda \mathrm{CDM}$. It shows a slight enhancement in very low redshift, suppression at intermediate redshift, and an upturn at high redshift, which may suggest that the $\Lambda \mathrm{CDM}$ model is insufficient to fully reconcile the observed data. Furthermore, the Bayesian model comparison provides evidence in favor of the PAge model over $\Lambda \mathrm{CDM}$, with Bayes factors of 9.17, 4.81, 1.01, and 2.71 in DESY5+DESI DR2+CC, DESY5+SDSS+CC, 
PantheonPlus+DESI DR2+CC,
PantheonPlus+SDSS+CC respectively. It gives nearly decisive evidence in favor of the PAge model over $\Lambda \mathrm{CDM}$ in the DESY5+DESI DR2+CC combination. Comparing PAge with $w_0w_a\mathrm{CDM}$, the Bayes factors are $-1.25,~0.32,~1.31,$ and $2.41$ for DESY5+DESI DR2+CC,
DESY5+SDSS+CC, PantheonPlus+DESI DR2+CC,
and PantheonPlus+SDSS+CC, respectively, showing at most weak evidence in favor of $w_{0}w_{a}\mathrm{CDM}$. The interpretation of these comparisons is influenced by the adopted prior width. The wider prior width for PAge parameters, the weaker the evidence in favor of PAge.

In future work, we aim to explore the data in more detail by performing a redshift-binned analysis. Motivated by the connection between the expansion history and the dark-energy equation of state, we will further investigate the occurrence of unphysical phantom behaviors~\cite{Caldwell:2003vq,Vikman:2004dc} using the PAge model. It should be noted that the PAge model is essentially a Taylor expansion, which may suffer from singularities as discussed in Ref.~\cite{Zhang:2016urt} when reconstructing the equation-of-state parameter. We are going to calibrate the Taylor expansion with alternatives beyond the early-matter-dominated Universe used in the PAge model, to avoid the singularities. These efforts may provide new insights into the nature of dark energy and further clarify the physical origin of the $H_0$ tension.

\begin{acknowledgments}
We thank Sheng-Han Zhou and Yi-Min Zhang for their helpful discussions. This work was supported by the National SKA Program of China (Grants No. 2022SKA0110200 and No. 2022SKA0110203), the National Natural Science Foundation of China (Grants No. 12533001, No. 12575049, No. 12473001, No. 11975072, No. 11875102, No. 11835009, No. 12422502, No. 12105344, No. 12235019, No. 12447101, No. 12073088, No. 11821505, No. 11991052, and No. 11947302), the China Manned Space Program (Grants No. CMS-CSST-2025-A01 and No. CMS-CSST-2025-A02), the National 111 Project (Grant No. B16009), and the National Key Research and Development Program of China (Grants No. 2021YFC2203004, No. 2021YFA0718304, and No. 2020YFC2201501).

\end{acknowledgments}

\section*{DATA AVAILABILITY}
The data that support the finding of this article are openly available \cite{Ling_ScienceDB_27772}.

\bibliography{Main}

\begin{thebibliography}{209}%
\makeatletter
\providecommand \@ifxundefined [1]{%
 \@ifx{#1\undefined}
}%
\providecommand \@ifnum [1]{%
 \ifnum #1\expandafter \@firstoftwo
 \else \expandafter \@secondoftwo
 \fi
}%
\providecommand \@ifx [1]{%
 \ifx #1\expandafter \@firstoftwo
 \else \expandafter \@secondoftwo
 \fi
}%
\providecommand \natexlab [1]{#1}%
\providecommand \enquote  [1]{``#1''}%
\providecommand \bibnamefont  [1]{#1}%
\providecommand \bibfnamefont [1]{#1}%
\providecommand \citenamefont [1]{#1}%
\providecommand \href@noop [0]{\@secondoftwo}%
\providecommand \href [0]{\begingroup \@sanitize@url \@href}%
\providecommand \@href[1]{\@@startlink{#1}\@@href}%
\providecommand \@@href[1]{\endgroup#1\@@endlink}%
\providecommand \@sanitize@url [0]{\catcode `\\12\catcode `\$12\catcode
  `\&12\catcode `\#12\catcode `\^12\catcode `\_12\catcode `\%12\relax}%
\providecommand \@@startlink[1]{}%
\providecommand \@@endlink[0]{}%
\providecommand \url  [0]{\begingroup\@sanitize@url \@url }%
\providecommand \@url [1]{\endgroup\@href {#1}{\urlprefix }}%
\providecommand \urlprefix  [0]{URL }%
\providecommand \Eprint [0]{\href }%
\providecommand \doibase [0]{http://dx.doi.org/}%
\providecommand \selectlanguage [0]{\@gobble}%
\providecommand \bibinfo  [0]{\@secondoftwo}%
\providecommand \bibfield  [0]{\@secondoftwo}%
\providecommand \translation [1]{[#1]}%
\providecommand \BibitemOpen [0]{}%
\providecommand \bibitemStop [0]{}%
\providecommand \bibitemNoStop [0]{.\EOS\space}%
\providecommand \EOS [0]{\spacefactor3000\relax}%
\providecommand \BibitemShut  [1]{\csname bibitem#1\endcsname}%
\let\auto@bib@innerbib\@empty
\bibitem [{\citenamefont {Riess}\ \emph {et~al.}(1998)\citenamefont {Riess}
  \emph {et~al.}}]{SupernovaSearchTeam:1998fmf}%
  \BibitemOpen
  \bibfield  {author} {\bibinfo {author} {\bibfnamefont {A.~G.}\ \bibnamefont
  {Riess}} \emph {et~al.} (\bibinfo {collaboration} {Supernova Search Team}),\
  }\href {\doibase 10.1086/300499} {\bibfield  {journal} {\bibinfo  {journal}
  {Astron. J.}\ }\textbf {\bibinfo {volume} {116}},\ \bibinfo {pages} {1009}
  (\bibinfo {year} {1998})},\ \Eprint {http://arxiv.org/abs/astro-ph/9805201}
  {arXiv:astro-ph/9805201} \BibitemShut {NoStop}%
\bibitem [{\citenamefont {Perlmutter}\ \emph {et~al.}(1999)\citenamefont
  {Perlmutter} \emph {et~al.}}]{SupernovaCosmologyProject:1998vns}%
  \BibitemOpen
  \bibfield  {author} {\bibinfo {author} {\bibfnamefont {S.}~\bibnamefont
  {Perlmutter}} \emph {et~al.} (\bibinfo {collaboration} {Supernova Cosmology
  Project}),\ }\href {\doibase 10.1086/307221} {\bibfield  {journal} {\bibinfo
  {journal} {Astrophys. J.}\ }\textbf {\bibinfo {volume} {517}},\ \bibinfo
  {pages} {565} (\bibinfo {year} {1999})},\ \Eprint
  {http://arxiv.org/abs/astro-ph/9812133} {arXiv:astro-ph/9812133} \BibitemShut
  {NoStop}%
\bibitem [{\citenamefont {Brout}\ \emph {et~al.}(2022)\citenamefont {Brout}
  \emph {et~al.}}]{Brout:2022vxf}%
  \BibitemOpen
  \bibfield  {author} {\bibinfo {author} {\bibfnamefont {D.}~\bibnamefont
  {Brout}} \emph {et~al.},\ }\href {\doibase 10.3847/1538-4357/ac8e04}
  {\bibfield  {journal} {\bibinfo  {journal} {Astrophys. J.}\ }\textbf
  {\bibinfo {volume} {938}},\ \bibinfo {pages} {110} (\bibinfo {year}
  {2022})},\ \Eprint {http://arxiv.org/abs/2202.04077} {arXiv:2202.04077
  [astro-ph.CO]} \BibitemShut {NoStop}%
\bibitem [{\citenamefont {Asgari}\ \emph {et~al.}(2021)\citenamefont {Asgari}
  \emph {et~al.}}]{KiDS:2020suj}%
  \BibitemOpen
  \bibfield  {author} {\bibinfo {author} {\bibfnamefont {M.}~\bibnamefont
  {Asgari}} \emph {et~al.} (\bibinfo {collaboration} {KiDS}),\ }\href {\doibase
  10.1051/0004-6361/202039070} {\bibfield  {journal} {\bibinfo  {journal}
  {Astron. Astrophys.}\ }\textbf {\bibinfo {volume} {645}},\ \bibinfo {pages}
  {A104} (\bibinfo {year} {2021})},\ \Eprint {http://arxiv.org/abs/2007.15633}
  {arXiv:2007.15633 [astro-ph.CO]} \BibitemShut {NoStop}%
\bibitem [{\citenamefont {Troxel}\ \emph {et~al.}(2018)\citenamefont {Troxel}
  \emph {et~al.}}]{DES:2017qwj}%
  \BibitemOpen
  \bibfield  {author} {\bibinfo {author} {\bibfnamefont {M.~A.}\ \bibnamefont
  {Troxel}} \emph {et~al.} (\bibinfo {collaboration} {DES}),\ }\href {\doibase
  10.1103/PhysRevD.98.043528} {\bibfield  {journal} {\bibinfo  {journal} {Phys.
  Rev. D}\ }\textbf {\bibinfo {volume} {98}},\ \bibinfo {pages} {043528}
  (\bibinfo {year} {2018})},\ \Eprint {http://arxiv.org/abs/1708.01538}
  {arXiv:1708.01538 [astro-ph.CO]} \BibitemShut {NoStop}%
\bibitem [{\citenamefont {Aiola}\ \emph {et~al.}(2020)\citenamefont {Aiola}
  \emph {et~al.}}]{ACT:2020gnv}%
  \BibitemOpen
  \bibfield  {author} {\bibinfo {author} {\bibfnamefont {S.}~\bibnamefont
  {Aiola}} \emph {et~al.} (\bibinfo {collaboration} {ACT}),\ }\href {\doibase
  10.1088/1475-7516/2020/12/047} {\bibfield  {journal} {\bibinfo  {journal}
  {JCAP}\ }\textbf {\bibinfo {volume} {12}},\ \bibinfo {pages} {047} (\bibinfo
  {year} {2020})},\ \Eprint {http://arxiv.org/abs/2007.07288} {arXiv:2007.07288
  [astro-ph.CO]} \BibitemShut {NoStop}%
\bibitem [{\citenamefont {Di~Valentino}\ \emph
  {et~al.}(2021{\natexlab{a}})\citenamefont {Di~Valentino} \emph
  {et~al.}}]{DiValentino:2020zio}%
  \BibitemOpen
  \bibfield  {author} {\bibinfo {author} {\bibfnamefont {E.}~\bibnamefont
  {Di~Valentino}} \emph {et~al.},\ }\href {\doibase
  10.1016/j.astropartphys.2021.102605} {\bibfield  {journal} {\bibinfo
  {journal} {Astropart. Phys.}\ }\textbf {\bibinfo {volume} {131}},\ \bibinfo
  {pages} {102605} (\bibinfo {year} {2021}{\natexlab{a}})},\ \Eprint
  {http://arxiv.org/abs/2008.11284} {arXiv:2008.11284 [astro-ph.CO]}
  \BibitemShut {NoStop}%
\bibitem [{\citenamefont {Perivolaropoulos}\ and\ \citenamefont
  {Skara}(2022)}]{Perivolaropoulos:2021jda}%
  \BibitemOpen
  \bibfield  {author} {\bibinfo {author} {\bibfnamefont {L.}~\bibnamefont
  {Perivolaropoulos}}\ and\ \bibinfo {author} {\bibfnamefont {F.}~\bibnamefont
  {Skara}},\ }\href {\doibase 10.1016/j.newar.2022.101659} {\bibfield
  {journal} {\bibinfo  {journal} {New Astron. Rev.}\ }\textbf {\bibinfo
  {volume} {95}},\ \bibinfo {pages} {101659} (\bibinfo {year} {2022})},\
  \Eprint {http://arxiv.org/abs/2105.05208} {arXiv:2105.05208 [astro-ph.CO]}
  \BibitemShut {NoStop}%
\bibitem [{\citenamefont {Shah}\ \emph {et~al.}(2021)\citenamefont {Shah},
  \citenamefont {Lemos},\ and\ \citenamefont {Lahav}}]{Shah:2021onj}%
  \BibitemOpen
  \bibfield  {author} {\bibinfo {author} {\bibfnamefont {P.}~\bibnamefont
  {Shah}}, \bibinfo {author} {\bibfnamefont {P.}~\bibnamefont {Lemos}}, \ and\
  \bibinfo {author} {\bibfnamefont {O.}~\bibnamefont {Lahav}},\ }\href
  {\doibase 10.1007/s00159-021-00137-4} {\bibfield  {journal} {\bibinfo
  {journal} {Astron. Astrophys. Rev.}\ }\textbf {\bibinfo {volume} {29}},\
  \bibinfo {pages} {9} (\bibinfo {year} {2021})},\ \Eprint
  {http://arxiv.org/abs/2109.01161} {arXiv:2109.01161 [astro-ph.CO]}
  \BibitemShut {NoStop}%
\bibitem [{\citenamefont {Verde}\ \emph {et~al.}(2019)\citenamefont {Verde},
  \citenamefont {Treu},\ and\ \citenamefont {Riess}}]{Verde:2019ivm}%
  \BibitemOpen
  \bibfield  {author} {\bibinfo {author} {\bibfnamefont {L.}~\bibnamefont
  {Verde}}, \bibinfo {author} {\bibfnamefont {T.}~\bibnamefont {Treu}}, \ and\
  \bibinfo {author} {\bibfnamefont {A.~G.}\ \bibnamefont {Riess}},\ }\href
  {\doibase 10.1038/s41550-019-0902-0} {\bibfield  {journal} {\bibinfo
  {journal} {Nature Astron.}\ }\textbf {\bibinfo {volume} {3}},\ \bibinfo
  {pages} {891} (\bibinfo {year} {2019})},\ \Eprint
  {http://arxiv.org/abs/1907.10625} {arXiv:1907.10625 [astro-ph.CO]}
  \BibitemShut {NoStop}%
\bibitem [{\citenamefont {Aghanim}\ \emph {et~al.}(2020)\citenamefont {Aghanim}
  \emph {et~al.}}]{Planck:2018vyg}%
  \BibitemOpen
  \bibfield  {author} {\bibinfo {author} {\bibfnamefont {N.}~\bibnamefont
  {Aghanim}} \emph {et~al.} (\bibinfo {collaboration} {Planck}),\ }\href
  {\doibase 10.1051/0004-6361/201833910} {\bibfield  {journal} {\bibinfo
  {journal} {Astron. Astrophys.}\ }\textbf {\bibinfo {volume} {641}},\ \bibinfo
  {pages} {A6} (\bibinfo {year} {2020})},\ \bibinfo {note} {[Erratum:
  Astron.Astrophys. 652, C4 (2021)]},\ \Eprint
  {http://arxiv.org/abs/1807.06209} {arXiv:1807.06209 [astro-ph.CO]}
  \BibitemShut {NoStop}%
\bibitem [{\citenamefont {Riess}\ \emph {et~al.}(2022)\citenamefont {Riess}
  \emph {et~al.}}]{Riess:2021jrx}%
  \BibitemOpen
  \bibfield  {author} {\bibinfo {author} {\bibfnamefont {A.~G.}\ \bibnamefont
  {Riess}} \emph {et~al.},\ }\href {\doibase 10.3847/2041-8213/ac5c5b}
  {\bibfield  {journal} {\bibinfo  {journal} {Astrophys. J. Lett.}\ }\textbf
  {\bibinfo {volume} {934}},\ \bibinfo {pages} {L7} (\bibinfo {year} {2022})},\
  \Eprint {http://arxiv.org/abs/2112.04510} {arXiv:2112.04510 [astro-ph.CO]}
  \BibitemShut {NoStop}%
\bibitem [{\citenamefont {Akrami}\ \emph {et~al.}(2020)\citenamefont {Akrami}
  \emph {et~al.}}]{Planck:2019evm}%
  \BibitemOpen
  \bibfield  {author} {\bibinfo {author} {\bibfnamefont {Y.}~\bibnamefont
  {Akrami}} \emph {et~al.} (\bibinfo {collaboration} {Planck}),\ }\href
  {\doibase 10.1051/0004-6361/201935201} {\bibfield  {journal} {\bibinfo
  {journal} {Astron. Astrophys.}\ }\textbf {\bibinfo {volume} {641}},\ \bibinfo
  {pages} {A7} (\bibinfo {year} {2020})},\ \Eprint
  {http://arxiv.org/abs/1906.02552} {arXiv:1906.02552 [astro-ph.CO]}
  \BibitemShut {NoStop}%
\bibitem [{\citenamefont {Zhang}\ \emph
  {et~al.}(2017{\natexlab{a}})\citenamefont {Zhang}, \citenamefont {Childress},
  \citenamefont {Davis}, \citenamefont {Karpenka}, \citenamefont {Lidman},
  \citenamefont {Schmidt},\ and\ \citenamefont {Smith}}]{Zhang:2017aqn}%
  \BibitemOpen
  \bibfield  {author} {\bibinfo {author} {\bibfnamefont {B.~R.}\ \bibnamefont
  {Zhang}}, \bibinfo {author} {\bibfnamefont {M.~J.}\ \bibnamefont
  {Childress}}, \bibinfo {author} {\bibfnamefont {T.~M.}\ \bibnamefont
  {Davis}}, \bibinfo {author} {\bibfnamefont {N.~V.}\ \bibnamefont {Karpenka}},
  \bibinfo {author} {\bibfnamefont {C.}~\bibnamefont {Lidman}}, \bibinfo
  {author} {\bibfnamefont {B.~P.}\ \bibnamefont {Schmidt}}, \ and\ \bibinfo
  {author} {\bibfnamefont {M.}~\bibnamefont {Smith}},\ }\href {\doibase
  10.1093/mnras/stx1600} {\bibfield  {journal} {\bibinfo  {journal} {Mon. Not.
  Roy. Astron. Soc.}\ }\textbf {\bibinfo {volume} {471}},\ \bibinfo {pages}
  {2254} (\bibinfo {year} {2017}{\natexlab{a}})},\ \Eprint
  {http://arxiv.org/abs/1706.07573} {arXiv:1706.07573 [astro-ph.CO]}
  \BibitemShut {NoStop}%
\bibitem [{\citenamefont {Di~Valentino}\ \emph
  {et~al.}(2021{\natexlab{b}})\citenamefont {Di~Valentino}, \citenamefont
  {Mena}, \citenamefont {Pan}, \citenamefont {Visinelli}, \citenamefont {Yang},
  \citenamefont {Melchiorri}, \citenamefont {Mota}, \citenamefont {Riess},\
  and\ \citenamefont {Silk}}]{DiValentino:2021izs}%
  \BibitemOpen
  \bibfield  {author} {\bibinfo {author} {\bibfnamefont {E.}~\bibnamefont
  {Di~Valentino}}, \bibinfo {author} {\bibfnamefont {O.}~\bibnamefont {Mena}},
  \bibinfo {author} {\bibfnamefont {S.}~\bibnamefont {Pan}}, \bibinfo {author}
  {\bibfnamefont {L.}~\bibnamefont {Visinelli}}, \bibinfo {author}
  {\bibfnamefont {W.}~\bibnamefont {Yang}}, \bibinfo {author} {\bibfnamefont
  {A.}~\bibnamefont {Melchiorri}}, \bibinfo {author} {\bibfnamefont {D.~F.}\
  \bibnamefont {Mota}}, \bibinfo {author} {\bibfnamefont {A.~G.}\ \bibnamefont
  {Riess}}, \ and\ \bibinfo {author} {\bibfnamefont {J.}~\bibnamefont {Silk}},\
  }\href {\doibase 10.1088/1361-6382/ac086d} {\bibfield  {journal} {\bibinfo
  {journal} {Class. Quant. Grav.}\ }\textbf {\bibinfo {volume} {38}},\ \bibinfo
  {pages} {153001} (\bibinfo {year} {2021}{\natexlab{b}})},\ \Eprint
  {http://arxiv.org/abs/2103.01183} {arXiv:2103.01183 [astro-ph.CO]}
  \BibitemShut {NoStop}%
\bibitem [{\citenamefont {Bernal}\ \emph {et~al.}(2016)\citenamefont {Bernal},
  \citenamefont {Verde},\ and\ \citenamefont {Riess}}]{Bernal:2016gxb}%
  \BibitemOpen
  \bibfield  {author} {\bibinfo {author} {\bibfnamefont {J.~L.}\ \bibnamefont
  {Bernal}}, \bibinfo {author} {\bibfnamefont {L.}~\bibnamefont {Verde}}, \
  and\ \bibinfo {author} {\bibfnamefont {A.~G.}\ \bibnamefont {Riess}},\ }\href
  {\doibase 10.1088/1475-7516/2016/10/019} {\bibfield  {journal} {\bibinfo
  {journal} {JCAP}\ }\textbf {\bibinfo {volume} {10}},\ \bibinfo {pages} {019}
  (\bibinfo {year} {2016})},\ \Eprint {http://arxiv.org/abs/1607.05617}
  {arXiv:1607.05617 [astro-ph.CO]} \BibitemShut {NoStop}%
\bibitem [{\citenamefont {Abdalla}\ \emph {et~al.}(2022)\citenamefont {Abdalla}
  \emph {et~al.}}]{Abdalla:2022yfr}%
  \BibitemOpen
  \bibfield  {author} {\bibinfo {author} {\bibfnamefont {E.}~\bibnamefont
  {Abdalla}} \emph {et~al.},\ }\href {\doibase 10.1016/j.jheap.2022.04.002}
  {\bibfield  {journal} {\bibinfo  {journal} {JHEAp}\ }\textbf {\bibinfo
  {volume} {34}},\ \bibinfo {pages} {49} (\bibinfo {year} {2022})},\ \Eprint
  {http://arxiv.org/abs/2203.06142} {arXiv:2203.06142 [astro-ph.CO]}
  \BibitemShut {NoStop}%
\bibitem [{\citenamefont {Di~Valentino}\ \emph {et~al.}(2025)\citenamefont
  {Di~Valentino} \emph {et~al.}}]{CosmoVerseNetwork:2025alb}%
  \BibitemOpen
  \bibfield  {author} {\bibinfo {author} {\bibfnamefont {E.}~\bibnamefont
  {Di~Valentino}} \emph {et~al.} (\bibinfo {collaboration} {CosmoVerse
  Network}),\ }\href {\doibase 10.1016/j.dark.2025.101965} {\bibfield
  {journal} {\bibinfo  {journal} {Phys. Dark Univ.}\ }\textbf {\bibinfo
  {volume} {49}},\ \bibinfo {pages} {101965} (\bibinfo {year} {2025})},\
  \Eprint {http://arxiv.org/abs/2504.01669} {arXiv:2504.01669 [astro-ph.CO]}
  \BibitemShut {NoStop}%
\bibitem [{\citenamefont {Poulin}\ \emph {et~al.}(2019)\citenamefont {Poulin},
  \citenamefont {Smith}, \citenamefont {Karwal},\ and\ \citenamefont
  {Kamionkowski}}]{Poulin:2018cxd}%
  \BibitemOpen
  \bibfield  {author} {\bibinfo {author} {\bibfnamefont {V.}~\bibnamefont
  {Poulin}}, \bibinfo {author} {\bibfnamefont {T.~L.}\ \bibnamefont {Smith}},
  \bibinfo {author} {\bibfnamefont {T.}~\bibnamefont {Karwal}}, \ and\ \bibinfo
  {author} {\bibfnamefont {M.}~\bibnamefont {Kamionkowski}},\ }\href {\doibase
  10.1103/PhysRevLett.122.221301} {\bibfield  {journal} {\bibinfo  {journal}
  {Phys. Rev. Lett.}\ }\textbf {\bibinfo {volume} {122}},\ \bibinfo {pages}
  {221301} (\bibinfo {year} {2019})},\ \Eprint
  {http://arxiv.org/abs/1811.04083} {arXiv:1811.04083 [astro-ph.CO]}
  \BibitemShut {NoStop}%
\bibitem [{\citenamefont {Karwal}\ and\ \citenamefont
  {Kamionkowski}(2016)}]{Karwal:2016vyq}%
  \BibitemOpen
  \bibfield  {author} {\bibinfo {author} {\bibfnamefont {T.}~\bibnamefont
  {Karwal}}\ and\ \bibinfo {author} {\bibfnamefont {M.}~\bibnamefont
  {Kamionkowski}},\ }\href {\doibase 10.1103/PhysRevD.94.103523} {\bibfield
  {journal} {\bibinfo  {journal} {Phys. Rev. D}\ }\textbf {\bibinfo {volume}
  {94}},\ \bibinfo {pages} {103523} (\bibinfo {year} {2016})},\ \Eprint
  {http://arxiv.org/abs/1608.01309} {arXiv:1608.01309 [astro-ph.CO]}
  \BibitemShut {NoStop}%
\bibitem [{\citenamefont {M{\"o}rtsell}\ and\ \citenamefont
  {Dhawan}(2018)}]{Mortsell:2018mfj}%
  \BibitemOpen
  \bibfield  {author} {\bibinfo {author} {\bibfnamefont {E.}~\bibnamefont
  {M{\"o}rtsell}}\ and\ \bibinfo {author} {\bibfnamefont {S.}~\bibnamefont
  {Dhawan}},\ }\href {\doibase 10.1088/1475-7516/2018/09/025} {\bibfield
  {journal} {\bibinfo  {journal} {JCAP}\ }\textbf {\bibinfo {volume} {09}},\
  \bibinfo {pages} {025} (\bibinfo {year} {2018})},\ \Eprint
  {http://arxiv.org/abs/1801.07260} {arXiv:1801.07260 [astro-ph.CO]}
  \BibitemShut {NoStop}%
\bibitem [{\citenamefont {Kamionkowski}\ \emph {et~al.}(2014)\citenamefont
  {Kamionkowski}, \citenamefont {Pradler},\ and\ \citenamefont
  {Walker}}]{Kamionkowski:2014zda}%
  \BibitemOpen
  \bibfield  {author} {\bibinfo {author} {\bibfnamefont {M.}~\bibnamefont
  {Kamionkowski}}, \bibinfo {author} {\bibfnamefont {J.}~\bibnamefont
  {Pradler}}, \ and\ \bibinfo {author} {\bibfnamefont {D.~G.~E.}\ \bibnamefont
  {Walker}},\ }\href {\doibase 10.1103/PhysRevLett.113.251302} {\bibfield
  {journal} {\bibinfo  {journal} {Phys. Rev. Lett.}\ }\textbf {\bibinfo
  {volume} {113}},\ \bibinfo {pages} {251302} (\bibinfo {year} {2014})},\
  \Eprint {http://arxiv.org/abs/1409.0549} {arXiv:1409.0549 [hep-ph]}
  \BibitemShut {NoStop}%
\bibitem [{\citenamefont {Berghaus}\ and\ \citenamefont
  {Karwal}(2020)}]{Berghaus:2019cls}%
  \BibitemOpen
  \bibfield  {author} {\bibinfo {author} {\bibfnamefont {K.~V.}\ \bibnamefont
  {Berghaus}}\ and\ \bibinfo {author} {\bibfnamefont {T.}~\bibnamefont
  {Karwal}},\ }\href {\doibase 10.1103/PhysRevD.101.083537} {\bibfield
  {journal} {\bibinfo  {journal} {Phys. Rev. D}\ }\textbf {\bibinfo {volume}
  {101}},\ \bibinfo {pages} {083537} (\bibinfo {year} {2020})},\ \Eprint
  {http://arxiv.org/abs/1911.06281} {arXiv:1911.06281 [astro-ph.CO]}
  \BibitemShut {NoStop}%
\bibitem [{\citenamefont {Zhang}\ \emph
  {et~al.}(2015{\natexlab{a}})\citenamefont {Zhang}, \citenamefont {Li},\ and\
  \citenamefont {Zhang}}]{Zhang:2014dxk}%
  \BibitemOpen
  \bibfield  {author} {\bibinfo {author} {\bibfnamefont {J.-F.}\ \bibnamefont
  {Zhang}}, \bibinfo {author} {\bibfnamefont {Y.-H.}\ \bibnamefont {Li}}, \
  and\ \bibinfo {author} {\bibfnamefont {X.}~\bibnamefont {Zhang}},\ }\href
  {\doibase 10.1016/j.physletb.2014.12.012} {\bibfield  {journal} {\bibinfo
  {journal} {Phys. Lett. B}\ }\textbf {\bibinfo {volume} {740}},\ \bibinfo
  {pages} {359} (\bibinfo {year} {2015}{\natexlab{a}})},\ \Eprint
  {http://arxiv.org/abs/1403.7028} {arXiv:1403.7028 [astro-ph.CO]} \BibitemShut
  {NoStop}%
\bibitem [{\citenamefont {Zhao}\ \emph
  {et~al.}(2017{\natexlab{a}})\citenamefont {Zhao}, \citenamefont {Li},
  \citenamefont {Zhang},\ and\ \citenamefont {Zhang}}]{Zhao:2016ecj}%
  \BibitemOpen
  \bibfield  {author} {\bibinfo {author} {\bibfnamefont {M.-M.}\ \bibnamefont
  {Zhao}}, \bibinfo {author} {\bibfnamefont {Y.-H.}\ \bibnamefont {Li}},
  \bibinfo {author} {\bibfnamefont {J.-F.}\ \bibnamefont {Zhang}}, \ and\
  \bibinfo {author} {\bibfnamefont {X.}~\bibnamefont {Zhang}},\ }\href
  {\doibase 10.1093/mnras/stx978} {\bibfield  {journal} {\bibinfo  {journal}
  {Mon. Not. Roy. Astron. Soc.}\ }\textbf {\bibinfo {volume} {469}},\ \bibinfo
  {pages} {1713} (\bibinfo {year} {2017}{\natexlab{a}})},\ \Eprint
  {http://arxiv.org/abs/1608.01219} {arXiv:1608.01219 [astro-ph.CO]}
  \BibitemShut {NoStop}%
\bibitem [{\citenamefont {Zhao}\ \emph
  {et~al.}(2017{\natexlab{b}})\citenamefont {Zhao}, \citenamefont {He},
  \citenamefont {Zhang},\ and\ \citenamefont {Zhang}}]{Zhao:2017urm}%
  \BibitemOpen
  \bibfield  {author} {\bibinfo {author} {\bibfnamefont {M.-M.}\ \bibnamefont
  {Zhao}}, \bibinfo {author} {\bibfnamefont {D.-Z.}\ \bibnamefont {He}},
  \bibinfo {author} {\bibfnamefont {J.-F.}\ \bibnamefont {Zhang}}, \ and\
  \bibinfo {author} {\bibfnamefont {X.}~\bibnamefont {Zhang}},\ }\href
  {\doibase 10.1103/PhysRevD.96.043520} {\bibfield  {journal} {\bibinfo
  {journal} {Phys. Rev. D}\ }\textbf {\bibinfo {volume} {96}},\ \bibinfo
  {pages} {043520} (\bibinfo {year} {2017}{\natexlab{b}})},\ \Eprint
  {http://arxiv.org/abs/1703.08456} {arXiv:1703.08456 [astro-ph.CO]}
  \BibitemShut {NoStop}%
\bibitem [{\citenamefont {Feng}\ \emph {et~al.}(2020)\citenamefont {Feng},
  \citenamefont {He}, \citenamefont {Li}, \citenamefont {Zhang},\ and\
  \citenamefont {Zhang}}]{Feng:2019jqa}%
  \BibitemOpen
  \bibfield  {author} {\bibinfo {author} {\bibfnamefont {L.}~\bibnamefont
  {Feng}}, \bibinfo {author} {\bibfnamefont {D.-Z.}\ \bibnamefont {He}},
  \bibinfo {author} {\bibfnamefont {H.-L.}\ \bibnamefont {Li}}, \bibinfo
  {author} {\bibfnamefont {J.-F.}\ \bibnamefont {Zhang}}, \ and\ \bibinfo
  {author} {\bibfnamefont {X.}~\bibnamefont {Zhang}},\ }\href {\doibase
  10.1007/s11433-019-1511-8} {\bibfield  {journal} {\bibinfo  {journal} {Sci.
  China Phys. Mech. Astron.}\ }\textbf {\bibinfo {volume} {63}},\ \bibinfo
  {pages} {290404} (\bibinfo {year} {2020})},\ \Eprint
  {http://arxiv.org/abs/1910.03872} {arXiv:1910.03872 [astro-ph.CO]}
  \BibitemShut {NoStop}%
\bibitem [{\citenamefont {Kreisch}\ \emph {et~al.}(2020)\citenamefont
  {Kreisch}, \citenamefont {Cyr-Racine},\ and\ \citenamefont
  {Dor{\'e}}}]{Kreisch:2019yzn}%
  \BibitemOpen
  \bibfield  {author} {\bibinfo {author} {\bibfnamefont {C.~D.}\ \bibnamefont
  {Kreisch}}, \bibinfo {author} {\bibfnamefont {F.-Y.}\ \bibnamefont
  {Cyr-Racine}}, \ and\ \bibinfo {author} {\bibfnamefont {O.}~\bibnamefont
  {Dor{\'e}}},\ }\href {\doibase 10.1103/PhysRevD.101.123505} {\bibfield
  {journal} {\bibinfo  {journal} {Phys. Rev. D}\ }\textbf {\bibinfo {volume}
  {101}},\ \bibinfo {pages} {123505} (\bibinfo {year} {2020})},\ \Eprint
  {http://arxiv.org/abs/1902.00534} {arXiv:1902.00534 [astro-ph.CO]}
  \BibitemShut {NoStop}%
\bibitem [{\citenamefont {Wyman}\ \emph {et~al.}(2014)\citenamefont {Wyman},
  \citenamefont {Rudd}, \citenamefont {Vanderveld},\ and\ \citenamefont
  {Hu}}]{Wyman:2013lza}%
  \BibitemOpen
  \bibfield  {author} {\bibinfo {author} {\bibfnamefont {M.}~\bibnamefont
  {Wyman}}, \bibinfo {author} {\bibfnamefont {D.~H.}\ \bibnamefont {Rudd}},
  \bibinfo {author} {\bibfnamefont {R.~A.}\ \bibnamefont {Vanderveld}}, \ and\
  \bibinfo {author} {\bibfnamefont {W.}~\bibnamefont {Hu}},\ }\href {\doibase
  10.1103/PhysRevLett.112.051302} {\bibfield  {journal} {\bibinfo  {journal}
  {Phys. Rev. Lett.}\ }\textbf {\bibinfo {volume} {112}},\ \bibinfo {pages}
  {051302} (\bibinfo {year} {2014})},\ \Eprint {http://arxiv.org/abs/1307.7715}
  {arXiv:1307.7715 [astro-ph.CO]} \BibitemShut {NoStop}%
\bibitem [{\citenamefont {Sakstein}\ and\ \citenamefont
  {Trodden}(2020)}]{Sakstein:2019fmf}%
  \BibitemOpen
  \bibfield  {author} {\bibinfo {author} {\bibfnamefont {J.}~\bibnamefont
  {Sakstein}}\ and\ \bibinfo {author} {\bibfnamefont {M.}~\bibnamefont
  {Trodden}},\ }\href {\doibase 10.1103/PhysRevLett.124.161301} {\bibfield
  {journal} {\bibinfo  {journal} {Phys. Rev. Lett.}\ }\textbf {\bibinfo
  {volume} {124}},\ \bibinfo {pages} {161301} (\bibinfo {year} {2020})},\
  \Eprint {http://arxiv.org/abs/1911.11760} {arXiv:1911.11760 [astro-ph.CO]}
  \BibitemShut {NoStop}%
\bibitem [{\citenamefont {Zhang}\ \emph
  {et~al.}(2014{\natexlab{a}})\citenamefont {Zhang}, \citenamefont {Li},\ and\
  \citenamefont {Zhang}}]{Zhang:2014nta}%
  \BibitemOpen
  \bibfield  {author} {\bibinfo {author} {\bibfnamefont {J.-F.}\ \bibnamefont
  {Zhang}}, \bibinfo {author} {\bibfnamefont {Y.-H.}\ \bibnamefont {Li}}, \
  and\ \bibinfo {author} {\bibfnamefont {X.}~\bibnamefont {Zhang}},\ }\href
  {\doibase 10.1140/epjc/s10052-014-2954-8} {\bibfield  {journal} {\bibinfo
  {journal} {Eur. Phys. J. C}\ }\textbf {\bibinfo {volume} {74}},\ \bibinfo
  {pages} {2954} (\bibinfo {year} {2014}{\natexlab{a}})},\ \Eprint
  {http://arxiv.org/abs/1404.3598} {arXiv:1404.3598 [astro-ph.CO]} \BibitemShut
  {NoStop}%
\bibitem [{\citenamefont {Sekiguchi}\ and\ \citenamefont
  {Takahashi}(2021)}]{Sekiguchi:2020teg}%
  \BibitemOpen
  \bibfield  {author} {\bibinfo {author} {\bibfnamefont {T.}~\bibnamefont
  {Sekiguchi}}\ and\ \bibinfo {author} {\bibfnamefont {T.}~\bibnamefont
  {Takahashi}},\ }\href {\doibase 10.1103/PhysRevD.103.083507} {\bibfield
  {journal} {\bibinfo  {journal} {Phys. Rev. D}\ }\textbf {\bibinfo {volume}
  {103}},\ \bibinfo {pages} {083507} (\bibinfo {year} {2021})},\ \Eprint
  {http://arxiv.org/abs/2007.03381} {arXiv:2007.03381 [astro-ph.CO]}
  \BibitemShut {NoStop}%
\bibitem [{\citenamefont {Chiang}\ and\ \citenamefont
  {Slosar}(2018)}]{Chiang:2018xpn}%
  \BibitemOpen
  \bibfield  {author} {\bibinfo {author} {\bibfnamefont {C.-T.}\ \bibnamefont
  {Chiang}}\ and\ \bibinfo {author} {\bibfnamefont {A.}~\bibnamefont
  {Slosar}},\ }\href@noop {} {\  (\bibinfo {year} {2018})},\ \Eprint
  {http://arxiv.org/abs/1811.03624} {arXiv:1811.03624 [astro-ph.CO]}
  \BibitemShut {NoStop}%
\bibitem [{\citenamefont {Mirpoorian}\ \emph {et~al.}(2025)\citenamefont
  {Mirpoorian}, \citenamefont {Jedamzik},\ and\ \citenamefont
  {Pogosian}}]{Mirpoorian:2024fka}%
  \BibitemOpen
  \bibfield  {author} {\bibinfo {author} {\bibfnamefont {S.~H.}\ \bibnamefont
  {Mirpoorian}}, \bibinfo {author} {\bibfnamefont {K.}~\bibnamefont
  {Jedamzik}}, \ and\ \bibinfo {author} {\bibfnamefont {L.}~\bibnamefont
  {Pogosian}},\ }\href {\doibase 10.1103/PhysRevD.111.083519} {\bibfield
  {journal} {\bibinfo  {journal} {Phys. Rev. D}\ }\textbf {\bibinfo {volume}
  {111}},\ \bibinfo {pages} {083519} (\bibinfo {year} {2025})},\ \Eprint
  {http://arxiv.org/abs/2411.16678} {arXiv:2411.16678 [astro-ph.CO]}
  \BibitemShut {NoStop}%
\bibitem [{\citenamefont {Jedamzik}\ and\ \citenamefont
  {Pogosian}(2020)}]{Jedamzik:2020krr}%
  \BibitemOpen
  \bibfield  {author} {\bibinfo {author} {\bibfnamefont {K.}~\bibnamefont
  {Jedamzik}}\ and\ \bibinfo {author} {\bibfnamefont {L.}~\bibnamefont
  {Pogosian}},\ }\href {\doibase 10.1103/PhysRevLett.125.181302} {\bibfield
  {journal} {\bibinfo  {journal} {Phys. Rev. Lett.}\ }\textbf {\bibinfo
  {volume} {125}},\ \bibinfo {pages} {181302} (\bibinfo {year} {2020})},\
  \Eprint {http://arxiv.org/abs/2004.09487} {arXiv:2004.09487 [astro-ph.CO]}
  \BibitemShut {NoStop}%
\bibitem [{\citenamefont {Jedamzik}\ \emph {et~al.}(2025)\citenamefont
  {Jedamzik}, \citenamefont {Pogosian},\ and\ \citenamefont
  {Abel}}]{Jedamzik:2025cax}%
  \BibitemOpen
  \bibfield  {author} {\bibinfo {author} {\bibfnamefont {K.}~\bibnamefont
  {Jedamzik}}, \bibinfo {author} {\bibfnamefont {L.}~\bibnamefont {Pogosian}},
  \ and\ \bibinfo {author} {\bibfnamefont {T.}~\bibnamefont {Abel}},\
  }\href@noop {} {\  (\bibinfo {year} {2025})},\ \Eprint
  {http://arxiv.org/abs/2503.09599} {arXiv:2503.09599 [astro-ph.CO]}
  \BibitemShut {NoStop}%
\bibitem [{\citenamefont {Jedamzik}\ and\ \citenamefont
  {Abel}(2013)}]{Jedamzik:2013gua}%
  \BibitemOpen
  \bibfield  {author} {\bibinfo {author} {\bibfnamefont {K.}~\bibnamefont
  {Jedamzik}}\ and\ \bibinfo {author} {\bibfnamefont {T.}~\bibnamefont
  {Abel}},\ }\href {\doibase 10.1088/1475-7516/2013/10/050} {\bibfield
  {journal} {\bibinfo  {journal} {JCAP}\ }\textbf {\bibinfo {volume} {10}},\
  \bibinfo {pages} {050} (\bibinfo {year} {2013})}\BibitemShut {NoStop}%
\bibitem [{\citenamefont {Forconi}\ \emph {et~al.}(2025)\citenamefont
  {Forconi}, \citenamefont {Favale},\ and\ \citenamefont
  {G{\'o}mez-Valent}}]{Forconi:2025cwp}%
  \BibitemOpen
  \bibfield  {author} {\bibinfo {author} {\bibfnamefont {M.}~\bibnamefont
  {Forconi}}, \bibinfo {author} {\bibfnamefont {A.}~\bibnamefont {Favale}}, \
  and\ \bibinfo {author} {\bibfnamefont {A.}~\bibnamefont {G{\'o}mez-Valent}},\
  }\href {\doibase 10.1103/rpf5-ldks} {\bibfield  {journal} {\bibinfo
  {journal} {Phys. Rev. D}\ }\textbf {\bibinfo {volume} {112}},\ \bibinfo
  {pages} {023517} (\bibinfo {year} {2025})},\ \Eprint
  {http://arxiv.org/abs/2501.11571} {arXiv:2501.11571 [astro-ph.CO]}
  \BibitemShut {NoStop}%
\bibitem [{\citenamefont {Sanchez}(2020)}]{Sanchez:2020vvb}%
  \BibitemOpen
  \bibfield  {author} {\bibinfo {author} {\bibfnamefont {A.~G.}\ \bibnamefont
  {Sanchez}},\ }\href {\doibase 10.1103/PhysRevD.102.123511} {\bibfield
  {journal} {\bibinfo  {journal} {Phys. Rev. D}\ }\textbf {\bibinfo {volume}
  {102}},\ \bibinfo {pages} {123511} (\bibinfo {year} {2020})},\ \Eprint
  {http://arxiv.org/abs/2002.07829} {arXiv:2002.07829 [astro-ph.CO]}
  \BibitemShut {NoStop}%
\bibitem [{\citenamefont {Hill}\ \emph {et~al.}(2020)\citenamefont {Hill},
  \citenamefont {McDonough}, \citenamefont {Toomey},\ and\ \citenamefont
  {Alexander}}]{Hill:2020osr}%
  \BibitemOpen
  \bibfield  {author} {\bibinfo {author} {\bibfnamefont {J.~C.}\ \bibnamefont
  {Hill}}, \bibinfo {author} {\bibfnamefont {E.}~\bibnamefont {McDonough}},
  \bibinfo {author} {\bibfnamefont {M.~W.}\ \bibnamefont {Toomey}}, \ and\
  \bibinfo {author} {\bibfnamefont {S.}~\bibnamefont {Alexander}},\ }\href
  {\doibase 10.1103/PhysRevD.102.043507} {\bibfield  {journal} {\bibinfo
  {journal} {Phys. Rev. D}\ }\textbf {\bibinfo {volume} {102}},\ \bibinfo
  {pages} {043507} (\bibinfo {year} {2020})},\ \Eprint
  {http://arxiv.org/abs/2003.07355} {arXiv:2003.07355 [astro-ph.CO]}
  \BibitemShut {NoStop}%
\bibitem [{\citenamefont {Krishnan}\ \emph {et~al.}(2020)\citenamefont
  {Krishnan}, \citenamefont {Colg{\'a}in}, \citenamefont {Ruchika},
  \citenamefont {Sen}, \citenamefont {Sheikh-Jabbari},\ and\ \citenamefont
  {Yang}}]{Krishnan:2020obg}%
  \BibitemOpen
  \bibfield  {author} {\bibinfo {author} {\bibfnamefont {C.}~\bibnamefont
  {Krishnan}}, \bibinfo {author} {\bibfnamefont {E.~{\'O}.}\ \bibnamefont
  {Colg{\'a}in}}, \bibinfo {author} {\bibnamefont {Ruchika}}, \bibinfo {author}
  {\bibfnamefont {A.~A.}\ \bibnamefont {Sen}}, \bibinfo {author} {\bibfnamefont
  {M.~M.}\ \bibnamefont {Sheikh-Jabbari}}, \ and\ \bibinfo {author}
  {\bibfnamefont {T.}~\bibnamefont {Yang}},\ }\href {\doibase
  10.1103/PhysRevD.102.103525} {\bibfield  {journal} {\bibinfo  {journal}
  {Phys. Rev. D}\ }\textbf {\bibinfo {volume} {102}},\ \bibinfo {pages}
  {103525} (\bibinfo {year} {2020})},\ \Eprint
  {http://arxiv.org/abs/2002.06044} {arXiv:2002.06044 [astro-ph.CO]}
  \BibitemShut {NoStop}%
\bibitem [{\citenamefont {Jedamzik}\ \emph {et~al.}(2021)\citenamefont
  {Jedamzik}, \citenamefont {Pogosian},\ and\ \citenamefont
  {Zhao}}]{Jedamzik:2020zmd}%
  \BibitemOpen
  \bibfield  {author} {\bibinfo {author} {\bibfnamefont {K.}~\bibnamefont
  {Jedamzik}}, \bibinfo {author} {\bibfnamefont {L.}~\bibnamefont {Pogosian}},
  \ and\ \bibinfo {author} {\bibfnamefont {G.-B.}\ \bibnamefont {Zhao}},\
  }\href {\doibase 10.1038/s42005-021-00628-x} {\bibfield  {journal} {\bibinfo
  {journal} {Commun. in Phys.}\ }\textbf {\bibinfo {volume} {4}},\ \bibinfo
  {pages} {123} (\bibinfo {year} {2021})},\ \Eprint
  {http://arxiv.org/abs/2010.04158} {arXiv:2010.04158 [astro-ph.CO]}
  \BibitemShut {NoStop}%
\bibitem [{\citenamefont {Vagnozzi}(2023)}]{Vagnozzi:2023nrq}%
  \BibitemOpen
  \bibfield  {author} {\bibinfo {author} {\bibfnamefont {S.}~\bibnamefont
  {Vagnozzi}},\ }\href {\doibase 10.3390/universe9090393} {\bibfield  {journal}
  {\bibinfo  {journal} {Universe}\ }\textbf {\bibinfo {volume} {9}},\ \bibinfo
  {pages} {393} (\bibinfo {year} {2023})},\ \Eprint
  {http://arxiv.org/abs/2308.16628} {arXiv:2308.16628 [astro-ph.CO]}
  \BibitemShut {NoStop}%
\bibitem [{\citenamefont {Dutta}\ \emph {et~al.}(2019)\citenamefont {Dutta},
  \citenamefont {Roy}, \citenamefont {Ruchika}, \citenamefont {Sen},\ and\
  \citenamefont {Sheikh-Jabbari}}]{Dutta:2019pio}%
  \BibitemOpen
  \bibfield  {author} {\bibinfo {author} {\bibfnamefont {K.}~\bibnamefont
  {Dutta}}, \bibinfo {author} {\bibfnamefont {A.}~\bibnamefont {Roy}}, \bibinfo
  {author} {\bibnamefont {Ruchika}}, \bibinfo {author} {\bibfnamefont {A.~A.}\
  \bibnamefont {Sen}}, \ and\ \bibinfo {author} {\bibfnamefont {M.~M.}\
  \bibnamefont {Sheikh-Jabbari}},\ }\href {\doibase
  10.1103/PhysRevD.100.103501} {\bibfield  {journal} {\bibinfo  {journal}
  {Phys. Rev. D}\ }\textbf {\bibinfo {volume} {100}},\ \bibinfo {pages}
  {103501} (\bibinfo {year} {2019})},\ \Eprint
  {http://arxiv.org/abs/1908.07267} {arXiv:1908.07267 [astro-ph.CO]}
  \BibitemShut {NoStop}%
\bibitem [{\citenamefont {Wang}\ \emph {et~al.}(2016)\citenamefont {Wang},
  \citenamefont {Abdalla}, \citenamefont {Atrio-Barandela},\ and\ \citenamefont
  {Pavon}}]{Wang:2016lxa}%
  \BibitemOpen
  \bibfield  {author} {\bibinfo {author} {\bibfnamefont {B.}~\bibnamefont
  {Wang}}, \bibinfo {author} {\bibfnamefont {E.}~\bibnamefont {Abdalla}},
  \bibinfo {author} {\bibfnamefont {F.}~\bibnamefont {Atrio-Barandela}}, \ and\
  \bibinfo {author} {\bibfnamefont {D.}~\bibnamefont {Pavon}},\ }\href
  {\doibase 10.1088/0034-4885/79/9/096901} {\bibfield  {journal} {\bibinfo
  {journal} {Rept. Prog. Phys.}\ }\textbf {\bibinfo {volume} {79}},\ \bibinfo
  {pages} {096901} (\bibinfo {year} {2016})},\ \Eprint
  {http://arxiv.org/abs/1603.08299} {arXiv:1603.08299 [astro-ph.CO]}
  \BibitemShut {NoStop}%
\bibitem [{\citenamefont {Agrawal}\ \emph {et~al.}(2023)\citenamefont
  {Agrawal}, \citenamefont {Cyr-Racine}, \citenamefont {Pinner},\ and\
  \citenamefont {Randall}}]{Agrawal:2019lmo}%
  \BibitemOpen
  \bibfield  {author} {\bibinfo {author} {\bibfnamefont {P.}~\bibnamefont
  {Agrawal}}, \bibinfo {author} {\bibfnamefont {F.-Y.}\ \bibnamefont
  {Cyr-Racine}}, \bibinfo {author} {\bibfnamefont {D.}~\bibnamefont {Pinner}},
  \ and\ \bibinfo {author} {\bibfnamefont {L.}~\bibnamefont {Randall}},\ }\href
  {\doibase 10.1016/j.dark.2023.101347} {\bibfield  {journal} {\bibinfo
  {journal} {Phys. Dark Univ.}\ }\textbf {\bibinfo {volume} {42}},\ \bibinfo
  {pages} {101347} (\bibinfo {year} {2023})},\ \Eprint
  {http://arxiv.org/abs/1904.01016} {arXiv:1904.01016 [astro-ph.CO]}
  \BibitemShut {NoStop}%
\bibitem [{\citenamefont {Li}\ and\ \citenamefont
  {Shafieloo}(2019)}]{Li:2019yem}%
  \BibitemOpen
  \bibfield  {author} {\bibinfo {author} {\bibfnamefont {X.}~\bibnamefont
  {Li}}\ and\ \bibinfo {author} {\bibfnamefont {A.}~\bibnamefont {Shafieloo}},\
  }\href {\doibase 10.3847/2041-8213/ab3e09} {\bibfield  {journal} {\bibinfo
  {journal} {Astrophys. J. Lett.}\ }\textbf {\bibinfo {volume} {883}},\
  \bibinfo {pages} {L3} (\bibinfo {year} {2019})},\ \Eprint
  {http://arxiv.org/abs/1906.08275} {arXiv:1906.08275 [astro-ph.CO]}
  \BibitemShut {NoStop}%
\bibitem [{\citenamefont {Li}(2004)}]{Li:2004rb}%
  \BibitemOpen
  \bibfield  {author} {\bibinfo {author} {\bibfnamefont {M.}~\bibnamefont
  {Li}},\ }\href {\doibase 10.1016/j.physletb.2004.10.014} {\bibfield
  {journal} {\bibinfo  {journal} {Phys. Lett. B}\ }\textbf {\bibinfo {volume}
  {603}},\ \bibinfo {pages} {1} (\bibinfo {year} {2004})},\ \Eprint
  {http://arxiv.org/abs/hep-th/0403127} {arXiv:hep-th/0403127} \BibitemShut
  {NoStop}%
\bibitem [{\citenamefont {Shapiro}\ and\ \citenamefont
  {Sola}(2002)}]{Shapiro:2000dz}%
  \BibitemOpen
  \bibfield  {author} {\bibinfo {author} {\bibfnamefont {I.~L.}\ \bibnamefont
  {Shapiro}}\ and\ \bibinfo {author} {\bibfnamefont {J.}~\bibnamefont {Sola}},\
  }\href {\doibase 10.1088/1126-6708/2002/02/006} {\bibfield  {journal}
  {\bibinfo  {journal} {JHEP}\ }\textbf {\bibinfo {volume} {02}},\ \bibinfo
  {pages} {006} (\bibinfo {year} {2002})},\ \Eprint
  {http://arxiv.org/abs/hep-th/0012227} {arXiv:hep-th/0012227} \BibitemShut
  {NoStop}%
\bibitem [{\citenamefont {Zhang}\ \emph {et~al.}(2012)\citenamefont {Zhang},
  \citenamefont {Li}, \citenamefont {Li}, \citenamefont {Zhang},\ and\
  \citenamefont {Li}}]{Zhang:2012uu}%
  \BibitemOpen
  \bibfield  {author} {\bibinfo {author} {\bibfnamefont {Z.}~\bibnamefont
  {Zhang}}, \bibinfo {author} {\bibfnamefont {S.}~\bibnamefont {Li}}, \bibinfo
  {author} {\bibfnamefont {X.-D.}\ \bibnamefont {Li}}, \bibinfo {author}
  {\bibfnamefont {X.}~\bibnamefont {Zhang}}, \ and\ \bibinfo {author}
  {\bibfnamefont {M.}~\bibnamefont {Li}},\ }\href {\doibase
  10.1088/1475-7516/2012/06/009} {\bibfield  {journal} {\bibinfo  {journal}
  {JCAP}\ }\textbf {\bibinfo {volume} {06}},\ \bibinfo {pages} {009} (\bibinfo
  {year} {2012})},\ \Eprint {http://arxiv.org/abs/1204.6135} {arXiv:1204.6135
  [astro-ph.CO]} \BibitemShut {NoStop}%
\bibitem [{\citenamefont {Li}\ \emph {et~al.}(2013)\citenamefont {Li},
  \citenamefont {Wang}, \citenamefont {Li},\ and\ \citenamefont
  {Zhang}}]{Li:2012spm}%
  \BibitemOpen
  \bibfield  {author} {\bibinfo {author} {\bibfnamefont {Y.-H.}\ \bibnamefont
  {Li}}, \bibinfo {author} {\bibfnamefont {S.}~\bibnamefont {Wang}}, \bibinfo
  {author} {\bibfnamefont {X.-D.}\ \bibnamefont {Li}}, \ and\ \bibinfo {author}
  {\bibfnamefont {X.}~\bibnamefont {Zhang}},\ }\href {\doibase
  10.1088/1475-7516/2013/02/033} {\bibfield  {journal} {\bibinfo  {journal}
  {JCAP}\ }\textbf {\bibinfo {volume} {02}},\ \bibinfo {pages} {033} (\bibinfo
  {year} {2013})},\ \Eprint {http://arxiv.org/abs/1207.6679} {arXiv:1207.6679
  [astro-ph.CO]} \BibitemShut {NoStop}%
\bibitem [{\citenamefont {Zhang}\ \emph
  {et~al.}(2015{\natexlab{b}})\citenamefont {Zhang}, \citenamefont {Zhao},
  \citenamefont {Li},\ and\ \citenamefont {Zhang}}]{Zhang:2015rha}%
  \BibitemOpen
  \bibfield  {author} {\bibinfo {author} {\bibfnamefont {J.-F.}\ \bibnamefont
  {Zhang}}, \bibinfo {author} {\bibfnamefont {M.-M.}\ \bibnamefont {Zhao}},
  \bibinfo {author} {\bibfnamefont {Y.-H.}\ \bibnamefont {Li}}, \ and\ \bibinfo
  {author} {\bibfnamefont {X.}~\bibnamefont {Zhang}},\ }\href {\doibase
  10.1088/1475-7516/2015/04/038} {\bibfield  {journal} {\bibinfo  {journal}
  {JCAP}\ }\textbf {\bibinfo {volume} {04}},\ \bibinfo {pages} {038} (\bibinfo
  {year} {2015}{\natexlab{b}})},\ \Eprint {http://arxiv.org/abs/1502.04028}
  {arXiv:1502.04028 [astro-ph.CO]} \BibitemShut {NoStop}%
\bibitem [{\citenamefont {Feng}\ and\ \citenamefont
  {Zhang}(2016)}]{Feng:2016djj}%
  \BibitemOpen
  \bibfield  {author} {\bibinfo {author} {\bibfnamefont {L.}~\bibnamefont
  {Feng}}\ and\ \bibinfo {author} {\bibfnamefont {X.}~\bibnamefont {Zhang}},\
  }\href {\doibase 10.1088/1475-7516/2016/08/072} {\bibfield  {journal}
  {\bibinfo  {journal} {JCAP}\ }\textbf {\bibinfo {volume} {08}},\ \bibinfo
  {pages} {072} (\bibinfo {year} {2016})},\ \Eprint
  {http://arxiv.org/abs/1607.05567} {arXiv:1607.05567 [astro-ph.CO]}
  \BibitemShut {NoStop}%
\bibitem [{\citenamefont {Aluri}\ \emph {et~al.}(2023)\citenamefont {Aluri}
  \emph {et~al.}}]{Aluri:2022hzs}%
  \BibitemOpen
  \bibfield  {author} {\bibinfo {author} {\bibfnamefont {P.~K.}\ \bibnamefont
  {Aluri}} \emph {et~al.},\ }\href {\doibase 10.1088/1361-6382/acbefc}
  {\bibfield  {journal} {\bibinfo  {journal} {Class. Quant. Grav.}\ }\textbf
  {\bibinfo {volume} {40}},\ \bibinfo {pages} {094001} (\bibinfo {year}
  {2023})},\ \Eprint {http://arxiv.org/abs/2207.05765} {arXiv:2207.05765
  [astro-ph.CO]} \BibitemShut {NoStop}%
\bibitem [{\citenamefont {Capozziello}\ and\ \citenamefont
  {De~Laurentis}(2011)}]{Capozziello:2011et}%
  \BibitemOpen
  \bibfield  {author} {\bibinfo {author} {\bibfnamefont {S.}~\bibnamefont
  {Capozziello}}\ and\ \bibinfo {author} {\bibfnamefont {M.}~\bibnamefont
  {De~Laurentis}},\ }\href {\doibase 10.1016/j.physrep.2011.09.003} {\bibfield
  {journal} {\bibinfo  {journal} {Phys. Rept.}\ }\textbf {\bibinfo {volume}
  {509}},\ \bibinfo {pages} {167} (\bibinfo {year} {2011})},\ \Eprint
  {http://arxiv.org/abs/1108.6266} {arXiv:1108.6266 [gr-qc]} \BibitemShut
  {NoStop}%
\bibitem [{\citenamefont {De~Felice}\ and\ \citenamefont
  {Tsujikawa}(2010)}]{DeFelice:2010aj}%
  \BibitemOpen
  \bibfield  {author} {\bibinfo {author} {\bibfnamefont {A.}~\bibnamefont
  {De~Felice}}\ and\ \bibinfo {author} {\bibfnamefont {S.}~\bibnamefont
  {Tsujikawa}},\ }\href {\doibase 10.12942/lrr-2010-3} {\bibfield  {journal}
  {\bibinfo  {journal} {Living Rev. Rel.}\ }\textbf {\bibinfo {volume} {13}},\
  \bibinfo {pages} {3} (\bibinfo {year} {2010})},\ \Eprint
  {http://arxiv.org/abs/1002.4928} {arXiv:1002.4928 [gr-qc]} \BibitemShut
  {NoStop}%
\bibitem [{\citenamefont {Mortonson}\ \emph {et~al.}(2009)\citenamefont
  {Mortonson}, \citenamefont {Hu},\ and\ \citenamefont
  {Huterer}}]{Mortonson:2009qq}%
  \BibitemOpen
  \bibfield  {author} {\bibinfo {author} {\bibfnamefont {M.~J.}\ \bibnamefont
  {Mortonson}}, \bibinfo {author} {\bibfnamefont {W.}~\bibnamefont {Hu}}, \
  and\ \bibinfo {author} {\bibfnamefont {D.}~\bibnamefont {Huterer}},\ }\href
  {\doibase 10.1103/PhysRevD.80.067301} {\bibfield  {journal} {\bibinfo
  {journal} {Phys. Rev. D}\ }\textbf {\bibinfo {volume} {80}},\ \bibinfo
  {pages} {067301} (\bibinfo {year} {2009})},\ \Eprint
  {http://arxiv.org/abs/0908.1408} {arXiv:0908.1408 [astro-ph.CO]} \BibitemShut
  {NoStop}%
\bibitem [{\citenamefont {Kazantzidis}\ and\ \citenamefont
  {Perivolaropoulos}(2019)}]{Kazantzidis:2019nuh}%
  \BibitemOpen
  \bibfield  {author} {\bibinfo {author} {\bibfnamefont {L.}~\bibnamefont
  {Kazantzidis}}\ and\ \bibinfo {author} {\bibfnamefont {L.}~\bibnamefont
  {Perivolaropoulos}},\ }\href@noop {} {\  (\bibinfo {year} {2019})},\ \Eprint
  {http://arxiv.org/abs/1907.03176} {arXiv:1907.03176 [astro-ph.CO]}
  \BibitemShut {NoStop}%
\bibitem [{\citenamefont {Benevento}\ \emph {et~al.}(2020)\citenamefont
  {Benevento}, \citenamefont {Hu},\ and\ \citenamefont
  {Raveri}}]{Benevento:2020fev}%
  \BibitemOpen
  \bibfield  {author} {\bibinfo {author} {\bibfnamefont {G.}~\bibnamefont
  {Benevento}}, \bibinfo {author} {\bibfnamefont {W.}~\bibnamefont {Hu}}, \
  and\ \bibinfo {author} {\bibfnamefont {M.}~\bibnamefont {Raveri}},\ }\href
  {\doibase 10.1103/PhysRevD.101.103517} {\bibfield  {journal} {\bibinfo
  {journal} {Phys. Rev. D}\ }\textbf {\bibinfo {volume} {101}},\ \bibinfo
  {pages} {103517} (\bibinfo {year} {2020})},\ \Eprint
  {http://arxiv.org/abs/2002.11707} {arXiv:2002.11707 [astro-ph.CO]}
  \BibitemShut {NoStop}%
\bibitem [{\citenamefont {Yang}\ \emph {et~al.}(2021)\citenamefont {Yang},
  \citenamefont {Di~Valentino}, \citenamefont {Pan}, \citenamefont {Wu},\ and\
  \citenamefont {Lu}}]{Yang:2021flj}%
  \BibitemOpen
  \bibfield  {author} {\bibinfo {author} {\bibfnamefont {W.}~\bibnamefont
  {Yang}}, \bibinfo {author} {\bibfnamefont {E.}~\bibnamefont {Di~Valentino}},
  \bibinfo {author} {\bibfnamefont {S.}~\bibnamefont {Pan}}, \bibinfo {author}
  {\bibfnamefont {Y.}~\bibnamefont {Wu}}, \ and\ \bibinfo {author}
  {\bibfnamefont {J.}~\bibnamefont {Lu}},\ }\href {\doibase
  10.1093/mnras/staa3914} {\bibfield  {journal} {\bibinfo  {journal} {Mon. Not.
  Roy. Astron. Soc.}\ }\textbf {\bibinfo {volume} {501}},\ \bibinfo {pages}
  {5845} (\bibinfo {year} {2021})},\ \Eprint {http://arxiv.org/abs/2101.02168}
  {arXiv:2101.02168 [astro-ph.CO]} \BibitemShut {NoStop}%
\bibitem [{\citenamefont {Wright}\ \emph {et~al.}(2025)\citenamefont {Wright}
  \emph {et~al.}}]{Wright:2025xka}%
  \BibitemOpen
  \bibfield  {author} {\bibinfo {author} {\bibfnamefont {A.~H.}\ \bibnamefont
  {Wright}} \emph {et~al.},\ }\href@noop {} {\  (\bibinfo {year} {2025})},\
  \Eprint {http://arxiv.org/abs/2503.19441} {arXiv:2503.19441 [astro-ph.CO]}
  \BibitemShut {NoStop}%
\bibitem [{\citenamefont {Guo}\ \emph {et~al.}(2019)\citenamefont {Guo},
  \citenamefont {Zhang},\ and\ \citenamefont {Zhang}}]{Guo:2018ans}%
  \BibitemOpen
  \bibfield  {author} {\bibinfo {author} {\bibfnamefont {R.-Y.}\ \bibnamefont
  {Guo}}, \bibinfo {author} {\bibfnamefont {J.-F.}\ \bibnamefont {Zhang}}, \
  and\ \bibinfo {author} {\bibfnamefont {X.}~\bibnamefont {Zhang}},\ }\href
  {\doibase 10.1088/1475-7516/2019/02/054} {\bibfield  {journal} {\bibinfo
  {journal} {JCAP}\ }\textbf {\bibinfo {volume} {02}},\ \bibinfo {pages} {054}
  (\bibinfo {year} {2019})},\ \Eprint {http://arxiv.org/abs/1809.02340}
  {arXiv:1809.02340 [astro-ph.CO]} \BibitemShut {NoStop}%
\bibitem [{\citenamefont {Gao}\ \emph {et~al.}(2021)\citenamefont {Gao},
  \citenamefont {Zhao}, \citenamefont {Xue},\ and\ \citenamefont
  {Zhang}}]{Gao:2021xnk}%
  \BibitemOpen
  \bibfield  {author} {\bibinfo {author} {\bibfnamefont {L.-Y.}\ \bibnamefont
  {Gao}}, \bibinfo {author} {\bibfnamefont {Z.-W.}\ \bibnamefont {Zhao}},
  \bibinfo {author} {\bibfnamefont {S.-S.}\ \bibnamefont {Xue}}, \ and\
  \bibinfo {author} {\bibfnamefont {X.}~\bibnamefont {Zhang}},\ }\href
  {\doibase 10.1088/1475-7516/2021/07/005} {\bibfield  {journal} {\bibinfo
  {journal} {JCAP}\ }\textbf {\bibinfo {volume} {07}},\ \bibinfo {pages} {005}
  (\bibinfo {year} {2021})},\ \Eprint {http://arxiv.org/abs/2101.10714}
  {arXiv:2101.10714 [astro-ph.CO]} \BibitemShut {NoStop}%
\bibitem [{\citenamefont {Heisenberg}\ \emph {et~al.}(2022)\citenamefont
  {Heisenberg}, \citenamefont {Villarrubia-Rojo},\ and\ \citenamefont
  {Zosso}}]{Heisenberg:2022gqk}%
  \BibitemOpen
  \bibfield  {author} {\bibinfo {author} {\bibfnamefont {L.}~\bibnamefont
  {Heisenberg}}, \bibinfo {author} {\bibfnamefont {H.}~\bibnamefont
  {Villarrubia-Rojo}}, \ and\ \bibinfo {author} {\bibfnamefont
  {J.}~\bibnamefont {Zosso}},\ }\href {\doibase 10.1103/PhysRevD.106.043503}
  {\bibfield  {journal} {\bibinfo  {journal} {Phys. Rev. D}\ }\textbf {\bibinfo
  {volume} {106}},\ \bibinfo {pages} {043503} (\bibinfo {year} {2022})},\
  \Eprint {http://arxiv.org/abs/2202.01202} {arXiv:2202.01202 [astro-ph.CO]}
  \BibitemShut {NoStop}%
\bibitem [{\citenamefont {Alestas}\ and\ \citenamefont
  {Perivolaropoulos}(2021)}]{Alestas:2021xes}%
  \BibitemOpen
  \bibfield  {author} {\bibinfo {author} {\bibfnamefont {G.}~\bibnamefont
  {Alestas}}\ and\ \bibinfo {author} {\bibfnamefont {L.}~\bibnamefont
  {Perivolaropoulos}},\ }\href {\doibase 10.1093/mnras/stab1070} {\bibfield
  {journal} {\bibinfo  {journal} {Mon. Not. Roy. Astron. Soc.}\ }\textbf
  {\bibinfo {volume} {504}},\ \bibinfo {pages} {3956} (\bibinfo {year}
  {2021})},\ \Eprint {http://arxiv.org/abs/2103.04045} {arXiv:2103.04045
  [astro-ph.CO]} \BibitemShut {NoStop}%
\bibitem [{\citenamefont {Birrer}\ \emph {et~al.}(2019)\citenamefont {Birrer}
  \emph {et~al.}}]{H0LiCOW:2018tyj}%
  \BibitemOpen
  \bibfield  {author} {\bibinfo {author} {\bibfnamefont {S.}~\bibnamefont
  {Birrer}} \emph {et~al.} (\bibinfo {collaboration} {H0LiCOW}),\ }\href
  {\doibase 10.1093/mnras/stz200} {\bibfield  {journal} {\bibinfo  {journal}
  {Mon. Not. Roy. Astron. Soc.}\ }\textbf {\bibinfo {volume} {484}},\ \bibinfo
  {pages} {4726} (\bibinfo {year} {2019})},\ \Eprint
  {http://arxiv.org/abs/1809.01274} {arXiv:1809.01274 [astro-ph.CO]}
  \BibitemShut {NoStop}%
\bibitem [{\citenamefont {Shajib}\ \emph {et~al.}(2020)\citenamefont {Shajib}
  \emph {et~al.}}]{DES:2019fny}%
  \BibitemOpen
  \bibfield  {author} {\bibinfo {author} {\bibfnamefont {A.~J.}\ \bibnamefont
  {Shajib}} \emph {et~al.} (\bibinfo {collaboration} {DES}),\ }\href {\doibase
  10.1093/mnras/staa828} {\bibfield  {journal} {\bibinfo  {journal} {Mon. Not.
  Roy. Astron. Soc.}\ }\textbf {\bibinfo {volume} {494}},\ \bibinfo {pages}
  {6072} (\bibinfo {year} {2020})},\ \Eprint {http://arxiv.org/abs/1910.06306}
  {arXiv:1910.06306 [astro-ph.CO]} \BibitemShut {NoStop}%
\bibitem [{\citenamefont {Guidorzi}\ \emph {et~al.}(2017)\citenamefont
  {Guidorzi} \emph {et~al.}}]{Guidorzi:2017ogy}%
  \BibitemOpen
  \bibfield  {author} {\bibinfo {author} {\bibfnamefont {C.}~\bibnamefont
  {Guidorzi}} \emph {et~al.},\ }\href {\doibase 10.3847/2041-8213/aaa009}
  {\bibfield  {journal} {\bibinfo  {journal} {Astrophys. J. Lett.}\ }\textbf
  {\bibinfo {volume} {851}},\ \bibinfo {pages} {L36} (\bibinfo {year}
  {2017})},\ \Eprint {http://arxiv.org/abs/1710.06426} {arXiv:1710.06426
  [astro-ph.CO]} \BibitemShut {NoStop}%
\bibitem [{\citenamefont {Song}\ \emph {et~al.}(2025)\citenamefont {Song},
  \citenamefont {Qi}, \citenamefont {Zhang},\ and\ \citenamefont
  {Zhang}}]{Song:2025ddm}%
  \BibitemOpen
  \bibfield  {author} {\bibinfo {author} {\bibfnamefont {J.-Y.}\ \bibnamefont
  {Song}}, \bibinfo {author} {\bibfnamefont {J.-Z.}\ \bibnamefont {Qi}},
  \bibinfo {author} {\bibfnamefont {J.-F.}\ \bibnamefont {Zhang}}, \ and\
  \bibinfo {author} {\bibfnamefont {X.}~\bibnamefont {Zhang}},\ }\href
  {\doibase 10.3847/2041-8213/add999} {\bibfield  {journal} {\bibinfo
  {journal} {Astrophys. J. Lett.}\ }\textbf {\bibinfo {volume} {985}},\
  \bibinfo {pages} {L44} (\bibinfo {year} {2025})},\ \Eprint
  {http://arxiv.org/abs/2503.10346} {arXiv:2503.10346 [astro-ph.CO]}
  \BibitemShut {NoStop}%
\bibitem [{\citenamefont {Jin}\ \emph {et~al.}(2025)\citenamefont {Jin},
  \citenamefont {Song}, \citenamefont {Sun}, \citenamefont {Xiao},
  \citenamefont {Wang}, \citenamefont {Wang}, \citenamefont {Zhang},\ and\
  \citenamefont {Zhang}}]{Jin:2025dvf}%
  \BibitemOpen
  \bibfield  {author} {\bibinfo {author} {\bibfnamefont {S.-J.}\ \bibnamefont
  {Jin}}, \bibinfo {author} {\bibfnamefont {J.-Y.}\ \bibnamefont {Song}},
  \bibinfo {author} {\bibfnamefont {T.-Y.}\ \bibnamefont {Sun}}, \bibinfo
  {author} {\bibfnamefont {S.-R.}\ \bibnamefont {Xiao}}, \bibinfo {author}
  {\bibfnamefont {H.}~\bibnamefont {Wang}}, \bibinfo {author} {\bibfnamefont
  {L.-F.}\ \bibnamefont {Wang}}, \bibinfo {author} {\bibfnamefont {J.-F.}\
  \bibnamefont {Zhang}}, \ and\ \bibinfo {author} {\bibfnamefont
  {X.}~\bibnamefont {Zhang}},\ }\href@noop {} {\  (\bibinfo {year} {2025})},\
  \Eprint {http://arxiv.org/abs/2507.12965} {arXiv:2507.12965 [astro-ph.CO]}
  \BibitemShut {NoStop}%
\bibitem [{\citenamefont {Pesce}\ \emph {et~al.}(2020)\citenamefont {Pesce}
  \emph {et~al.}}]{Pesce:2020xfe}%
  \BibitemOpen
  \bibfield  {author} {\bibinfo {author} {\bibfnamefont {D.~W.}\ \bibnamefont
  {Pesce}} \emph {et~al.},\ }\href {\doibase 10.3847/2041-8213/ab75f0}
  {\bibfield  {journal} {\bibinfo  {journal} {Astrophys. J. Lett.}\ }\textbf
  {\bibinfo {volume} {891}},\ \bibinfo {pages} {L1} (\bibinfo {year} {2020})},\
  \Eprint {http://arxiv.org/abs/2001.09213} {arXiv:2001.09213 [astro-ph.CO]}
  \BibitemShut {NoStop}%
\bibitem [{\citenamefont {Freedman}(2021)}]{Freedman:2021ahq}%
  \BibitemOpen
  \bibfield  {author} {\bibinfo {author} {\bibfnamefont {W.~L.}\ \bibnamefont
  {Freedman}},\ }\href {\doibase 10.3847/1538-4357/ac0e95} {\bibfield
  {journal} {\bibinfo  {journal} {Astrophys. J.}\ }\textbf {\bibinfo {volume}
  {919}},\ \bibinfo {pages} {16} (\bibinfo {year} {2021})},\ \Eprint
  {http://arxiv.org/abs/2106.15656} {arXiv:2106.15656 [astro-ph.CO]}
  \BibitemShut {NoStop}%
\bibitem [{\citenamefont {Blakeslee}\ \emph {et~al.}(2021)\citenamefont
  {Blakeslee}, \citenamefont {Jensen}, \citenamefont {Ma}, \citenamefont
  {Milne},\ and\ \citenamefont {Greene}}]{Blakeslee:2021rqi}%
  \BibitemOpen
  \bibfield  {author} {\bibinfo {author} {\bibfnamefont {J.~P.}\ \bibnamefont
  {Blakeslee}}, \bibinfo {author} {\bibfnamefont {J.~B.}\ \bibnamefont
  {Jensen}}, \bibinfo {author} {\bibfnamefont {C.-P.}\ \bibnamefont {Ma}},
  \bibinfo {author} {\bibfnamefont {P.~A.}\ \bibnamefont {Milne}}, \ and\
  \bibinfo {author} {\bibfnamefont {J.~E.}\ \bibnamefont {Greene}},\ }\href
  {\doibase 10.3847/1538-4357/abe86a} {\bibfield  {journal} {\bibinfo
  {journal} {Astrophys. J.}\ }\textbf {\bibinfo {volume} {911}},\ \bibinfo
  {pages} {65} (\bibinfo {year} {2021})},\ \Eprint
  {http://arxiv.org/abs/2101.02221} {arXiv:2101.02221 [astro-ph.CO]}
  \BibitemShut {NoStop}%
\bibitem [{\citenamefont {Huang}\ \emph {et~al.}(2019)\citenamefont {Huang},
  \citenamefont {Riess}, \citenamefont {Yuan}, \citenamefont {Macri},
  \citenamefont {Zakamska}, \citenamefont {Casertano}, \citenamefont
  {Whitelock}, \citenamefont {Hoffmann}, \citenamefont {Filippenko},\ and\
  \citenamefont {Scolnic}}]{Huang:2019yhh}%
  \BibitemOpen
  \bibfield  {author} {\bibinfo {author} {\bibfnamefont {C.~D.}\ \bibnamefont
  {Huang}}, \bibinfo {author} {\bibfnamefont {A.~G.}\ \bibnamefont {Riess}},
  \bibinfo {author} {\bibfnamefont {W.}~\bibnamefont {Yuan}}, \bibinfo {author}
  {\bibfnamefont {L.~M.}\ \bibnamefont {Macri}}, \bibinfo {author}
  {\bibfnamefont {N.~L.}\ \bibnamefont {Zakamska}}, \bibinfo {author}
  {\bibfnamefont {S.}~\bibnamefont {Casertano}}, \bibinfo {author}
  {\bibfnamefont {P.~A.}\ \bibnamefont {Whitelock}}, \bibinfo {author}
  {\bibfnamefont {S.~L.}\ \bibnamefont {Hoffmann}}, \bibinfo {author}
  {\bibfnamefont {A.~V.}\ \bibnamefont {Filippenko}}, \ and\ \bibinfo {author}
  {\bibfnamefont {D.}~\bibnamefont {Scolnic}},\ }\href {\doibase
  10.3847/1538-4357/ab5dbd} {\  (\bibinfo {year} {2019}),\
  10.3847/1538-4357/ab5dbd},\ \Eprint {http://arxiv.org/abs/1908.10883}
  {arXiv:1908.10883 [astro-ph.CO]} \BibitemShut {NoStop}%
\bibitem [{\citenamefont {Aubourg}\ \emph {et~al.}(2015)\citenamefont {Aubourg}
  \emph {et~al.}}]{BOSS:2014hhw}%
  \BibitemOpen
  \bibfield  {author} {\bibinfo {author} {\bibfnamefont {{\'E}.}~\bibnamefont
  {Aubourg}} \emph {et~al.} (\bibinfo {collaboration} {BOSS}),\ }\href
  {\doibase 10.1103/PhysRevD.92.123516} {\bibfield  {journal} {\bibinfo
  {journal} {Phys. Rev. D}\ }\textbf {\bibinfo {volume} {92}},\ \bibinfo
  {pages} {123516} (\bibinfo {year} {2015})},\ \Eprint
  {http://arxiv.org/abs/1411.1074} {arXiv:1411.1074 [astro-ph.CO]} \BibitemShut
  {NoStop}%
\bibitem [{\citenamefont {Cuesta}\ \emph {et~al.}(2015)\citenamefont {Cuesta},
  \citenamefont {Verde}, \citenamefont {Riess},\ and\ \citenamefont
  {Jimenez}}]{Cuesta:2014asa}%
  \BibitemOpen
  \bibfield  {author} {\bibinfo {author} {\bibfnamefont {A.~J.}\ \bibnamefont
  {Cuesta}}, \bibinfo {author} {\bibfnamefont {L.}~\bibnamefont {Verde}},
  \bibinfo {author} {\bibfnamefont {A.}~\bibnamefont {Riess}}, \ and\ \bibinfo
  {author} {\bibfnamefont {R.}~\bibnamefont {Jimenez}},\ }\href {\doibase
  10.1093/mnras/stv261} {\bibfield  {journal} {\bibinfo  {journal} {Mon. Not.
  Roy. Astron. Soc.}\ }\textbf {\bibinfo {volume} {448}},\ \bibinfo {pages}
  {3463} (\bibinfo {year} {2015})},\ \Eprint {http://arxiv.org/abs/1411.1094}
  {arXiv:1411.1094 [astro-ph.CO]} \BibitemShut {NoStop}%
\bibitem [{\citenamefont {Verde}\ \emph {et~al.}(2017)\citenamefont {Verde},
  \citenamefont {Bernal}, \citenamefont {Heavens},\ and\ \citenamefont
  {Jimenez}}]{Verde:2016ccp}%
  \BibitemOpen
  \bibfield  {author} {\bibinfo {author} {\bibfnamefont {L.}~\bibnamefont
  {Verde}}, \bibinfo {author} {\bibfnamefont {J.~L.}\ \bibnamefont {Bernal}},
  \bibinfo {author} {\bibfnamefont {A.~F.}\ \bibnamefont {Heavens}}, \ and\
  \bibinfo {author} {\bibfnamefont {R.}~\bibnamefont {Jimenez}},\ }\href
  {\doibase 10.1093/mnras/stx116} {\bibfield  {journal} {\bibinfo  {journal}
  {Mon. Not. Roy. Astron. Soc.}\ }\textbf {\bibinfo {volume} {467}},\ \bibinfo
  {pages} {731} (\bibinfo {year} {2017})},\ \Eprint
  {http://arxiv.org/abs/1607.05297} {arXiv:1607.05297 [astro-ph.CO]}
  \BibitemShut {NoStop}%
\bibitem [{\citenamefont {Lemos}\ \emph {et~al.}(2019)\citenamefont {Lemos},
  \citenamefont {Lee}, \citenamefont {Efstathiou},\ and\ \citenamefont
  {Gratton}}]{Lemos:2018smw}%
  \BibitemOpen
  \bibfield  {author} {\bibinfo {author} {\bibfnamefont {P.}~\bibnamefont
  {Lemos}}, \bibinfo {author} {\bibfnamefont {E.}~\bibnamefont {Lee}}, \bibinfo
  {author} {\bibfnamefont {G.}~\bibnamefont {Efstathiou}}, \ and\ \bibinfo
  {author} {\bibfnamefont {S.}~\bibnamefont {Gratton}},\ }\href {\doibase
  10.1093/mnras/sty3082} {\bibfield  {journal} {\bibinfo  {journal} {Mon. Not.
  Roy. Astron. Soc.}\ }\textbf {\bibinfo {volume} {483}},\ \bibinfo {pages}
  {4803} (\bibinfo {year} {2019})},\ \Eprint {http://arxiv.org/abs/1806.06781}
  {arXiv:1806.06781 [astro-ph.CO]} \BibitemShut {NoStop}%
\bibitem [{\citenamefont {Yu}\ \emph {et~al.}(2018)\citenamefont {Yu},
  \citenamefont {Ratra},\ and\ \citenamefont {Wang}}]{Yu:2017iju}%
  \BibitemOpen
  \bibfield  {author} {\bibinfo {author} {\bibfnamefont {H.}~\bibnamefont
  {Yu}}, \bibinfo {author} {\bibfnamefont {B.}~\bibnamefont {Ratra}}, \ and\
  \bibinfo {author} {\bibfnamefont {F.-Y.}\ \bibnamefont {Wang}},\ }\href
  {\doibase 10.3847/1538-4357/aab0a2} {\bibfield  {journal} {\bibinfo
  {journal} {Astrophys. J.}\ }\textbf {\bibinfo {volume} {856}},\ \bibinfo
  {pages} {3} (\bibinfo {year} {2018})},\ \Eprint
  {http://arxiv.org/abs/1711.03437} {arXiv:1711.03437 [astro-ph.CO]}
  \BibitemShut {NoStop}%
\bibitem [{\citenamefont {G{\'o}mez-Valent}\ and\ \citenamefont
  {Amendola}(2018)}]{Gomez-Valent:2018hwc}%
  \BibitemOpen
  \bibfield  {author} {\bibinfo {author} {\bibfnamefont {A.}~\bibnamefont
  {G{\'o}mez-Valent}}\ and\ \bibinfo {author} {\bibfnamefont {L.}~\bibnamefont
  {Amendola}},\ }\href {\doibase 10.1088/1475-7516/2018/04/051} {\bibfield
  {journal} {\bibinfo  {journal} {JCAP}\ }\textbf {\bibinfo {volume} {04}},\
  \bibinfo {pages} {051} (\bibinfo {year} {2018})},\ \Eprint
  {http://arxiv.org/abs/1802.01505} {arXiv:1802.01505 [astro-ph.CO]}
  \BibitemShut {NoStop}%
\bibitem [{\citenamefont {Haridasu}\ \emph {et~al.}(2018)\citenamefont
  {Haridasu}, \citenamefont {Lukovi{\'c}}, \citenamefont {Moresco},\ and\
  \citenamefont {Vittorio}}]{Haridasu:2018gqm}%
  \BibitemOpen
  \bibfield  {author} {\bibinfo {author} {\bibfnamefont {B.~S.}\ \bibnamefont
  {Haridasu}}, \bibinfo {author} {\bibfnamefont {V.~V.}\ \bibnamefont
  {Lukovi{\'c}}}, \bibinfo {author} {\bibfnamefont {M.}~\bibnamefont
  {Moresco}}, \ and\ \bibinfo {author} {\bibfnamefont {N.}~\bibnamefont
  {Vittorio}},\ }\href {\doibase 10.1088/1475-7516/2018/10/015} {\bibfield
  {journal} {\bibinfo  {journal} {JCAP}\ }\textbf {\bibinfo {volume} {10}},\
  \bibinfo {pages} {015} (\bibinfo {year} {2018})},\ \Eprint
  {http://arxiv.org/abs/1805.03595} {arXiv:1805.03595 [astro-ph.CO]}
  \BibitemShut {NoStop}%
\bibitem [{\citenamefont {G{\'o}mez-Valent}\ \emph {et~al.}(2024)\citenamefont
  {G{\'o}mez-Valent}, \citenamefont {Favale}, \citenamefont {Migliaccio},\ and\
  \citenamefont {Sen}}]{Gomez-Valent:2023uof}%
  \BibitemOpen
  \bibfield  {author} {\bibinfo {author} {\bibfnamefont {A.}~\bibnamefont
  {G{\'o}mez-Valent}}, \bibinfo {author} {\bibfnamefont {A.}~\bibnamefont
  {Favale}}, \bibinfo {author} {\bibfnamefont {M.}~\bibnamefont {Migliaccio}},
  \ and\ \bibinfo {author} {\bibfnamefont {A.~A.}\ \bibnamefont {Sen}},\ }\href
  {\doibase 10.1103/PhysRevD.109.023525} {\bibfield  {journal} {\bibinfo
  {journal} {Phys. Rev. D}\ }\textbf {\bibinfo {volume} {109}},\ \bibinfo
  {pages} {023525} (\bibinfo {year} {2024})},\ \Eprint
  {http://arxiv.org/abs/2309.07795} {arXiv:2309.07795 [astro-ph.CO]}
  \BibitemShut {NoStop}%
\bibitem [{\citenamefont {Percival}\ \emph {et~al.}(2010)\citenamefont
  {Percival} \emph {et~al.}}]{SDSS:2009ocz}%
  \BibitemOpen
  \bibfield  {author} {\bibinfo {author} {\bibfnamefont {W.~J.}\ \bibnamefont
  {Percival}} \emph {et~al.} (\bibinfo {collaboration} {SDSS}),\ }\href
  {\doibase 10.1111/j.1365-2966.2009.15812.x} {\bibfield  {journal} {\bibinfo
  {journal} {Mon. Not. Roy. Astron. Soc.}\ }\textbf {\bibinfo {volume} {401}},\
  \bibinfo {pages} {2148} (\bibinfo {year} {2010})},\ \Eprint
  {http://arxiv.org/abs/0907.1660} {arXiv:0907.1660 [astro-ph.CO]} \BibitemShut
  {NoStop}%
\bibitem [{\citenamefont {Heavens}\ \emph {et~al.}(2014)\citenamefont
  {Heavens}, \citenamefont {Jimenez},\ and\ \citenamefont
  {Verde}}]{Heavens:2014rja}%
  \BibitemOpen
  \bibfield  {author} {\bibinfo {author} {\bibfnamefont {A.}~\bibnamefont
  {Heavens}}, \bibinfo {author} {\bibfnamefont {R.}~\bibnamefont {Jimenez}}, \
  and\ \bibinfo {author} {\bibfnamefont {L.}~\bibnamefont {Verde}},\ }\href
  {\doibase 10.1103/PhysRevLett.113.241302} {\bibfield  {journal} {\bibinfo
  {journal} {Phys. Rev. Lett.}\ }\textbf {\bibinfo {volume} {113}},\ \bibinfo
  {pages} {241302} (\bibinfo {year} {2014})},\ \Eprint
  {http://arxiv.org/abs/1409.6217} {arXiv:1409.6217 [astro-ph.CO]} \BibitemShut
  {NoStop}%
\bibitem [{\citenamefont {Abbott}\ \emph {et~al.}(2018)\citenamefont {Abbott}
  \emph {et~al.}}]{DES:2017txv}%
  \BibitemOpen
  \bibfield  {author} {\bibinfo {author} {\bibfnamefont {T.~M.~C.}\
  \bibnamefont {Abbott}} \emph {et~al.} (\bibinfo {collaboration} {DES}),\
  }\href {\doibase 10.1093/mnras/sty1939} {\bibfield  {journal} {\bibinfo
  {journal} {Mon. Not. Roy. Astron. Soc.}\ }\textbf {\bibinfo {volume} {480}},\
  \bibinfo {pages} {3879} (\bibinfo {year} {2018})},\ \Eprint
  {http://arxiv.org/abs/1711.00403} {arXiv:1711.00403 [astro-ph.CO]}
  \BibitemShut {NoStop}%
\bibitem [{\citenamefont {Alam}\ \emph {et~al.}(2021)\citenamefont {Alam} \emph
  {et~al.}}]{eBOSS:2020yzd}%
  \BibitemOpen
  \bibfield  {author} {\bibinfo {author} {\bibfnamefont {S.}~\bibnamefont
  {Alam}} \emph {et~al.} (\bibinfo {collaboration} {eBOSS}),\ }\href {\doibase
  10.1103/PhysRevD.103.083533} {\bibfield  {journal} {\bibinfo  {journal}
  {Phys. Rev. D}\ }\textbf {\bibinfo {volume} {103}},\ \bibinfo {pages}
  {083533} (\bibinfo {year} {2021})},\ \Eprint
  {http://arxiv.org/abs/2007.08991} {arXiv:2007.08991 [astro-ph.CO]}
  \BibitemShut {NoStop}%
\bibitem [{\citenamefont {Camilleri}\ \emph {et~al.}(2025)\citenamefont
  {Camilleri} \emph {et~al.}}]{DES:2024ywx}%
  \BibitemOpen
  \bibfield  {author} {\bibinfo {author} {\bibfnamefont {R.}~\bibnamefont
  {Camilleri}} \emph {et~al.} (\bibinfo {collaboration} {DES}),\ }\href
  {\doibase 10.1093/mnras/staf122} {\bibfield  {journal} {\bibinfo  {journal}
  {Mon. Not. Roy. Astron. Soc.}\ }\textbf {\bibinfo {volume} {537}},\ \bibinfo
  {pages} {1818} (\bibinfo {year} {2025})},\ \Eprint
  {http://arxiv.org/abs/2406.05049} {arXiv:2406.05049 [astro-ph.CO]}
  \BibitemShut {NoStop}%
\bibitem [{\citenamefont {Adame}\ \emph {et~al.}(2025)\citenamefont {Adame}
  \emph {et~al.}}]{DESI:2024mwx}%
  \BibitemOpen
  \bibfield  {author} {\bibinfo {author} {\bibfnamefont {A.~G.}\ \bibnamefont
  {Adame}} \emph {et~al.} (\bibinfo {collaboration} {DESI}),\ }\href {\doibase
  10.1088/1475-7516/2025/02/021} {\bibfield  {journal} {\bibinfo  {journal}
  {JCAP}\ }\textbf {\bibinfo {volume} {02}},\ \bibinfo {pages} {021} (\bibinfo
  {year} {2025})},\ \Eprint {http://arxiv.org/abs/2404.03002} {arXiv:2404.03002
  [astro-ph.CO]} \BibitemShut {NoStop}%
\bibitem [{\citenamefont {Macaulay}\ \emph {et~al.}(2019)\citenamefont
  {Macaulay} \emph {et~al.}}]{DES:2018rjw}%
  \BibitemOpen
  \bibfield  {author} {\bibinfo {author} {\bibfnamefont {E.}~\bibnamefont
  {Macaulay}} \emph {et~al.} (\bibinfo {collaboration} {DES}),\ }\href
  {\doibase 10.1093/mnras/stz978} {\bibfield  {journal} {\bibinfo  {journal}
  {Mon. Not. Roy. Astron. Soc.}\ }\textbf {\bibinfo {volume} {486}},\ \bibinfo
  {pages} {2184} (\bibinfo {year} {2019})},\ \Eprint
  {http://arxiv.org/abs/1811.02376} {arXiv:1811.02376 [astro-ph.CO]}
  \BibitemShut {NoStop}%
\bibitem [{\citenamefont {Aylor}\ \emph {et~al.}(2019)\citenamefont {Aylor},
  \citenamefont {Joy}, \citenamefont {Knox}, \citenamefont {Millea},
  \citenamefont {Raghunathan},\ and\ \citenamefont {Wu}}]{Aylor:2018drw}%
  \BibitemOpen
  \bibfield  {author} {\bibinfo {author} {\bibfnamefont {K.}~\bibnamefont
  {Aylor}}, \bibinfo {author} {\bibfnamefont {M.}~\bibnamefont {Joy}}, \bibinfo
  {author} {\bibfnamefont {L.}~\bibnamefont {Knox}}, \bibinfo {author}
  {\bibfnamefont {M.}~\bibnamefont {Millea}}, \bibinfo {author} {\bibfnamefont
  {S.}~\bibnamefont {Raghunathan}}, \ and\ \bibinfo {author} {\bibfnamefont
  {W.~L.~K.}\ \bibnamefont {Wu}},\ }\href {\doibase 10.3847/1538-4357/ab0898}
  {\bibfield  {journal} {\bibinfo  {journal} {Astrophys. J.}\ }\textbf
  {\bibinfo {volume} {874}},\ \bibinfo {pages} {4} (\bibinfo {year} {2019})},\
  \Eprint {http://arxiv.org/abs/1811.00537} {arXiv:1811.00537 [astro-ph.CO]}
  \BibitemShut {NoStop}%
\bibitem [{\citenamefont {Barua}\ and\ \citenamefont
  {Desai}(2025)}]{Barua:2024gei}%
  \BibitemOpen
  \bibfield  {author} {\bibinfo {author} {\bibfnamefont {S.}~\bibnamefont
  {Barua}}\ and\ \bibinfo {author} {\bibfnamefont {S.}~\bibnamefont {Desai}},\
  }\href {\doibase 10.1140/epjc/s10052-025-14219-5} {\bibfield  {journal}
  {\bibinfo  {journal} {Eur. Phys. J. C}\ }\textbf {\bibinfo {volume} {85}},\
  \bibinfo {pages} {470} (\bibinfo {year} {2025})},\ \Eprint
  {http://arxiv.org/abs/2412.19240} {arXiv:2412.19240 [astro-ph.CO]}
  \BibitemShut {NoStop}%
\bibitem [{\citenamefont {Luongo}\ and\ \citenamefont
  {Muccino}(2024)}]{Luongo:2024fww}%
  \BibitemOpen
  \bibfield  {author} {\bibinfo {author} {\bibfnamefont {O.}~\bibnamefont
  {Luongo}}\ and\ \bibinfo {author} {\bibfnamefont {M.}~\bibnamefont
  {Muccino}},\ }\href {\doibase 10.1051/0004-6361/202450512} {\bibfield
  {journal} {\bibinfo  {journal} {Astron. Astrophys.}\ }\textbf {\bibinfo
  {volume} {690}},\ \bibinfo {pages} {A40} (\bibinfo {year} {2024})},\ \Eprint
  {http://arxiv.org/abs/2404.07070} {arXiv:2404.07070 [astro-ph.CO]}
  \BibitemShut {NoStop}%
\bibitem [{\citenamefont {Wojtak}\ and\ \citenamefont
  {Agnello}(2019)}]{Wojtak:2019tkz}%
  \BibitemOpen
  \bibfield  {author} {\bibinfo {author} {\bibfnamefont {R.}~\bibnamefont
  {Wojtak}}\ and\ \bibinfo {author} {\bibfnamefont {A.}~\bibnamefont
  {Agnello}},\ }\href {\doibase 10.1093/mnras/stz1163} {\bibfield  {journal}
  {\bibinfo  {journal} {Mon. Not. Roy. Astron. Soc.}\ }\textbf {\bibinfo
  {volume} {486}},\ \bibinfo {pages} {5046} (\bibinfo {year} {2019})},\ \Eprint
  {http://arxiv.org/abs/1908.02401} {arXiv:1908.02401 [astro-ph.CO]}
  \BibitemShut {NoStop}%
\bibitem [{\citenamefont {Taubenberger}\ \emph {et~al.}(2019)\citenamefont
  {Taubenberger}, \citenamefont {Suyu}, \citenamefont {Komatsu}, \citenamefont
  {Jee}, \citenamefont {Birrer}, \citenamefont {Bonvin}, \citenamefont
  {Courbin}, \citenamefont {Rusu}, \citenamefont {Shajib},\ and\ \citenamefont
  {Wong}}]{Taubenberger:2019qna}%
  \BibitemOpen
  \bibfield  {author} {\bibinfo {author} {\bibfnamefont {S.}~\bibnamefont
  {Taubenberger}}, \bibinfo {author} {\bibfnamefont {S.~H.}\ \bibnamefont
  {Suyu}}, \bibinfo {author} {\bibfnamefont {E.}~\bibnamefont {Komatsu}},
  \bibinfo {author} {\bibfnamefont {I.}~\bibnamefont {Jee}}, \bibinfo {author}
  {\bibfnamefont {S.}~\bibnamefont {Birrer}}, \bibinfo {author} {\bibfnamefont
  {V.}~\bibnamefont {Bonvin}}, \bibinfo {author} {\bibfnamefont
  {F.}~\bibnamefont {Courbin}}, \bibinfo {author} {\bibfnamefont {C.~E.}\
  \bibnamefont {Rusu}}, \bibinfo {author} {\bibfnamefont {A.~J.}\ \bibnamefont
  {Shajib}}, \ and\ \bibinfo {author} {\bibfnamefont {K.~C.}\ \bibnamefont
  {Wong}},\ }\href {\doibase 10.1051/0004-6361/201935980} {\bibfield  {journal}
  {\bibinfo  {journal} {Astron. Astrophys.}\ }\textbf {\bibinfo {volume}
  {628}},\ \bibinfo {pages} {L7} (\bibinfo {year} {2019})},\ \Eprint
  {http://arxiv.org/abs/1905.12496} {arXiv:1905.12496 [astro-ph.CO]}
  \BibitemShut {NoStop}%
\bibitem [{\citenamefont {Wang}\ \emph {et~al.}(2022)\citenamefont {Wang},
  \citenamefont {Zhang}, \citenamefont {He}, \citenamefont {Zhang},\ and\
  \citenamefont {Zhang}}]{Wang:2021kxc}%
  \BibitemOpen
  \bibfield  {author} {\bibinfo {author} {\bibfnamefont {L.-F.}\ \bibnamefont
  {Wang}}, \bibinfo {author} {\bibfnamefont {J.-H.}\ \bibnamefont {Zhang}},
  \bibinfo {author} {\bibfnamefont {D.-Z.}\ \bibnamefont {He}}, \bibinfo
  {author} {\bibfnamefont {J.-F.}\ \bibnamefont {Zhang}}, \ and\ \bibinfo
  {author} {\bibfnamefont {X.}~\bibnamefont {Zhang}},\ }\href {\doibase
  10.1093/mnras/stac1468} {\bibfield  {journal} {\bibinfo  {journal} {Mon. Not.
  Roy. Astron. Soc.}\ }\textbf {\bibinfo {volume} {514}},\ \bibinfo {pages}
  {1433} (\bibinfo {year} {2022})},\ \Eprint {http://arxiv.org/abs/2102.09331}
  {arXiv:2102.09331 [astro-ph.CO]} \BibitemShut {NoStop}%
\bibitem [{\citenamefont {Gong}\ \emph {et~al.}(2024)\citenamefont {Gong},
  \citenamefont {Liu},\ and\ \citenamefont {Wang}}]{Gong:2024yne}%
  \BibitemOpen
  \bibfield  {author} {\bibinfo {author} {\bibfnamefont {X.}~\bibnamefont
  {Gong}}, \bibinfo {author} {\bibfnamefont {T.}~\bibnamefont {Liu}}, \ and\
  \bibinfo {author} {\bibfnamefont {J.}~\bibnamefont {Wang}},\ }\href {\doibase
  10.1140/epjc/s10052-024-13259-7} {\bibfield  {journal} {\bibinfo  {journal}
  {Eur. Phys. J. C}\ }\textbf {\bibinfo {volume} {84}},\ \bibinfo {pages} {873}
  (\bibinfo {year} {2024})}\BibitemShut {NoStop}%
\bibitem [{\citenamefont {Moresco}(2024)}]{Moresco:2024wmr}%
  \BibitemOpen
  \bibfield  {author} {\bibinfo {author} {\bibfnamefont {M.}~\bibnamefont
  {Moresco}},\ }\href@noop {} {\  (\bibinfo {year} {2024})},\ \Eprint
  {http://arxiv.org/abs/2412.01994} {arXiv:2412.01994 [astro-ph.CO]}
  \BibitemShut {NoStop}%
\bibitem [{\citenamefont {Vagnozzi}\ \emph {et~al.}(2022)\citenamefont
  {Vagnozzi}, \citenamefont {Pacucci},\ and\ \citenamefont
  {Loeb}}]{Vagnozzi:2021tjv}%
  \BibitemOpen
  \bibfield  {author} {\bibinfo {author} {\bibfnamefont {S.}~\bibnamefont
  {Vagnozzi}}, \bibinfo {author} {\bibfnamefont {F.}~\bibnamefont {Pacucci}}, \
  and\ \bibinfo {author} {\bibfnamefont {A.}~\bibnamefont {Loeb}},\ }\href
  {\doibase 10.1016/j.jheap.2022.07.004} {\bibfield  {journal} {\bibinfo
  {journal} {JHEAp}\ }\textbf {\bibinfo {volume} {36}},\ \bibinfo {pages} {27}
  (\bibinfo {year} {2022})},\ \Eprint {http://arxiv.org/abs/2105.10421}
  {arXiv:2105.10421 [astro-ph.CO]} \BibitemShut {NoStop}%
\bibitem [{\citenamefont {Jimenez}\ \emph {et~al.}(2019)\citenamefont
  {Jimenez}, \citenamefont {Cimatti}, \citenamefont {Verde}, \citenamefont
  {Moresco},\ and\ \citenamefont {Wandelt}}]{Jimenez:2019onw}%
  \BibitemOpen
  \bibfield  {author} {\bibinfo {author} {\bibfnamefont {R.}~\bibnamefont
  {Jimenez}}, \bibinfo {author} {\bibfnamefont {A.}~\bibnamefont {Cimatti}},
  \bibinfo {author} {\bibfnamefont {L.}~\bibnamefont {Verde}}, \bibinfo
  {author} {\bibfnamefont {M.}~\bibnamefont {Moresco}}, \ and\ \bibinfo
  {author} {\bibfnamefont {B.}~\bibnamefont {Wandelt}},\ }\href {\doibase
  10.1088/1475-7516/2019/03/043} {\bibfield  {journal} {\bibinfo  {journal}
  {JCAP}\ }\textbf {\bibinfo {volume} {03}},\ \bibinfo {pages} {043} (\bibinfo
  {year} {2019})},\ \Eprint {http://arxiv.org/abs/1902.07081} {arXiv:1902.07081
  [astro-ph.CO]} \BibitemShut {NoStop}%
\bibitem [{\citenamefont {Guo}\ \emph {et~al.}(2025)\citenamefont {Guo},
  \citenamefont {Wang}, \citenamefont {Cao}, \citenamefont {Biesiada},
  \citenamefont {Liu}, \citenamefont {Lian}, \citenamefont {Jiang},
  \citenamefont {Mu},\ and\ \citenamefont {Cheng}}]{Guo:2024pkx}%
  \BibitemOpen
  \bibfield  {author} {\bibinfo {author} {\bibfnamefont {W.}~\bibnamefont
  {Guo}}, \bibinfo {author} {\bibfnamefont {Q.}~\bibnamefont {Wang}}, \bibinfo
  {author} {\bibfnamefont {S.}~\bibnamefont {Cao}}, \bibinfo {author}
  {\bibfnamefont {M.}~\bibnamefont {Biesiada}}, \bibinfo {author}
  {\bibfnamefont {T.}~\bibnamefont {Liu}}, \bibinfo {author} {\bibfnamefont
  {Y.}~\bibnamefont {Lian}}, \bibinfo {author} {\bibfnamefont {X.}~\bibnamefont
  {Jiang}}, \bibinfo {author} {\bibfnamefont {C.}~\bibnamefont {Mu}}, \ and\
  \bibinfo {author} {\bibfnamefont {D.}~\bibnamefont {Cheng}},\ }\href
  {\doibase 10.3847/2041-8213/ada37f} {\bibfield  {journal} {\bibinfo
  {journal} {Astrophys. J. Lett.}\ }\textbf {\bibinfo {volume} {978}},\
  \bibinfo {pages} {L33} (\bibinfo {year} {2025})},\ \Eprint
  {http://arxiv.org/abs/2412.13045} {arXiv:2412.13045 [astro-ph.CO]}
  \BibitemShut {NoStop}%
\bibitem [{\citenamefont {Favale}\ \emph {et~al.}(2023)\citenamefont {Favale},
  \citenamefont {G{\'o}mez-Valent},\ and\ \citenamefont
  {Migliaccio}}]{Favale:2023lnp}%
  \BibitemOpen
  \bibfield  {author} {\bibinfo {author} {\bibfnamefont {A.}~\bibnamefont
  {Favale}}, \bibinfo {author} {\bibfnamefont {A.}~\bibnamefont
  {G{\'o}mez-Valent}}, \ and\ \bibinfo {author} {\bibfnamefont
  {M.}~\bibnamefont {Migliaccio}},\ }\href {\doibase 10.1093/mnras/stad1621}
  {\bibfield  {journal} {\bibinfo  {journal} {Mon. Not. Roy. Astron. Soc.}\
  }\textbf {\bibinfo {volume} {523}},\ \bibinfo {pages} {3406} (\bibinfo {year}
  {2023})},\ \Eprint {http://arxiv.org/abs/2301.09591} {arXiv:2301.09591
  [astro-ph.CO]} \BibitemShut {NoStop}%
\bibitem [{\citenamefont {Cai}\ \emph {et~al.}(2022{\natexlab{a}})\citenamefont
  {Cai}, \citenamefont {Guo}, \citenamefont {Wang}, \citenamefont {Yu},\ and\
  \citenamefont {Zhou}}]{Cai:2021weh}%
  \BibitemOpen
  \bibfield  {author} {\bibinfo {author} {\bibfnamefont {R.-G.}\ \bibnamefont
  {Cai}}, \bibinfo {author} {\bibfnamefont {Z.-K.}\ \bibnamefont {Guo}},
  \bibinfo {author} {\bibfnamefont {S.-J.}\ \bibnamefont {Wang}}, \bibinfo
  {author} {\bibfnamefont {W.-W.}\ \bibnamefont {Yu}}, \ and\ \bibinfo {author}
  {\bibfnamefont {Y.}~\bibnamefont {Zhou}},\ }\href {\doibase
  10.1103/PhysRevD.105.L021301} {\bibfield  {journal} {\bibinfo  {journal}
  {Phys. Rev. D}\ }\textbf {\bibinfo {volume} {105}},\ \bibinfo {pages}
  {L021301} (\bibinfo {year} {2022}{\natexlab{a}})},\ \Eprint
  {http://arxiv.org/abs/2107.13286} {arXiv:2107.13286 [astro-ph.CO]}
  \BibitemShut {NoStop}%
\bibitem [{\citenamefont {Pourojaghi}\ \emph {et~al.}(2025)\citenamefont
  {Pourojaghi}, \citenamefont {Malekjani},\ and\ \citenamefont
  {Davari}}]{Pourojaghi:2024bxa}%
  \BibitemOpen
  \bibfield  {author} {\bibinfo {author} {\bibfnamefont {S.}~\bibnamefont
  {Pourojaghi}}, \bibinfo {author} {\bibfnamefont {M.}~\bibnamefont
  {Malekjani}}, \ and\ \bibinfo {author} {\bibfnamefont {Z.}~\bibnamefont
  {Davari}},\ }\href {\doibase 10.1093/mnras/staf037} {\bibfield  {journal}
  {\bibinfo  {journal} {Mon. Not. Roy. Astron. Soc.}\ }\textbf {\bibinfo
  {volume} {537}},\ \bibinfo {pages} {436} (\bibinfo {year} {2025})},\ \Eprint
  {http://arxiv.org/abs/2408.10704} {arXiv:2408.10704 [astro-ph.CO]}
  \BibitemShut {NoStop}%
\bibitem [{\citenamefont {Cattoen}\ and\ \citenamefont
  {Visser}(2007)}]{Cattoen:2007sk}%
  \BibitemOpen
  \bibfield  {author} {\bibinfo {author} {\bibfnamefont {C.}~\bibnamefont
  {Cattoen}}\ and\ \bibinfo {author} {\bibfnamefont {M.}~\bibnamefont
  {Visser}},\ }\href {\doibase 10.1088/0264-9381/24/23/018} {\bibfield
  {journal} {\bibinfo  {journal} {Class. Quant. Grav.}\ }\textbf {\bibinfo
  {volume} {24}},\ \bibinfo {pages} {5985} (\bibinfo {year} {2007})},\ \Eprint
  {http://arxiv.org/abs/0710.1887} {arXiv:0710.1887 [gr-qc]} \BibitemShut
  {NoStop}%
\bibitem [{\citenamefont {Zhang}\ \emph
  {et~al.}(2017{\natexlab{b}})\citenamefont {Zhang}, \citenamefont {Li},\ and\
  \citenamefont {Xia}}]{Zhang:2016urt}%
  \BibitemOpen
  \bibfield  {author} {\bibinfo {author} {\bibfnamefont {M.-J.}\ \bibnamefont
  {Zhang}}, \bibinfo {author} {\bibfnamefont {H.}~\bibnamefont {Li}}, \ and\
  \bibinfo {author} {\bibfnamefont {J.-Q.}\ \bibnamefont {Xia}},\ }\href
  {\doibase 10.1140/epjc/s10052-017-5005-4} {\bibfield  {journal} {\bibinfo
  {journal} {Eur. Phys. J. C}\ }\textbf {\bibinfo {volume} {77}},\ \bibinfo
  {pages} {434} (\bibinfo {year} {2017}{\natexlab{b}})},\ \Eprint
  {http://arxiv.org/abs/1601.01758} {arXiv:1601.01758 [astro-ph.CO]}
  \BibitemShut {NoStop}%
\bibitem [{\citenamefont {Chiba}\ and\ \citenamefont
  {Nakamura}(1998)}]{Chiba:1998tc}%
  \BibitemOpen
  \bibfield  {author} {\bibinfo {author} {\bibfnamefont {T.}~\bibnamefont
  {Chiba}}\ and\ \bibinfo {author} {\bibfnamefont {T.}~\bibnamefont
  {Nakamura}},\ }\href {\doibase 10.1143/PTP.100.1077} {\bibfield  {journal}
  {\bibinfo  {journal} {Prog. Theor. Phys.}\ }\textbf {\bibinfo {volume}
  {100}},\ \bibinfo {pages} {1077} (\bibinfo {year} {1998})},\ \Eprint
  {http://arxiv.org/abs/astro-ph/9808022} {arXiv:astro-ph/9808022} \BibitemShut
  {NoStop}%
\bibitem [{\citenamefont {Capozziello}\ \emph {et~al.}(2011)\citenamefont
  {Capozziello}, \citenamefont {Lazkoz},\ and\ \citenamefont
  {Salzano}}]{Capozziello:2011tj}%
  \BibitemOpen
  \bibfield  {author} {\bibinfo {author} {\bibfnamefont {S.}~\bibnamefont
  {Capozziello}}, \bibinfo {author} {\bibfnamefont {R.}~\bibnamefont {Lazkoz}},
  \ and\ \bibinfo {author} {\bibfnamefont {V.}~\bibnamefont {Salzano}},\ }\href
  {\doibase 10.1103/PhysRevD.84.124061} {\bibfield  {journal} {\bibinfo
  {journal} {Phys. Rev. D}\ }\textbf {\bibinfo {volume} {84}},\ \bibinfo
  {pages} {124061} (\bibinfo {year} {2011})},\ \Eprint
  {http://arxiv.org/abs/1104.3096} {arXiv:1104.3096 [astro-ph.CO]} \BibitemShut
  {NoStop}%
\bibitem [{\citenamefont {Aviles}\ \emph {et~al.}(2012)\citenamefont {Aviles},
  \citenamefont {Gruber}, \citenamefont {Luongo},\ and\ \citenamefont
  {Quevedo}}]{Aviles:2012ay}%
  \BibitemOpen
  \bibfield  {author} {\bibinfo {author} {\bibfnamefont {A.}~\bibnamefont
  {Aviles}}, \bibinfo {author} {\bibfnamefont {C.}~\bibnamefont {Gruber}},
  \bibinfo {author} {\bibfnamefont {O.}~\bibnamefont {Luongo}}, \ and\ \bibinfo
  {author} {\bibfnamefont {H.}~\bibnamefont {Quevedo}},\ }\href {\doibase
  10.1103/PhysRevD.86.123516} {\bibfield  {journal} {\bibinfo  {journal} {Phys.
  Rev. D}\ }\textbf {\bibinfo {volume} {86}},\ \bibinfo {pages} {123516}
  (\bibinfo {year} {2012})},\ \Eprint {http://arxiv.org/abs/1204.2007}
  {arXiv:1204.2007 [astro-ph.CO]} \BibitemShut {NoStop}%
\bibitem [{\citenamefont {Jiang}\ \emph {et~al.}(2024)\citenamefont {Jiang},
  \citenamefont {Pedrotti}, \citenamefont {da~Costa},\ and\ \citenamefont
  {Vagnozzi}}]{Jiang:2024xnu}%
  \BibitemOpen
  \bibfield  {author} {\bibinfo {author} {\bibfnamefont {J.-Q.}\ \bibnamefont
  {Jiang}}, \bibinfo {author} {\bibfnamefont {D.}~\bibnamefont {Pedrotti}},
  \bibinfo {author} {\bibfnamefont {S.~S.}\ \bibnamefont {da~Costa}}, \ and\
  \bibinfo {author} {\bibfnamefont {S.}~\bibnamefont {Vagnozzi}},\ }\href
  {\doibase 10.1103/PhysRevD.110.123519} {\bibfield  {journal} {\bibinfo
  {journal} {Phys. Rev. D}\ }\textbf {\bibinfo {volume} {110}},\ \bibinfo
  {pages} {123519} (\bibinfo {year} {2024})},\ \Eprint
  {http://arxiv.org/abs/2408.02365} {arXiv:2408.02365 [astro-ph.CO]}
  \BibitemShut {NoStop}%
\bibitem [{\citenamefont {He}\ \emph {et~al.}(2022)\citenamefont {He},
  \citenamefont {Pan}, \citenamefont {Shi}, \citenamefont {Li}, \citenamefont
  {Cao},\ and\ \citenamefont {Cheng}}]{He:2021rzc}%
  \BibitemOpen
  \bibfield  {author} {\bibinfo {author} {\bibfnamefont {Y.}~\bibnamefont
  {He}}, \bibinfo {author} {\bibfnamefont {Y.}~\bibnamefont {Pan}}, \bibinfo
  {author} {\bibfnamefont {D.}~\bibnamefont {Shi}}, \bibinfo {author}
  {\bibfnamefont {J.}~\bibnamefont {Li}}, \bibinfo {author} {\bibfnamefont
  {S.}~\bibnamefont {Cao}}, \ and\ \bibinfo {author} {\bibfnamefont
  {W.}~\bibnamefont {Cheng}},\ }\href {\doibase 10.1088/1674-4527/ac77e3}
  {\bibfield  {journal} {\bibinfo  {journal} {Res. Astron. Astrophys.}\
  }\textbf {\bibinfo {volume} {22}},\ \bibinfo {pages} {085016} (\bibinfo
  {year} {2022})},\ \Eprint {http://arxiv.org/abs/2112.14477} {arXiv:2112.14477
  [astro-ph.CO]} \BibitemShut {NoStop}%
\bibitem [{\citenamefont {Mukherjee}\ and\ \citenamefont
  {Sen}(2024)}]{Mukherjee:2024ryz}%
  \BibitemOpen
  \bibfield  {author} {\bibinfo {author} {\bibfnamefont {P.}~\bibnamefont
  {Mukherjee}}\ and\ \bibinfo {author} {\bibfnamefont {A.~A.}\ \bibnamefont
  {Sen}},\ }\href {\doibase 10.1103/PhysRevD.110.123502} {\bibfield  {journal}
  {\bibinfo  {journal} {Phys. Rev. D}\ }\textbf {\bibinfo {volume} {110}},\
  \bibinfo {pages} {123502} (\bibinfo {year} {2024})},\ \Eprint
  {http://arxiv.org/abs/2405.19178} {arXiv:2405.19178 [astro-ph.CO]}
  \BibitemShut {NoStop}%
\bibitem [{\citenamefont {Yang}\ \emph {et~al.}(2023)\citenamefont {Yang},
  \citenamefont {Lu}, \citenamefont {Qian},\ and\ \citenamefont
  {Cao}}]{Yang:2022jkf}%
  \BibitemOpen
  \bibfield  {author} {\bibinfo {author} {\bibfnamefont {Y.}~\bibnamefont
  {Yang}}, \bibinfo {author} {\bibfnamefont {X.}~\bibnamefont {Lu}}, \bibinfo
  {author} {\bibfnamefont {L.}~\bibnamefont {Qian}}, \ and\ \bibinfo {author}
  {\bibfnamefont {S.}~\bibnamefont {Cao}},\ }\href {\doibase
  10.1093/mnras/stac3617} {\bibfield  {journal} {\bibinfo  {journal} {Mon. Not.
  Roy. Astron. Soc.}\ }\textbf {\bibinfo {volume} {519}},\ \bibinfo {pages}
  {4938} (\bibinfo {year} {2023})},\ \Eprint {http://arxiv.org/abs/2204.01020}
  {arXiv:2204.01020 [astro-ph.CO]} \BibitemShut {NoStop}%
\bibitem [{\citenamefont {Banerjee}\ \emph {et~al.}(2021)\citenamefont
  {Banerjee}, \citenamefont {Colg{\'a}in}, \citenamefont {Sasaki},
  \citenamefont {Sheikh-Jabbari},\ and\ \citenamefont
  {Yang}}]{Banerjee:2020bjq}%
  \BibitemOpen
  \bibfield  {author} {\bibinfo {author} {\bibfnamefont {A.}~\bibnamefont
  {Banerjee}}, \bibinfo {author} {\bibfnamefont {E.~{\'O}.}\ \bibnamefont
  {Colg{\'a}in}}, \bibinfo {author} {\bibfnamefont {M.}~\bibnamefont {Sasaki}},
  \bibinfo {author} {\bibfnamefont {M.~M.}\ \bibnamefont {Sheikh-Jabbari}}, \
  and\ \bibinfo {author} {\bibfnamefont {T.}~\bibnamefont {Yang}},\ }\href
  {\doibase 10.1016/j.physletb.2021.136366} {\bibfield  {journal} {\bibinfo
  {journal} {Phys. Lett. B}\ }\textbf {\bibinfo {volume} {818}},\ \bibinfo
  {pages} {136366} (\bibinfo {year} {2021})},\ \Eprint
  {http://arxiv.org/abs/2009.04109} {arXiv:2009.04109 [astro-ph.CO]}
  \BibitemShut {NoStop}%
\bibitem [{\citenamefont {Cai}\ \emph {et~al.}(2022{\natexlab{b}})\citenamefont
  {Cai}, \citenamefont {Guo}, \citenamefont {Wang}, \citenamefont {Yu},\ and\
  \citenamefont {Zhou}}]{Cai:2022dkh}%
  \BibitemOpen
  \bibfield  {author} {\bibinfo {author} {\bibfnamefont {R.-G.}\ \bibnamefont
  {Cai}}, \bibinfo {author} {\bibfnamefont {Z.-K.}\ \bibnamefont {Guo}},
  \bibinfo {author} {\bibfnamefont {S.-J.}\ \bibnamefont {Wang}}, \bibinfo
  {author} {\bibfnamefont {W.-W.}\ \bibnamefont {Yu}}, \ and\ \bibinfo {author}
  {\bibfnamefont {Y.}~\bibnamefont {Zhou}},\ }\href {\doibase
  10.1103/PhysRevD.106.063519} {\bibfield  {journal} {\bibinfo  {journal}
  {Phys. Rev. D}\ }\textbf {\bibinfo {volume} {106}},\ \bibinfo {pages}
  {063519} (\bibinfo {year} {2022}{\natexlab{b}})},\ \Eprint
  {http://arxiv.org/abs/2202.12214} {arXiv:2202.12214 [astro-ph.CO]}
  \BibitemShut {NoStop}%
\bibitem [{\citenamefont {{\'O}~Colg{\'a}in}\ and\ \citenamefont
  {Sheikh-Jabbari}(2021)}]{OColgain:2021pyh}%
  \BibitemOpen
  \bibfield  {author} {\bibinfo {author} {\bibfnamefont {E.}~\bibnamefont
  {{\'O}~Colg{\'a}in}}\ and\ \bibinfo {author} {\bibfnamefont {M.~M.}\
  \bibnamefont {Sheikh-Jabbari}},\ }\href {\doibase
  10.1140/epjc/s10052-021-09708-2} {\bibfield  {journal} {\bibinfo  {journal}
  {Eur. Phys. J. C}\ }\textbf {\bibinfo {volume} {81}},\ \bibinfo {pages} {892}
  (\bibinfo {year} {2021})},\ \Eprint {http://arxiv.org/abs/2101.08565}
  {arXiv:2101.08565 [astro-ph.CO]} \BibitemShut {NoStop}%
\bibitem [{\citenamefont {Kjerrgren}\ and\ \citenamefont
  {Mortsell}(2022)}]{Kjerrgren:2021zuo}%
  \BibitemOpen
  \bibfield  {author} {\bibinfo {author} {\bibfnamefont {A.~A.}\ \bibnamefont
  {Kjerrgren}}\ and\ \bibinfo {author} {\bibfnamefont {E.}~\bibnamefont
  {Mortsell}},\ }\href {\doibase 10.1093/mnras/stac1978} {\bibfield  {journal}
  {\bibinfo  {journal} {Mon. Not. Roy. Astron. Soc.}\ }\textbf {\bibinfo
  {volume} {518}},\ \bibinfo {pages} {585} (\bibinfo {year} {2022})},\ \Eprint
  {http://arxiv.org/abs/2106.11317} {arXiv:2106.11317 [astro-ph.CO]}
  \BibitemShut {NoStop}%
\bibitem [{\citenamefont {Huang}(2020)}]{Huang:2020mub}%
  \BibitemOpen
  \bibfield  {author} {\bibinfo {author} {\bibfnamefont {Z.}~\bibnamefont
  {Huang}},\ }\href {\doibase 10.3847/2041-8213/ab8011} {\bibfield  {journal}
  {\bibinfo  {journal} {Astrophys. J. Lett.}\ }\textbf {\bibinfo {volume}
  {892}},\ \bibinfo {pages} {L28} (\bibinfo {year} {2020})},\ \Eprint
  {http://arxiv.org/abs/2001.06926} {arXiv:2001.06926 [astro-ph.CO]}
  \BibitemShut {NoStop}%
\bibitem [{\citenamefont {Luo}\ \emph {et~al.}(2020)\citenamefont {Luo},
  \citenamefont {Huang}, \citenamefont {Qian},\ and\ \citenamefont
  {Huang}}]{Luo:2020ufj}%
  \BibitemOpen
  \bibfield  {author} {\bibinfo {author} {\bibfnamefont {X.}~\bibnamefont
  {Luo}}, \bibinfo {author} {\bibfnamefont {Z.}~\bibnamefont {Huang}}, \bibinfo
  {author} {\bibfnamefont {Q.}~\bibnamefont {Qian}}, \ and\ \bibinfo {author}
  {\bibfnamefont {L.}~\bibnamefont {Huang}},\ }\href {\doibase
  10.3847/1538-4357/abc25f} {\bibfield  {journal} {\bibinfo  {journal}
  {Astrophys. J.}\ }\textbf {\bibinfo {volume} {905}},\ \bibinfo {pages} {53}
  (\bibinfo {year} {2020})},\ \Eprint {http://arxiv.org/abs/2008.00487}
  {arXiv:2008.00487 [astro-ph.CO]} \BibitemShut {NoStop}%
\bibitem [{\citenamefont {Huang}\ \emph
  {et~al.}(2021{\natexlab{a}})\citenamefont {Huang}, \citenamefont {Huang},
  \citenamefont {Huang}, \citenamefont {Li}, \citenamefont {Li},\ and\
  \citenamefont {Zhou}}]{Huang:2021aku}%
  \BibitemOpen
  \bibfield  {author} {\bibinfo {author} {\bibfnamefont {L.}~\bibnamefont
  {Huang}}, \bibinfo {author} {\bibfnamefont {Z.-Q.}\ \bibnamefont {Huang}},
  \bibinfo {author} {\bibfnamefont {Z.}~\bibnamefont {Huang}}, \bibinfo
  {author} {\bibfnamefont {Z.-Y.}\ \bibnamefont {Li}}, \bibinfo {author}
  {\bibfnamefont {Z.}~\bibnamefont {Li}}, \ and\ \bibinfo {author}
  {\bibfnamefont {H.}~\bibnamefont {Zhou}},\ }\href {\doibase
  10.1088/1674-4527/21/11/277} {\bibfield  {journal} {\bibinfo  {journal} {Res.
  Astron. Astrophys.}\ }\textbf {\bibinfo {volume} {21}},\ \bibinfo {pages}
  {277} (\bibinfo {year} {2021}{\natexlab{a}})},\ \Eprint
  {http://arxiv.org/abs/2108.03959} {arXiv:2108.03959 [astro-ph.CO]}
  \BibitemShut {NoStop}%
\bibitem [{\citenamefont {Huang}\ \emph
  {et~al.}(2021{\natexlab{b}})\citenamefont {Huang}, \citenamefont {Huang},
  \citenamefont {Luo}, \citenamefont {He},\ and\ \citenamefont
  {Fang}}]{Huang:2020evj}%
  \BibitemOpen
  \bibfield  {author} {\bibinfo {author} {\bibfnamefont {L.}~\bibnamefont
  {Huang}}, \bibinfo {author} {\bibfnamefont {Z.}~\bibnamefont {Huang}},
  \bibinfo {author} {\bibfnamefont {X.}~\bibnamefont {Luo}}, \bibinfo {author}
  {\bibfnamefont {X.}~\bibnamefont {He}}, \ and\ \bibinfo {author}
  {\bibfnamefont {Y.}~\bibnamefont {Fang}},\ }\href {\doibase
  10.1103/PhysRevD.103.123521} {\bibfield  {journal} {\bibinfo  {journal}
  {Phys. Rev. D}\ }\textbf {\bibinfo {volume} {103}},\ \bibinfo {pages}
  {123521} (\bibinfo {year} {2021}{\natexlab{b}})},\ \Eprint
  {http://arxiv.org/abs/2012.02474} {arXiv:2012.02474 [astro-ph.CO]}
  \BibitemShut {NoStop}%
\bibitem [{\citenamefont {Huang}\ \emph {et~al.}(2022)\citenamefont {Huang},
  \citenamefont {Huang}, \citenamefont {Zhou},\ and\ \citenamefont
  {Li}}]{Huang:2021tvo}%
  \BibitemOpen
  \bibfield  {author} {\bibinfo {author} {\bibfnamefont {L.}~\bibnamefont
  {Huang}}, \bibinfo {author} {\bibfnamefont {Z.}~\bibnamefont {Huang}},
  \bibinfo {author} {\bibfnamefont {H.}~\bibnamefont {Zhou}}, \ and\ \bibinfo
  {author} {\bibfnamefont {Z.}~\bibnamefont {Li}},\ }\href {\doibase
  10.1007/s11433-021-1838-1} {\bibfield  {journal} {\bibinfo  {journal} {Sci.
  China Phys. Mech. Astron.}\ }\textbf {\bibinfo {volume} {65}},\ \bibinfo
  {pages} {239512} (\bibinfo {year} {2022})},\ \Eprint
  {http://arxiv.org/abs/2110.08498} {arXiv:2110.08498 [astro-ph.CO]}
  \BibitemShut {NoStop}%
\bibitem [{\citenamefont {Li}\ \emph {et~al.}(2022)\citenamefont {Li},
  \citenamefont {Huang},\ and\ \citenamefont {Wang}}]{Li:2022inq}%
  \BibitemOpen
  \bibfield  {author} {\bibinfo {author} {\bibfnamefont {Z.}~\bibnamefont
  {Li}}, \bibinfo {author} {\bibfnamefont {L.}~\bibnamefont {Huang}}, \ and\
  \bibinfo {author} {\bibfnamefont {J.}~\bibnamefont {Wang}},\ }\href {\doibase
  10.1093/mnras/stac2735} {\bibfield  {journal} {\bibinfo  {journal} {Mon. Not.
  Roy. Astron. Soc.}\ }\textbf {\bibinfo {volume} {517}},\ \bibinfo {pages}
  {1901} (\bibinfo {year} {2022})},\ \Eprint {http://arxiv.org/abs/2210.02816}
  {arXiv:2210.02816 [astro-ph.CO]} \BibitemShut {NoStop}%
\bibitem [{\citenamefont {Huang}\ \emph
  {et~al.}(2025{\natexlab{a}})\citenamefont {Huang}, \citenamefont {Wang},\
  and\ \citenamefont {Yu}}]{Huang:2024erq}%
  \BibitemOpen
  \bibfield  {author} {\bibinfo {author} {\bibfnamefont {L.}~\bibnamefont
  {Huang}}, \bibinfo {author} {\bibfnamefont {S.-J.}\ \bibnamefont {Wang}}, \
  and\ \bibinfo {author} {\bibfnamefont {W.-W.}\ \bibnamefont {Yu}},\ }\href
  {\doibase 10.1007/s11433-024-2528-8} {\bibfield  {journal} {\bibinfo
  {journal} {Sci. China Phys. Mech. Astron.}\ }\textbf {\bibinfo {volume}
  {68}},\ \bibinfo {pages} {220413} (\bibinfo {year} {2025}{\natexlab{a}})},\
  \Eprint {http://arxiv.org/abs/2401.14170} {arXiv:2401.14170 [astro-ph.CO]}
  \BibitemShut {NoStop}%
\bibitem [{\citenamefont {Abdul~Karim}\ \emph {et~al.}(2025)\citenamefont
  {Abdul~Karim} \emph {et~al.}}]{DESI:2025zgx}%
  \BibitemOpen
  \bibfield  {author} {\bibinfo {author} {\bibfnamefont {M.}~\bibnamefont
  {Abdul~Karim}} \emph {et~al.} (\bibinfo {collaboration} {DESI}),\ }\href
  {\doibase 10.1103/tr6y-kpc6} {\bibfield  {journal} {\bibinfo  {journal}
  {Phys. Rev. D}\ }\textbf {\bibinfo {volume} {112}},\ \bibinfo {pages}
  {083515} (\bibinfo {year} {2025})},\ \Eprint
  {http://arxiv.org/abs/2503.14738} {arXiv:2503.14738 [astro-ph.CO]}
  \BibitemShut {NoStop}%
\bibitem [{\citenamefont {Malekjani}\ \emph {et~al.}(2025)\citenamefont
  {Malekjani}, \citenamefont {Davari},\ and\ \citenamefont
  {Pourojaghi}}]{Malekjani:2024bgi}%
  \BibitemOpen
  \bibfield  {author} {\bibinfo {author} {\bibfnamefont {M.}~\bibnamefont
  {Malekjani}}, \bibinfo {author} {\bibfnamefont {Z.}~\bibnamefont {Davari}}, \
  and\ \bibinfo {author} {\bibfnamefont {S.}~\bibnamefont {Pourojaghi}}
  (\bibinfo {collaboration} {DESI}),\ }\href {\doibase
  10.1103/PhysRevD.111.083547} {\bibfield  {journal} {\bibinfo  {journal}
  {Phys. Rev. D}\ }\textbf {\bibinfo {volume} {111}},\ \bibinfo {pages}
  {083547} (\bibinfo {year} {2025})},\ \Eprint
  {http://arxiv.org/abs/2407.09767} {arXiv:2407.09767 [astro-ph.CO]}
  \BibitemShut {NoStop}%
\bibitem [{\citenamefont {Colg{\'a}in}\ and\ \citenamefont
  {Sheikh-Jabbari}(2024)}]{Colgain:2024mtg}%
  \BibitemOpen
  \bibfield  {author} {\bibinfo {author} {\bibfnamefont {E.~{\'O}.}\
  \bibnamefont {Colg{\'a}in}}\ and\ \bibinfo {author} {\bibfnamefont {M.~M.}\
  \bibnamefont {Sheikh-Jabbari}},\ }\href@noop {} {\  (\bibinfo {year}
  {2024})},\ \Eprint {http://arxiv.org/abs/2412.12905} {arXiv:2412.12905
  [astro-ph.CO]} \BibitemShut {NoStop}%
\bibitem [{\citenamefont {Jia}\ \emph {et~al.}(2025)\citenamefont {Jia},
  \citenamefont {Hu}, \citenamefont {Yi},\ and\ \citenamefont
  {Wang}}]{Jia:2024wix}%
  \BibitemOpen
  \bibfield  {author} {\bibinfo {author} {\bibfnamefont {X.~D.}\ \bibnamefont
  {Jia}}, \bibinfo {author} {\bibfnamefont {J.~P.}\ \bibnamefont {Hu}},
  \bibinfo {author} {\bibfnamefont {S.~X.}\ \bibnamefont {Yi}}, \ and\ \bibinfo
  {author} {\bibfnamefont {F.~Y.}\ \bibnamefont {Wang}},\ }\href {\doibase
  10.3847/2041-8213/ada94d} {\bibfield  {journal} {\bibinfo  {journal}
  {Astrophys. J. Lett.}\ }\textbf {\bibinfo {volume} {979}},\ \bibinfo {pages}
  {L34} (\bibinfo {year} {2025})},\ \Eprint {http://arxiv.org/abs/2406.02019}
  {arXiv:2406.02019 [astro-ph.CO]} \BibitemShut {NoStop}%
\bibitem [{\citenamefont {Wang}\ \emph
  {et~al.}(2024{\natexlab{a}})\citenamefont {Wang}, \citenamefont {Lin},
  \citenamefont {Ding},\ and\ \citenamefont {Hu}}]{Wang:2024pui}%
  \BibitemOpen
  \bibfield  {author} {\bibinfo {author} {\bibfnamefont {Z.}~\bibnamefont
  {Wang}}, \bibinfo {author} {\bibfnamefont {S.}~\bibnamefont {Lin}}, \bibinfo
  {author} {\bibfnamefont {Z.}~\bibnamefont {Ding}}, \ and\ \bibinfo {author}
  {\bibfnamefont {B.}~\bibnamefont {Hu}},\ }\href {\doibase
  10.1093/mnras/stae2309} {\bibfield  {journal} {\bibinfo  {journal} {Mon. Not.
  Roy. Astron. Soc.}\ }\textbf {\bibinfo {volume} {534}},\ \bibinfo {pages}
  {3869} (\bibinfo {year} {2024}{\natexlab{a}})},\ \Eprint
  {http://arxiv.org/abs/2405.02168} {arXiv:2405.02168 [astro-ph.CO]}
  \BibitemShut {NoStop}%
\bibitem [{\citenamefont {Huang}\ \emph
  {et~al.}(2025{\natexlab{b}})\citenamefont {Huang}, \citenamefont {Cai},\ and\
  \citenamefont {Wang}}]{Huang:2025som}%
  \BibitemOpen
  \bibfield  {author} {\bibinfo {author} {\bibfnamefont {L.}~\bibnamefont
  {Huang}}, \bibinfo {author} {\bibfnamefont {R.-G.}\ \bibnamefont {Cai}}, \
  and\ \bibinfo {author} {\bibfnamefont {S.-J.}\ \bibnamefont {Wang}},\ }\href
  {\doibase 10.1007/s11433-025-2754-5} {\bibfield  {journal} {\bibinfo
  {journal} {Sci. China Phys. Mech. Astron.}\ }\textbf {\bibinfo {volume}
  {68}},\ \bibinfo {pages} {100413} (\bibinfo {year} {2025}{\natexlab{b}})},\
  \Eprint {http://arxiv.org/abs/2502.04212} {arXiv:2502.04212 [astro-ph.CO]}
  \BibitemShut {NoStop}%
\bibitem [{\citenamefont {Cort{\^e}s}\ and\ \citenamefont
  {Liddle}(2024)}]{Cortes:2024lgw}%
  \BibitemOpen
  \bibfield  {author} {\bibinfo {author} {\bibfnamefont {M.}~\bibnamefont
  {Cort{\^e}s}}\ and\ \bibinfo {author} {\bibfnamefont {A.~R.}\ \bibnamefont
  {Liddle}},\ }\href {\doibase 10.1088/1475-7516/2024/12/007} {\bibfield
  {journal} {\bibinfo  {journal} {JCAP}\ }\textbf {\bibinfo {volume} {12}},\
  \bibinfo {pages} {007} (\bibinfo {year} {2024})},\ \Eprint
  {http://arxiv.org/abs/2404.08056} {arXiv:2404.08056 [astro-ph.CO]}
  \BibitemShut {NoStop}%
\bibitem [{\citenamefont {Efstathiou}(2025)}]{Efstathiou:2024xcq}%
  \BibitemOpen
  \bibfield  {author} {\bibinfo {author} {\bibfnamefont {G.}~\bibnamefont
  {Efstathiou}},\ }\href {\doibase 10.1093/mnras/staf301} {\bibfield  {journal}
  {\bibinfo  {journal} {Mon. Not. Roy. Astron. Soc.}\ }\textbf {\bibinfo
  {volume} {538}},\ \bibinfo {pages} {875} (\bibinfo {year} {2025})},\ \Eprint
  {http://arxiv.org/abs/2408.07175} {arXiv:2408.07175 [astro-ph.CO]}
  \BibitemShut {NoStop}%
\bibitem [{\citenamefont {Abreu}\ and\ \citenamefont
  {Turner}(2025)}]{Abreu:2025zng}%
  \BibitemOpen
  \bibfield  {author} {\bibinfo {author} {\bibfnamefont {M.~L.}\ \bibnamefont
  {Abreu}}\ and\ \bibinfo {author} {\bibfnamefont {M.~S.}\ \bibnamefont
  {Turner}},\ }\href@noop {} {\  (\bibinfo {year} {2025})},\ \Eprint
  {http://arxiv.org/abs/2502.08876} {arXiv:2502.08876 [astro-ph.CO]}
  \BibitemShut {NoStop}%
\bibitem [{\citenamefont {Pang}\ \emph
  {et~al.}(2025{\natexlab{a}})\citenamefont {Pang}, \citenamefont {Zhang},\
  and\ \citenamefont {Huang}}]{Pang:2024qyh}%
  \BibitemOpen
  \bibfield  {author} {\bibinfo {author} {\bibfnamefont {Y.-H.}\ \bibnamefont
  {Pang}}, \bibinfo {author} {\bibfnamefont {X.}~\bibnamefont {Zhang}}, \ and\
  \bibinfo {author} {\bibfnamefont {Q.-G.}\ \bibnamefont {Huang}},\ }\href
  {\doibase 10.1103/PhysRevD.111.123504} {\bibfield  {journal} {\bibinfo
  {journal} {Phys. Rev. D}\ }\textbf {\bibinfo {volume} {111}},\ \bibinfo
  {pages} {123504} (\bibinfo {year} {2025}{\natexlab{a}})},\ \Eprint
  {http://arxiv.org/abs/2408.14787} {arXiv:2408.14787 [astro-ph.CO]}
  \BibitemShut {NoStop}%
\bibitem [{\citenamefont {Fikri}\ \emph {et~al.}(2025)\citenamefont {Fikri},
  \citenamefont {Elkhateeb}, \citenamefont {Lashin},\ and\ \citenamefont
  {El~Hanafy}}]{Fikri:2024klc}%
  \BibitemOpen
  \bibfield  {author} {\bibinfo {author} {\bibfnamefont {R.}~\bibnamefont
  {Fikri}}, \bibinfo {author} {\bibfnamefont {E.}~\bibnamefont {Elkhateeb}},
  \bibinfo {author} {\bibfnamefont {E.~I.}\ \bibnamefont {Lashin}}, \ and\
  \bibinfo {author} {\bibfnamefont {W.}~\bibnamefont {El~Hanafy}},\ }\href
  {\doibase 10.1016/j.aop.2025.170190} {\bibfield  {journal} {\bibinfo
  {journal} {Annals Phys.}\ }\textbf {\bibinfo {volume} {481}},\ \bibinfo
  {pages} {170190} (\bibinfo {year} {2025})},\ \Eprint
  {http://arxiv.org/abs/2411.19362} {arXiv:2411.19362 [astro-ph.CO]}
  \BibitemShut {NoStop}%
\bibitem [{\citenamefont {Feng}\ \emph
  {et~al.}(2025{\natexlab{a}})\citenamefont {Feng}, \citenamefont {Li},
  \citenamefont {Du}, \citenamefont {Zhang},\ and\ \citenamefont
  {Zhang}}]{Feng:2025mlo}%
  \BibitemOpen
  \bibfield  {author} {\bibinfo {author} {\bibfnamefont {L.}~\bibnamefont
  {Feng}}, \bibinfo {author} {\bibfnamefont {T.-N.}\ \bibnamefont {Li}},
  \bibinfo {author} {\bibfnamefont {G.-H.}\ \bibnamefont {Du}}, \bibinfo
  {author} {\bibfnamefont {J.-F.}\ \bibnamefont {Zhang}}, \ and\ \bibinfo
  {author} {\bibfnamefont {X.}~\bibnamefont {Zhang}},\ }\href {\doibase
  10.1016/j.dark.2025.101935} {\bibfield  {journal} {\bibinfo  {journal} {Phys.
  Dark Univ.}\ }\textbf {\bibinfo {volume} {48}},\ \bibinfo {pages} {101935}
  (\bibinfo {year} {2025}{\natexlab{a}})},\ \Eprint
  {http://arxiv.org/abs/2503.10423} {arXiv:2503.10423 [astro-ph.CO]}
  \BibitemShut {NoStop}%
\bibitem [{\citenamefont {Du}\ \emph {et~al.}(2025{\natexlab{a}})\citenamefont
  {Du}, \citenamefont {Wu}, \citenamefont {Li},\ and\ \citenamefont
  {Zhang}}]{Du:2024pai}%
  \BibitemOpen
  \bibfield  {author} {\bibinfo {author} {\bibfnamefont {G.-H.}\ \bibnamefont
  {Du}}, \bibinfo {author} {\bibfnamefont {P.-J.}\ \bibnamefont {Wu}}, \bibinfo
  {author} {\bibfnamefont {T.-N.}\ \bibnamefont {Li}}, \ and\ \bibinfo {author}
  {\bibfnamefont {X.}~\bibnamefont {Zhang}},\ }\href {\doibase
  10.1140/epjc/s10052-025-14094-0} {\bibfield  {journal} {\bibinfo  {journal}
  {Eur. Phys. J. C}\ }\textbf {\bibinfo {volume} {85}},\ \bibinfo {pages} {392}
  (\bibinfo {year} {2025}{\natexlab{a}})},\ \Eprint
  {http://arxiv.org/abs/2407.15640} {arXiv:2407.15640 [astro-ph.CO]}
  \BibitemShut {NoStop}%
\bibitem [{\citenamefont {Li}\ \emph {et~al.}(2025{\natexlab{a}})\citenamefont
  {Li}, \citenamefont {Li}, \citenamefont {Du}, \citenamefont {Wu},
  \citenamefont {Feng}, \citenamefont {Zhang},\ and\ \citenamefont
  {Zhang}}]{Li:2024qus}%
  \BibitemOpen
  \bibfield  {author} {\bibinfo {author} {\bibfnamefont {T.-N.}\ \bibnamefont
  {Li}}, \bibinfo {author} {\bibfnamefont {Y.-H.}\ \bibnamefont {Li}}, \bibinfo
  {author} {\bibfnamefont {G.-H.}\ \bibnamefont {Du}}, \bibinfo {author}
  {\bibfnamefont {P.-J.}\ \bibnamefont {Wu}}, \bibinfo {author} {\bibfnamefont
  {L.}~\bibnamefont {Feng}}, \bibinfo {author} {\bibfnamefont {J.-F.}\
  \bibnamefont {Zhang}}, \ and\ \bibinfo {author} {\bibfnamefont
  {X.}~\bibnamefont {Zhang}},\ }\href {\doibase
  10.1140/epjc/s10052-025-14279-7} {\bibfield  {journal} {\bibinfo  {journal}
  {Eur. Phys. J. C}\ }\textbf {\bibinfo {volume} {85}},\ \bibinfo {pages} {608}
  (\bibinfo {year} {2025}{\natexlab{a}})},\ \Eprint
  {http://arxiv.org/abs/2411.08639} {arXiv:2411.08639 [astro-ph.CO]}
  \BibitemShut {NoStop}%
\bibitem [{\citenamefont {Li}\ \emph {et~al.}(2024)\citenamefont {Li},
  \citenamefont {Wu}, \citenamefont {Du}, \citenamefont {Jin}, \citenamefont
  {Li}, \citenamefont {Zhang},\ and\ \citenamefont {Zhang}}]{Li:2024qso}%
  \BibitemOpen
  \bibfield  {author} {\bibinfo {author} {\bibfnamefont {T.-N.}\ \bibnamefont
  {Li}}, \bibinfo {author} {\bibfnamefont {P.-J.}\ \bibnamefont {Wu}}, \bibinfo
  {author} {\bibfnamefont {G.-H.}\ \bibnamefont {Du}}, \bibinfo {author}
  {\bibfnamefont {S.-J.}\ \bibnamefont {Jin}}, \bibinfo {author} {\bibfnamefont
  {H.-L.}\ \bibnamefont {Li}}, \bibinfo {author} {\bibfnamefont {J.-F.}\
  \bibnamefont {Zhang}}, \ and\ \bibinfo {author} {\bibfnamefont
  {X.}~\bibnamefont {Zhang}},\ }\href {\doibase 10.3847/1538-4357/ad87f0}
  {\bibfield  {journal} {\bibinfo  {journal} {Astrophys. J.}\ }\textbf
  {\bibinfo {volume} {976}},\ \bibinfo {pages} {1} (\bibinfo {year} {2024})},\
  \Eprint {http://arxiv.org/abs/2407.14934} {arXiv:2407.14934 [astro-ph.CO]}
  \BibitemShut {NoStop}%
\bibitem [{\citenamefont {Li}\ \emph {et~al.}(2025{\natexlab{b}})\citenamefont
  {Li}, \citenamefont {Du}, \citenamefont {Li}, \citenamefont {Wu},
  \citenamefont {Jin}, \citenamefont {Zhang},\ and\ \citenamefont
  {Zhang}}]{Li:2025owk}%
  \BibitemOpen
  \bibfield  {author} {\bibinfo {author} {\bibfnamefont {T.-N.}\ \bibnamefont
  {Li}}, \bibinfo {author} {\bibfnamefont {G.-H.}\ \bibnamefont {Du}}, \bibinfo
  {author} {\bibfnamefont {Y.-H.}\ \bibnamefont {Li}}, \bibinfo {author}
  {\bibfnamefont {P.-J.}\ \bibnamefont {Wu}}, \bibinfo {author} {\bibfnamefont
  {S.-J.}\ \bibnamefont {Jin}}, \bibinfo {author} {\bibfnamefont {J.-F.}\
  \bibnamefont {Zhang}}, \ and\ \bibinfo {author} {\bibfnamefont
  {X.}~\bibnamefont {Zhang}},\ }\href@noop {} {\  (\bibinfo {year}
  {2025}{\natexlab{b}})},\ \Eprint {http://arxiv.org/abs/2501.07361}
  {arXiv:2501.07361 [astro-ph.CO]} \BibitemShut {NoStop}%
\bibitem [{\citenamefont {Du}\ \emph {et~al.}(2025{\natexlab{b}})\citenamefont
  {Du}, \citenamefont {Li}, \citenamefont {Wu}, \citenamefont {Feng},
  \citenamefont {Zhou}, \citenamefont {Zhang},\ and\ \citenamefont
  {Zhang}}]{Du:2025iow}%
  \BibitemOpen
  \bibfield  {author} {\bibinfo {author} {\bibfnamefont {G.-H.}\ \bibnamefont
  {Du}}, \bibinfo {author} {\bibfnamefont {T.-N.}\ \bibnamefont {Li}}, \bibinfo
  {author} {\bibfnamefont {P.-J.}\ \bibnamefont {Wu}}, \bibinfo {author}
  {\bibfnamefont {L.}~\bibnamefont {Feng}}, \bibinfo {author} {\bibfnamefont
  {S.-H.}\ \bibnamefont {Zhou}}, \bibinfo {author} {\bibfnamefont {J.-F.}\
  \bibnamefont {Zhang}}, \ and\ \bibinfo {author} {\bibfnamefont
  {X.}~\bibnamefont {Zhang}},\ }\href@noop {} {\  (\bibinfo {year}
  {2025}{\natexlab{b}})},\ \Eprint {http://arxiv.org/abs/2501.10785}
  {arXiv:2501.10785 [astro-ph.CO]} \BibitemShut {NoStop}%
\bibitem [{\citenamefont {Pang}\ \emph
  {et~al.}(2025{\natexlab{b}})\citenamefont {Pang}, \citenamefont {Zhang},\
  and\ \citenamefont {Huang}}]{Pang:2025lvh}%
  \BibitemOpen
  \bibfield  {author} {\bibinfo {author} {\bibfnamefont {Y.-H.}\ \bibnamefont
  {Pang}}, \bibinfo {author} {\bibfnamefont {X.}~\bibnamefont {Zhang}}, \ and\
  \bibinfo {author} {\bibfnamefont {Q.-G.}\ \bibnamefont {Huang}},\ }\href
  {\doibase 10.1007/s11433-025-2713-8} {\bibfield  {journal} {\bibinfo
  {journal} {Sci. China Phys. Mech. Astron.}\ }\textbf {\bibinfo {volume}
  {68}},\ \bibinfo {pages} {280410} (\bibinfo {year} {2025}{\natexlab{b}})},\
  \Eprint {http://arxiv.org/abs/2503.21600} {arXiv:2503.21600 [astro-ph.CO]}
  \BibitemShut {NoStop}%
\bibitem [{\citenamefont {Wang}\ \emph
  {et~al.}(2025{\natexlab{a}})\citenamefont {Wang}, \citenamefont {Peng},\ and\
  \citenamefont {Piao}}]{Wang:2025ljj}%
  \BibitemOpen
  \bibfield  {author} {\bibinfo {author} {\bibfnamefont {H.}~\bibnamefont
  {Wang}}, \bibinfo {author} {\bibfnamefont {Z.-Y.}\ \bibnamefont {Peng}}, \
  and\ \bibinfo {author} {\bibfnamefont {Y.-S.}\ \bibnamefont {Piao}},\
  }\href@noop {} {\  (\bibinfo {year} {2025}{\natexlab{a}})},\ \Eprint
  {http://arxiv.org/abs/2503.23918} {arXiv:2503.23918 [astro-ph.CO]}
  \BibitemShut {NoStop}%
\bibitem [{\citenamefont {Wang}\ and\ \citenamefont
  {Piao}(2024)}]{Wang:2024dka}%
  \BibitemOpen
  \bibfield  {author} {\bibinfo {author} {\bibfnamefont {H.}~\bibnamefont
  {Wang}}\ and\ \bibinfo {author} {\bibfnamefont {Y.-S.}\ \bibnamefont
  {Piao}},\ }\href@noop {} {\  (\bibinfo {year} {2024})},\ \Eprint
  {http://arxiv.org/abs/2404.18579} {arXiv:2404.18579 [astro-ph.CO]}
  \BibitemShut {NoStop}%
\bibitem [{\citenamefont {Lodha}\ \emph {et~al.}(2025)\citenamefont {Lodha}
  \emph {et~al.}}]{DESI:2025fii}%
  \BibitemOpen
  \bibfield  {author} {\bibinfo {author} {\bibfnamefont {K.}~\bibnamefont
  {Lodha}} \emph {et~al.} (\bibinfo {collaboration} {DESI}),\ }\href {\doibase
  10.1103/w4c6-1r5j} {\bibfield  {journal} {\bibinfo  {journal} {Phys. Rev. D}\
  }\textbf {\bibinfo {volume} {112}},\ \bibinfo {pages} {083511} (\bibinfo
  {year} {2025})},\ \Eprint {http://arxiv.org/abs/2503.14743} {arXiv:2503.14743
  [astro-ph.CO]} \BibitemShut {NoStop}%
\bibitem [{\citenamefont {Pan}\ \emph {et~al.}(2025)\citenamefont {Pan},
  \citenamefont {Paul}, \citenamefont {Saridakis},\ and\ \citenamefont
  {Yang}}]{Pan:2025qwy}%
  \BibitemOpen
  \bibfield  {author} {\bibinfo {author} {\bibfnamefont {S.}~\bibnamefont
  {Pan}}, \bibinfo {author} {\bibfnamefont {S.}~\bibnamefont {Paul}}, \bibinfo
  {author} {\bibfnamefont {E.~N.}\ \bibnamefont {Saridakis}}, \ and\ \bibinfo
  {author} {\bibfnamefont {W.}~\bibnamefont {Yang}},\ }\href@noop {} {\
  (\bibinfo {year} {2025})},\ \Eprint {http://arxiv.org/abs/2504.00994}
  {arXiv:2504.00994 [astro-ph.CO]} \BibitemShut {NoStop}%
\bibitem [{\citenamefont {You}\ \emph {et~al.}(2025)\citenamefont {You},
  \citenamefont {Wang},\ and\ \citenamefont {Yang}}]{You:2025uon}%
  \BibitemOpen
  \bibfield  {author} {\bibinfo {author} {\bibfnamefont {C.}~\bibnamefont
  {You}}, \bibinfo {author} {\bibfnamefont {D.}~\bibnamefont {Wang}}, \ and\
  \bibinfo {author} {\bibfnamefont {T.}~\bibnamefont {Yang}},\ }\href {\doibase
  10.1103/f6v7-n9fr} {\bibfield  {journal} {\bibinfo  {journal} {Phys. Rev. D}\
  }\textbf {\bibinfo {volume} {112}},\ \bibinfo {pages} {043503} (\bibinfo
  {year} {2025})},\ \Eprint {http://arxiv.org/abs/2504.00985} {arXiv:2504.00985
  [astro-ph.CO]} \BibitemShut {NoStop}%
\bibitem [{\citenamefont {Pan}\ and\ \citenamefont {Ye}(2025)}]{Pan:2025psn}%
  \BibitemOpen
  \bibfield  {author} {\bibinfo {author} {\bibfnamefont {J.}~\bibnamefont
  {Pan}}\ and\ \bibinfo {author} {\bibfnamefont {G.}~\bibnamefont {Ye}},\
  }\href@noop {} {\  (\bibinfo {year} {2025})},\ \Eprint
  {http://arxiv.org/abs/2503.19898} {arXiv:2503.19898 [astro-ph.CO]}
  \BibitemShut {NoStop}%
\bibitem [{\citenamefont {Silva}\ \emph {et~al.}(2025)\citenamefont {Silva},
  \citenamefont {Sabogal}, \citenamefont {Scherer}, \citenamefont {Nunes},
  \citenamefont {Di~Valentino},\ and\ \citenamefont {Kumar}}]{Silva:2025hxw}%
  \BibitemOpen
  \bibfield  {author} {\bibinfo {author} {\bibfnamefont {E.}~\bibnamefont
  {Silva}}, \bibinfo {author} {\bibfnamefont {M.~A.}\ \bibnamefont {Sabogal}},
  \bibinfo {author} {\bibfnamefont {M.}~\bibnamefont {Scherer}}, \bibinfo
  {author} {\bibfnamefont {R.~C.}\ \bibnamefont {Nunes}}, \bibinfo {author}
  {\bibfnamefont {E.}~\bibnamefont {Di~Valentino}}, \ and\ \bibinfo {author}
  {\bibfnamefont {S.}~\bibnamefont {Kumar}},\ }\href {\doibase
  10.1103/qqc6-76z4} {\bibfield  {journal} {\bibinfo  {journal} {Phys. Rev. D}\
  }\textbf {\bibinfo {volume} {111}},\ \bibinfo {pages} {123511} (\bibinfo
  {year} {2025})},\ \Eprint {http://arxiv.org/abs/2503.23225} {arXiv:2503.23225
  [astro-ph.CO]} \BibitemShut {NoStop}%
\bibitem [{\citenamefont {Qiang}\ \emph {et~al.}(2025)\citenamefont {Qiang},
  \citenamefont {Jia},\ and\ \citenamefont {Wei}}]{Qiang:2025cxp}%
  \BibitemOpen
  \bibfield  {author} {\bibinfo {author} {\bibfnamefont {D.-C.}\ \bibnamefont
  {Qiang}}, \bibinfo {author} {\bibfnamefont {J.-Y.}\ \bibnamefont {Jia}}, \
  and\ \bibinfo {author} {\bibfnamefont {H.}~\bibnamefont {Wei}},\ }\href@noop
  {} {\  (\bibinfo {year} {2025})},\ \Eprint {http://arxiv.org/abs/2507.09981}
  {arXiv:2507.09981 [astro-ph.CO]} \BibitemShut {NoStop}%
\bibitem [{\citenamefont {Cai}\ \emph {et~al.}(2025)\citenamefont {Cai},
  \citenamefont {Ren}, \citenamefont {Qiu}, \citenamefont {Li},\ and\
  \citenamefont {Zhang}}]{Cai:2025mas}%
  \BibitemOpen
  \bibfield  {author} {\bibinfo {author} {\bibfnamefont {Y.}~\bibnamefont
  {Cai}}, \bibinfo {author} {\bibfnamefont {X.}~\bibnamefont {Ren}}, \bibinfo
  {author} {\bibfnamefont {T.}~\bibnamefont {Qiu}}, \bibinfo {author}
  {\bibfnamefont {M.}~\bibnamefont {Li}}, \ and\ \bibinfo {author}
  {\bibfnamefont {X.}~\bibnamefont {Zhang}},\ }\href@noop {} {\  (\bibinfo
  {year} {2025})},\ \Eprint {http://arxiv.org/abs/2505.24732} {arXiv:2505.24732
  [astro-ph.CO]} \BibitemShut {NoStop}%
\bibitem [{\citenamefont {Cline}\ and\ \citenamefont
  {Muralidharan}(2025)}]{Cline:2025sbt}%
  \BibitemOpen
  \bibfield  {author} {\bibinfo {author} {\bibfnamefont {J.~M.}\ \bibnamefont
  {Cline}}\ and\ \bibinfo {author} {\bibfnamefont {V.}~\bibnamefont
  {Muralidharan}},\ }\href {\doibase 10.1103/8z2m-nbv6} {\bibfield  {journal}
  {\bibinfo  {journal} {Phys. Rev. D}\ }\textbf {\bibinfo {volume} {112}},\
  \bibinfo {pages} {063539} (\bibinfo {year} {2025})},\ \Eprint
  {http://arxiv.org/abs/2506.13047} {arXiv:2506.13047 [astro-ph.CO]}
  \BibitemShut {NoStop}%
\bibitem [{\citenamefont {Rodrigues}\ \emph {et~al.}(2025)\citenamefont
  {Rodrigues}, \citenamefont {de~Souza},\ and\ \citenamefont
  {Alcaniz}}]{Rodrigues:2025tfg}%
  \BibitemOpen
  \bibfield  {author} {\bibinfo {author} {\bibfnamefont {G.}~\bibnamefont
  {Rodrigues}}, \bibinfo {author} {\bibfnamefont {R.}~\bibnamefont {de~Souza}},
  \ and\ \bibinfo {author} {\bibfnamefont {J.}~\bibnamefont {Alcaniz}},\
  }\href@noop {} {\  (\bibinfo {year} {2025})},\ \Eprint
  {http://arxiv.org/abs/2506.22373} {arXiv:2506.22373 [astro-ph.CO]}
  \BibitemShut {NoStop}%
\bibitem [{\citenamefont {Li}\ and\ \citenamefont {Wang}(2025)}]{Li:2025ops}%
  \BibitemOpen
  \bibfield  {author} {\bibinfo {author} {\bibfnamefont {J.-X.}\ \bibnamefont
  {Li}}\ and\ \bibinfo {author} {\bibfnamefont {S.}~\bibnamefont {Wang}},\
  }\href@noop {} {\  (\bibinfo {year} {2025})},\ \Eprint
  {http://arxiv.org/abs/2506.22953} {arXiv:2506.22953 [astro-ph.CO]}
  \BibitemShut {NoStop}%
\bibitem [{\citenamefont {Bhattacharjee}\ \emph {et~al.}(2025)\citenamefont
  {Bhattacharjee}, \citenamefont {Halder}, \citenamefont {de~Haro},
  \citenamefont {Pan},\ and\ \citenamefont
  {Saridakis}}]{Bhattacharjee:2025xeb}%
  \BibitemOpen
  \bibfield  {author} {\bibinfo {author} {\bibfnamefont {S.}~\bibnamefont
  {Bhattacharjee}}, \bibinfo {author} {\bibfnamefont {S.}~\bibnamefont
  {Halder}}, \bibinfo {author} {\bibfnamefont {J.}~\bibnamefont {de~Haro}},
  \bibinfo {author} {\bibfnamefont {S.}~\bibnamefont {Pan}}, \ and\ \bibinfo
  {author} {\bibfnamefont {E.~N.}\ \bibnamefont {Saridakis}},\ }\href@noop {}
  {\  (\bibinfo {year} {2025})},\ \Eprint {http://arxiv.org/abs/2507.15575}
  {arXiv:2507.15575 [astro-ph.CO]} \BibitemShut {NoStop}%
\bibitem [{\citenamefont {Gialamas}\ \emph
  {et~al.}(2025{\natexlab{a}})\citenamefont {Gialamas}, \citenamefont
  {H{\"u}tsi}, \citenamefont {Raidal}, \citenamefont {Urrutia}, \citenamefont
  {Vasar},\ and\ \citenamefont {Veerm{\"a}e}}]{Gialamas:2025pwv}%
  \BibitemOpen
  \bibfield  {author} {\bibinfo {author} {\bibfnamefont {I.~D.}\ \bibnamefont
  {Gialamas}}, \bibinfo {author} {\bibfnamefont {G.}~\bibnamefont {H{\"u}tsi}},
  \bibinfo {author} {\bibfnamefont {M.}~\bibnamefont {Raidal}}, \bibinfo
  {author} {\bibfnamefont {J.}~\bibnamefont {Urrutia}}, \bibinfo {author}
  {\bibfnamefont {M.}~\bibnamefont {Vasar}}, \ and\ \bibinfo {author}
  {\bibfnamefont {H.}~\bibnamefont {Veerm{\"a}e}},\ }\href {\doibase
  10.1103/kdqc-y37v} {\bibfield  {journal} {\bibinfo  {journal} {Phys. Rev. D}\
  }\textbf {\bibinfo {volume} {112}},\ \bibinfo {pages} {063551} (\bibinfo
  {year} {2025}{\natexlab{a}})},\ \Eprint {http://arxiv.org/abs/2506.21542}
  {arXiv:2506.21542 [astro-ph.CO]} \BibitemShut {NoStop}%
\bibitem [{\citenamefont {Wang}\ and\ \citenamefont
  {Piao}(2025)}]{Wang:2025dtk}%
  \BibitemOpen
  \bibfield  {author} {\bibinfo {author} {\bibfnamefont {H.}~\bibnamefont
  {Wang}}\ and\ \bibinfo {author} {\bibfnamefont {Y.-S.}\ \bibnamefont
  {Piao}},\ }\href@noop {} {\  (\bibinfo {year} {2025})},\ \Eprint
  {http://arxiv.org/abs/2506.04306} {arXiv:2506.04306 [gr-qc]} \BibitemShut
  {NoStop}%
\bibitem [{\citenamefont {Li}\ \emph {et~al.}(2025{\natexlab{c}})\citenamefont
  {Li}, \citenamefont {Zhang}, \citenamefont {Yao}, \citenamefont {Wu},
  \citenamefont {Zhang},\ and\ \citenamefont {Zhang}}]{Li:2025eqh}%
  \BibitemOpen
  \bibfield  {author} {\bibinfo {author} {\bibfnamefont {T.-N.}\ \bibnamefont
  {Li}}, \bibinfo {author} {\bibfnamefont {Y.-M.}\ \bibnamefont {Zhang}},
  \bibinfo {author} {\bibfnamefont {Y.-H.}\ \bibnamefont {Yao}}, \bibinfo
  {author} {\bibfnamefont {P.-J.}\ \bibnamefont {Wu}}, \bibinfo {author}
  {\bibfnamefont {J.-F.}\ \bibnamefont {Zhang}}, \ and\ \bibinfo {author}
  {\bibfnamefont {X.}~\bibnamefont {Zhang}},\ }\href@noop {} {\  (\bibinfo
  {year} {2025}{\natexlab{c}})},\ \Eprint {http://arxiv.org/abs/2506.09819}
  {arXiv:2506.09819 [astro-ph.CO]} \BibitemShut {NoStop}%
\bibitem [{\citenamefont {Li}\ and\ \citenamefont {Zhang}(2025)}]{Li:2025ula}%
  \BibitemOpen
  \bibfield  {author} {\bibinfo {author} {\bibfnamefont {Y.-H.}\ \bibnamefont
  {Li}}\ and\ \bibinfo {author} {\bibfnamefont {X.}~\bibnamefont {Zhang}},\
  }\href@noop {} {\  (\bibinfo {year} {2025})},\ \Eprint
  {http://arxiv.org/abs/2506.18477} {arXiv:2506.18477 [astro-ph.CO]}
  \BibitemShut {NoStop}%
\bibitem [{\citenamefont {Li}\ \emph {et~al.}(2025{\natexlab{d}})\citenamefont
  {Li}, \citenamefont {Wu}, \citenamefont {Du}, \citenamefont {Yao},
  \citenamefont {Zhang},\ and\ \citenamefont {Zhang}}]{Li:2025dwz}%
  \BibitemOpen
  \bibfield  {author} {\bibinfo {author} {\bibfnamefont {T.-N.}\ \bibnamefont
  {Li}}, \bibinfo {author} {\bibfnamefont {P.-J.}\ \bibnamefont {Wu}}, \bibinfo
  {author} {\bibfnamefont {G.-H.}\ \bibnamefont {Du}}, \bibinfo {author}
  {\bibfnamefont {Y.-H.}\ \bibnamefont {Yao}}, \bibinfo {author} {\bibfnamefont
  {J.-F.}\ \bibnamefont {Zhang}}, \ and\ \bibinfo {author} {\bibfnamefont
  {X.}~\bibnamefont {Zhang}},\ }\href {\doibase 10.1016/j.dark.2025.102068}
  {\bibfield  {journal} {\bibinfo  {journal} {Phys. Dark Univ.}\ }\textbf
  {\bibinfo {volume} {50}},\ \bibinfo {pages} {102068} (\bibinfo {year}
  {2025}{\natexlab{d}})},\ \Eprint {http://arxiv.org/abs/2507.07798}
  {arXiv:2507.07798 [astro-ph.CO]} \BibitemShut {NoStop}%
\bibitem [{\citenamefont {Li}\ \emph {et~al.}(2025{\natexlab{e}})\citenamefont
  {Li}, \citenamefont {Du}, \citenamefont {Wu}, \citenamefont {Qi},
  \citenamefont {Zhang},\ and\ \citenamefont {Zhang}}]{Li:2025htp}%
  \BibitemOpen
  \bibfield  {author} {\bibinfo {author} {\bibfnamefont {T.-N.}\ \bibnamefont
  {Li}}, \bibinfo {author} {\bibfnamefont {G.-H.}\ \bibnamefont {Du}}, \bibinfo
  {author} {\bibfnamefont {P.-J.}\ \bibnamefont {Wu}}, \bibinfo {author}
  {\bibfnamefont {J.-Z.}\ \bibnamefont {Qi}}, \bibinfo {author} {\bibfnamefont
  {J.-F.}\ \bibnamefont {Zhang}}, \ and\ \bibinfo {author} {\bibfnamefont
  {X.}~\bibnamefont {Zhang}},\ }\href@noop {} {\  (\bibinfo {year}
  {2025}{\natexlab{e}})},\ \Eprint {http://arxiv.org/abs/2507.13811}
  {arXiv:2507.13811 [astro-ph.CO]} \BibitemShut {NoStop}%
\bibitem [{\citenamefont {Gonz{\'a}lez-Fuentes}\ and\ \citenamefont
  {G{\'o}mez-Valent}(2025)}]{Gonzalez-Fuentes:2025lei}%
  \BibitemOpen
  \bibfield  {author} {\bibinfo {author} {\bibfnamefont {A.}~\bibnamefont
  {Gonz{\'a}lez-Fuentes}}\ and\ \bibinfo {author} {\bibfnamefont
  {A.}~\bibnamefont {G{\'o}mez-Valent}},\ }\href@noop {} {\  (\bibinfo {year}
  {2025})},\ \Eprint {http://arxiv.org/abs/2506.11758} {arXiv:2506.11758
  [astro-ph.CO]} \BibitemShut {NoStop}%
\bibitem [{\citenamefont {G{\'o}mez-Valent}\ and\ \citenamefont
  {Sol{\`a}~Peracaula}(2025)}]{Gomez-Valent:2024ejh}%
  \BibitemOpen
  \bibfield  {author} {\bibinfo {author} {\bibfnamefont {A.}~\bibnamefont
  {G{\'o}mez-Valent}}\ and\ \bibinfo {author} {\bibfnamefont {J.}~\bibnamefont
  {Sol{\`a}~Peracaula}},\ }\href {\doibase 10.1016/j.physletb.2025.139391}
  {\bibfield  {journal} {\bibinfo  {journal} {Phys. Lett. B}\ }\textbf
  {\bibinfo {volume} {864}},\ \bibinfo {pages} {139391} (\bibinfo {year}
  {2025})},\ \Eprint {http://arxiv.org/abs/2412.15124} {arXiv:2412.15124
  [astro-ph.CO]} \BibitemShut {NoStop}%
\bibitem [{\citenamefont {Wang}\ \emph
  {et~al.}(2025{\natexlab{b}})\citenamefont {Wang}, \citenamefont {Cai},
  \citenamefont {Guo},\ and\ \citenamefont {Wang}}]{Wang:2025znm}%
  \BibitemOpen
  \bibfield  {author} {\bibinfo {author} {\bibfnamefont {J.-Q.}\ \bibnamefont
  {Wang}}, \bibinfo {author} {\bibfnamefont {R.-G.}\ \bibnamefont {Cai}},
  \bibinfo {author} {\bibfnamefont {Z.-K.}\ \bibnamefont {Guo}}, \ and\
  \bibinfo {author} {\bibfnamefont {S.-J.}\ \bibnamefont {Wang}},\ }\href@noop
  {} {\  (\bibinfo {year} {2025}{\natexlab{b}})},\ \Eprint
  {http://arxiv.org/abs/2508.01759} {arXiv:2508.01759 [astro-ph.CO]}
  \BibitemShut {NoStop}%
\bibitem [{\citenamefont {Du}\ \emph {et~al.}(2025{\natexlab{c}})\citenamefont
  {Du}, \citenamefont {Li}, \citenamefont {Wu}, \citenamefont {Zhang},\ and\
  \citenamefont {Zhang}}]{Du:2025xes}%
  \BibitemOpen
  \bibfield  {author} {\bibinfo {author} {\bibfnamefont {G.-H.}\ \bibnamefont
  {Du}}, \bibinfo {author} {\bibfnamefont {T.-N.}\ \bibnamefont {Li}}, \bibinfo
  {author} {\bibfnamefont {P.-J.}\ \bibnamefont {Wu}}, \bibinfo {author}
  {\bibfnamefont {J.-F.}\ \bibnamefont {Zhang}}, \ and\ \bibinfo {author}
  {\bibfnamefont {X.}~\bibnamefont {Zhang}},\ }\href@noop {} {\  (\bibinfo
  {year} {2025}{\natexlab{c}})},\ \Eprint {http://arxiv.org/abs/2507.16589}
  {arXiv:2507.16589 [astro-ph.CO]} \BibitemShut {NoStop}%
\bibitem [{\citenamefont {Feng}\ \emph
  {et~al.}(2025{\natexlab{b}})\citenamefont {Feng}, \citenamefont {Han},
  \citenamefont {Zhang},\ and\ \citenamefont {Zhang}}]{Feng:2025wbz}%
  \BibitemOpen
  \bibfield  {author} {\bibinfo {author} {\bibfnamefont {L.}~\bibnamefont
  {Feng}}, \bibinfo {author} {\bibfnamefont {T.}~\bibnamefont {Han}}, \bibinfo
  {author} {\bibfnamefont {J.-F.}\ \bibnamefont {Zhang}}, \ and\ \bibinfo
  {author} {\bibfnamefont {X.}~\bibnamefont {Zhang}},\ }\href@noop {} {\
  (\bibinfo {year} {2025}{\natexlab{b}})},\ \Eprint
  {http://arxiv.org/abs/2507.17315} {arXiv:2507.17315 [astro-ph.CO]}
  \BibitemShut {NoStop}%
\bibitem [{\citenamefont {Wang}\ \emph
  {et~al.}(2024{\natexlab{b}})\citenamefont {Wang}, \citenamefont {Liu},
  \citenamefont {Yu},\ and\ \citenamefont {Wu}}]{Wang:2024nsi}%
  \BibitemOpen
  \bibfield  {author} {\bibinfo {author} {\bibfnamefont {B.}~\bibnamefont
  {Wang}}, \bibinfo {author} {\bibfnamefont {Y.}~\bibnamefont {Liu}}, \bibinfo
  {author} {\bibfnamefont {H.}~\bibnamefont {Yu}}, \ and\ \bibinfo {author}
  {\bibfnamefont {P.}~\bibnamefont {Wu}},\ }\href {\doibase
  10.3847/1538-4357/ad1ab5} {\bibfield  {journal} {\bibinfo  {journal}
  {Astrophys. J.}\ }\textbf {\bibinfo {volume} {962}},\ \bibinfo {pages} {103}
  (\bibinfo {year} {2024}{\natexlab{b}})},\ \Eprint
  {http://arxiv.org/abs/2401.01540} {arXiv:2401.01540 [astro-ph.CO]}
  \BibitemShut {NoStop}%
\bibitem [{\citenamefont {Visser}(2004)}]{Visser:2003vq}%
  \BibitemOpen
  \bibfield  {author} {\bibinfo {author} {\bibfnamefont {M.}~\bibnamefont
  {Visser}},\ }\href {\doibase 10.1088/0264-9381/21/11/006} {\bibfield
  {journal} {\bibinfo  {journal} {Class. Quant. Grav.}\ }\textbf {\bibinfo
  {volume} {21}},\ \bibinfo {pages} {2603} (\bibinfo {year} {2004})},\ \Eprint
  {http://arxiv.org/abs/gr-qc/0309109} {arXiv:gr-qc/0309109} \BibitemShut
  {NoStop}%
\bibitem [{\citenamefont {Hu}\ and\ \citenamefont {Wang}(2022)}]{Hu:2022udt}%
  \BibitemOpen
  \bibfield  {author} {\bibinfo {author} {\bibfnamefont {J.~P.}\ \bibnamefont
  {Hu}}\ and\ \bibinfo {author} {\bibfnamefont {F.~Y.}\ \bibnamefont {Wang}},\
  }\href {\doibase 10.1051/0004-6361/202142162} {\bibfield  {journal} {\bibinfo
   {journal} {Astron. Astrophys.}\ }\textbf {\bibinfo {volume} {661}},\
  \bibinfo {pages} {A71} (\bibinfo {year} {2022})},\ \Eprint
  {http://arxiv.org/abs/2202.09075} {arXiv:2202.09075 [astro-ph.CO]}
  \BibitemShut {NoStop}%
\bibitem [{\citenamefont {Scolnic}\ \emph {et~al.}(2022)\citenamefont {Scolnic}
  \emph {et~al.}}]{Scolnic:2021amr}%
  \BibitemOpen
  \bibfield  {author} {\bibinfo {author} {\bibfnamefont {D.}~\bibnamefont
  {Scolnic}} \emph {et~al.},\ }\href {\doibase 10.3847/1538-4357/ac8b7a}
  {\bibfield  {journal} {\bibinfo  {journal} {Astrophys. J.}\ }\textbf
  {\bibinfo {volume} {938}},\ \bibinfo {pages} {113} (\bibinfo {year}
  {2022})},\ \Eprint {http://arxiv.org/abs/2112.03863} {arXiv:2112.03863
  [astro-ph.CO]} \BibitemShut {NoStop}%
\bibitem [{\citenamefont {Abbott}\ \emph {et~al.}(2024)\citenamefont {Abbott}
  \emph {et~al.}}]{DES:2024jxu}%
  \BibitemOpen
  \bibfield  {author} {\bibinfo {author} {\bibfnamefont {T.~M.~C.}\
  \bibnamefont {Abbott}} \emph {et~al.} (\bibinfo {collaboration} {DES}),\
  }\href {\doibase 10.3847/2041-8213/ad6f9f} {\bibfield  {journal} {\bibinfo
  {journal} {Astrophys. J. Lett.}\ }\textbf {\bibinfo {volume} {973}},\
  \bibinfo {pages} {L14} (\bibinfo {year} {2024})},\ \Eprint
  {http://arxiv.org/abs/2401.02929} {arXiv:2401.02929 [astro-ph.CO]}
  \BibitemShut {NoStop}%
\bibitem [{\citenamefont {Vincenzi}\ \emph {et~al.}(2024)\citenamefont
  {Vincenzi} \emph {et~al.}}]{DES:2024hip}%
  \BibitemOpen
  \bibfield  {author} {\bibinfo {author} {\bibfnamefont {M.}~\bibnamefont
  {Vincenzi}} \emph {et~al.} (\bibinfo {collaboration} {DES}),\ }\href
  {\doibase 10.3847/1538-4357/ad5e6c} {\bibfield  {journal} {\bibinfo
  {journal} {Astrophys. J.}\ }\textbf {\bibinfo {volume} {975}},\ \bibinfo
  {pages} {86} (\bibinfo {year} {2024})},\ \Eprint
  {http://arxiv.org/abs/2401.02945} {arXiv:2401.02945 [astro-ph.CO]}
  \BibitemShut {NoStop}%
\bibitem [{\citenamefont {S{\'a}nchez}\ \emph {et~al.}(2024)\citenamefont
  {S{\'a}nchez} \emph {et~al.}}]{DES:2024upw}%
  \BibitemOpen
  \bibfield  {author} {\bibinfo {author} {\bibfnamefont {B.~O.}\ \bibnamefont
  {S{\'a}nchez}} \emph {et~al.} (\bibinfo {collaboration} {DES}),\ }\href
  {\doibase 10.3847/1538-4357/ad739a} {\bibfield  {journal} {\bibinfo
  {journal} {Astrophys. J.}\ }\textbf {\bibinfo {volume} {975}},\ \bibinfo
  {pages} {5} (\bibinfo {year} {2024})},\ \Eprint
  {http://arxiv.org/abs/2406.05046} {arXiv:2406.05046 [astro-ph.CO]}
  \BibitemShut {NoStop}%
\bibitem [{\citenamefont {Hicken}\ \emph {et~al.}(2009)\citenamefont {Hicken},
  \citenamefont {Challis}, \citenamefont {Jha}, \citenamefont {Kirsher},
  \citenamefont {Matheson}, \citenamefont {Modjaz}, \citenamefont {Rest},\ and\
  \citenamefont {Wood-Vasey}}]{Hicken:2009df}%
  \BibitemOpen
  \bibfield  {author} {\bibinfo {author} {\bibfnamefont {M.}~\bibnamefont
  {Hicken}}, \bibinfo {author} {\bibfnamefont {P.}~\bibnamefont {Challis}},
  \bibinfo {author} {\bibfnamefont {S.}~\bibnamefont {Jha}}, \bibinfo {author}
  {\bibfnamefont {R.~P.}\ \bibnamefont {Kirsher}}, \bibinfo {author}
  {\bibfnamefont {T.}~\bibnamefont {Matheson}}, \bibinfo {author}
  {\bibfnamefont {M.}~\bibnamefont {Modjaz}}, \bibinfo {author} {\bibfnamefont
  {A.}~\bibnamefont {Rest}}, \ and\ \bibinfo {author} {\bibfnamefont {W.~M.}\
  \bibnamefont {Wood-Vasey}},\ }\href {\doibase 10.1088/0004-637X/700/1/331}
  {\bibfield  {journal} {\bibinfo  {journal} {Astrophys. J.}\ }\textbf
  {\bibinfo {volume} {700}},\ \bibinfo {pages} {331} (\bibinfo {year}
  {2009})},\ \Eprint {http://arxiv.org/abs/0901.4787} {arXiv:0901.4787
  [astro-ph.CO]} \BibitemShut {NoStop}%
\bibitem [{\citenamefont {Hicken}\ \emph {et~al.}(2012)\citenamefont {Hicken}
  \emph {et~al.}}]{Hicken:2012zr}%
  \BibitemOpen
  \bibfield  {author} {\bibinfo {author} {\bibfnamefont {M.}~\bibnamefont
  {Hicken}} \emph {et~al.},\ }\href {\doibase 10.1088/0067-0049/200/2/12}
  {\bibfield  {journal} {\bibinfo  {journal} {Astrophys. J. Suppl.}\ }\textbf
  {\bibinfo {volume} {200}},\ \bibinfo {pages} {12} (\bibinfo {year} {2012})},\
  \Eprint {http://arxiv.org/abs/1205.4493} {arXiv:1205.4493 [astro-ph.CO]}
  \BibitemShut {NoStop}%
\bibitem [{\citenamefont {Krisciunas}\ \emph {et~al.}(2017)\citenamefont
  {Krisciunas} \emph {et~al.}}]{Krisciunas:2017yoe}%
  \BibitemOpen
  \bibfield  {author} {\bibinfo {author} {\bibfnamefont {K.}~\bibnamefont
  {Krisciunas}} \emph {et~al.},\ }\href {\doibase 10.3847/1538-3881/aa8df0}
  {\bibfield  {journal} {\bibinfo  {journal} {Astron. J.}\ }\textbf {\bibinfo
  {volume} {154}},\ \bibinfo {pages} {211} (\bibinfo {year} {2017})},\ \Eprint
  {http://arxiv.org/abs/1709.05146} {arXiv:1709.05146 [astro-ph.IM]}
  \BibitemShut {NoStop}%
\bibitem [{\citenamefont {Foley}\ \emph {et~al.}(2018)\citenamefont {Foley}
  \emph {et~al.}}]{Foley:2017zdq}%
  \BibitemOpen
  \bibfield  {author} {\bibinfo {author} {\bibfnamefont {R.~J.}\ \bibnamefont
  {Foley}} \emph {et~al.},\ }\href {\doibase 10.1093/mnras/stx3136} {\bibfield
  {journal} {\bibinfo  {journal} {Mon. Not. Roy. Astron. Soc.}\ }\textbf
  {\bibinfo {volume} {475}},\ \bibinfo {pages} {193} (\bibinfo {year}
  {2018})},\ \Eprint {http://arxiv.org/abs/1711.02474} {arXiv:1711.02474
  [astro-ph.HE]} \BibitemShut {NoStop}%
\bibitem [{\citenamefont {Beutler}\ \emph {et~al.}(2011)\citenamefont
  {Beutler}, \citenamefont {Blake}, \citenamefont {Colless}, \citenamefont
  {Jones}, \citenamefont {Staveley-Smith}, \citenamefont {Campbell},
  \citenamefont {Parker}, \citenamefont {Saunders},\ and\ \citenamefont
  {Watson}}]{Beutler:2011hx}%
  \BibitemOpen
  \bibfield  {author} {\bibinfo {author} {\bibfnamefont {F.}~\bibnamefont
  {Beutler}}, \bibinfo {author} {\bibfnamefont {C.}~\bibnamefont {Blake}},
  \bibinfo {author} {\bibfnamefont {M.}~\bibnamefont {Colless}}, \bibinfo
  {author} {\bibfnamefont {D.~H.}\ \bibnamefont {Jones}}, \bibinfo {author}
  {\bibfnamefont {L.}~\bibnamefont {Staveley-Smith}}, \bibinfo {author}
  {\bibfnamefont {L.}~\bibnamefont {Campbell}}, \bibinfo {author}
  {\bibfnamefont {Q.}~\bibnamefont {Parker}}, \bibinfo {author} {\bibfnamefont
  {W.}~\bibnamefont {Saunders}}, \ and\ \bibinfo {author} {\bibfnamefont
  {F.}~\bibnamefont {Watson}},\ }\href {\doibase
  10.1111/j.1365-2966.2011.19250.x} {\bibfield  {journal} {\bibinfo  {journal}
  {Mon. Not. Roy. Astron. Soc.}\ }\textbf {\bibinfo {volume} {416}},\ \bibinfo
  {pages} {3017} (\bibinfo {year} {2011})},\ \Eprint
  {http://arxiv.org/abs/1106.3366} {arXiv:1106.3366 [astro-ph.CO]} \BibitemShut
  {NoStop}%
\bibitem [{\citenamefont {Ross}\ \emph {et~al.}(2015)\citenamefont {Ross},
  \citenamefont {Samushia}, \citenamefont {Howlett}, \citenamefont {Percival},
  \citenamefont {Burden},\ and\ \citenamefont {Manera}}]{Ross:2014qpa}%
  \BibitemOpen
  \bibfield  {author} {\bibinfo {author} {\bibfnamefont {A.~J.}\ \bibnamefont
  {Ross}}, \bibinfo {author} {\bibfnamefont {L.}~\bibnamefont {Samushia}},
  \bibinfo {author} {\bibfnamefont {C.}~\bibnamefont {Howlett}}, \bibinfo
  {author} {\bibfnamefont {W.~J.}\ \bibnamefont {Percival}}, \bibinfo {author}
  {\bibfnamefont {A.}~\bibnamefont {Burden}}, \ and\ \bibinfo {author}
  {\bibfnamefont {M.}~\bibnamefont {Manera}},\ }\href {\doibase
  10.1093/mnras/stv154} {\bibfield  {journal} {\bibinfo  {journal} {Mon. Not.
  Roy. Astron. Soc.}\ }\textbf {\bibinfo {volume} {449}},\ \bibinfo {pages}
  {835} (\bibinfo {year} {2015})},\ \Eprint {http://arxiv.org/abs/1409.3242}
  {arXiv:1409.3242 [astro-ph.CO]} \BibitemShut {NoStop}%
\bibitem [{\citenamefont {Jimenez}\ \emph {et~al.}(2003)\citenamefont
  {Jimenez}, \citenamefont {Verde}, \citenamefont {Treu},\ and\ \citenamefont
  {Stern}}]{Jimenez:2003iv}%
  \BibitemOpen
  \bibfield  {author} {\bibinfo {author} {\bibfnamefont {R.}~\bibnamefont
  {Jimenez}}, \bibinfo {author} {\bibfnamefont {L.}~\bibnamefont {Verde}},
  \bibinfo {author} {\bibfnamefont {T.}~\bibnamefont {Treu}}, \ and\ \bibinfo
  {author} {\bibfnamefont {D.}~\bibnamefont {Stern}},\ }\href {\doibase
  10.1086/376595} {\bibfield  {journal} {\bibinfo  {journal} {Astrophys. J.}\
  }\textbf {\bibinfo {volume} {593}},\ \bibinfo {pages} {622} (\bibinfo {year}
  {2003})},\ \Eprint {http://arxiv.org/abs/astro-ph/0302560}
  {arXiv:astro-ph/0302560} \BibitemShut {NoStop}%
\bibitem [{\citenamefont {Simon}\ \emph {et~al.}(2005)\citenamefont {Simon},
  \citenamefont {Verde},\ and\ \citenamefont {Jimenez}}]{Simon:2004tf}%
  \BibitemOpen
  \bibfield  {author} {\bibinfo {author} {\bibfnamefont {J.}~\bibnamefont
  {Simon}}, \bibinfo {author} {\bibfnamefont {L.}~\bibnamefont {Verde}}, \ and\
  \bibinfo {author} {\bibfnamefont {R.}~\bibnamefont {Jimenez}},\ }\href
  {\doibase 10.1103/PhysRevD.71.123001} {\bibfield  {journal} {\bibinfo
  {journal} {Phys. Rev. D}\ }\textbf {\bibinfo {volume} {71}},\ \bibinfo
  {pages} {123001} (\bibinfo {year} {2005})},\ \Eprint
  {http://arxiv.org/abs/astro-ph/0412269} {arXiv:astro-ph/0412269} \BibitemShut
  {NoStop}%
\bibitem [{\citenamefont {Stern}\ \emph {et~al.}(2010)\citenamefont {Stern},
  \citenamefont {Jimenez}, \citenamefont {Verde}, \citenamefont
  {Kamionkowski},\ and\ \citenamefont {Stanford}}]{Stern:2009ep}%
  \BibitemOpen
  \bibfield  {author} {\bibinfo {author} {\bibfnamefont {D.}~\bibnamefont
  {Stern}}, \bibinfo {author} {\bibfnamefont {R.}~\bibnamefont {Jimenez}},
  \bibinfo {author} {\bibfnamefont {L.}~\bibnamefont {Verde}}, \bibinfo
  {author} {\bibfnamefont {M.}~\bibnamefont {Kamionkowski}}, \ and\ \bibinfo
  {author} {\bibfnamefont {S.~A.}\ \bibnamefont {Stanford}},\ }\href {\doibase
  10.1088/1475-7516/2010/02/008} {\bibfield  {journal} {\bibinfo  {journal}
  {JCAP}\ }\textbf {\bibinfo {volume} {02}},\ \bibinfo {pages} {008} (\bibinfo
  {year} {2010})},\ \Eprint {http://arxiv.org/abs/0907.3149} {arXiv:0907.3149
  [astro-ph.CO]} \BibitemShut {NoStop}%
\bibitem [{\citenamefont {Moresco}\ \emph {et~al.}(2012)\citenamefont {Moresco}
  \emph {et~al.}}]{Moresco:2012jh}%
  \BibitemOpen
  \bibfield  {author} {\bibinfo {author} {\bibfnamefont {M.}~\bibnamefont
  {Moresco}} \emph {et~al.},\ }\href {\doibase 10.1088/1475-7516/2012/08/006}
  {\bibfield  {journal} {\bibinfo  {journal} {JCAP}\ }\textbf {\bibinfo
  {volume} {08}},\ \bibinfo {pages} {006} (\bibinfo {year} {2012})},\ \Eprint
  {http://arxiv.org/abs/1201.3609} {arXiv:1201.3609 [astro-ph.CO]} \BibitemShut
  {NoStop}%
\bibitem [{\citenamefont {Zhang}\ \emph
  {et~al.}(2014{\natexlab{b}})\citenamefont {Zhang}, \citenamefont {Zhang},
  \citenamefont {Yuan}, \citenamefont {Zhang},\ and\ \citenamefont
  {Sun}}]{Zhang:2012mp}%
  \BibitemOpen
  \bibfield  {author} {\bibinfo {author} {\bibfnamefont {C.}~\bibnamefont
  {Zhang}}, \bibinfo {author} {\bibfnamefont {H.}~\bibnamefont {Zhang}},
  \bibinfo {author} {\bibfnamefont {S.}~\bibnamefont {Yuan}}, \bibinfo {author}
  {\bibfnamefont {T.-J.}\ \bibnamefont {Zhang}}, \ and\ \bibinfo {author}
  {\bibfnamefont {Y.-C.}\ \bibnamefont {Sun}},\ }\href {\doibase
  10.1088/1674-4527/14/10/002} {\bibfield  {journal} {\bibinfo  {journal} {Res.
  Astron. Astrophys.}\ }\textbf {\bibinfo {volume} {14}},\ \bibinfo {pages}
  {1221} (\bibinfo {year} {2014}{\natexlab{b}})},\ \Eprint
  {http://arxiv.org/abs/1207.4541} {arXiv:1207.4541 [astro-ph.CO]} \BibitemShut
  {NoStop}%
\bibitem [{\citenamefont {Moresco}(2015)}]{Moresco:2015cya}%
  \BibitemOpen
  \bibfield  {author} {\bibinfo {author} {\bibfnamefont {M.}~\bibnamefont
  {Moresco}},\ }\href {\doibase 10.1093/mnrasl/slv037} {\bibfield  {journal}
  {\bibinfo  {journal} {Mon. Not. Roy. Astron. Soc.}\ }\textbf {\bibinfo
  {volume} {450}},\ \bibinfo {pages} {L16} (\bibinfo {year} {2015})},\ \Eprint
  {http://arxiv.org/abs/1503.01116} {arXiv:1503.01116 [astro-ph.CO]}
  \BibitemShut {NoStop}%
\bibitem [{\citenamefont {Moresco}\ \emph {et~al.}(2016)\citenamefont
  {Moresco}, \citenamefont {Pozzetti}, \citenamefont {Cimatti}, \citenamefont
  {Jimenez}, \citenamefont {Maraston}, \citenamefont {Verde}, \citenamefont
  {Thomas}, \citenamefont {Citro}, \citenamefont {Tojeiro},\ and\ \citenamefont
  {Wilkinson}}]{Moresco:2016mzx}%
  \BibitemOpen
  \bibfield  {author} {\bibinfo {author} {\bibfnamefont {M.}~\bibnamefont
  {Moresco}}, \bibinfo {author} {\bibfnamefont {L.}~\bibnamefont {Pozzetti}},
  \bibinfo {author} {\bibfnamefont {A.}~\bibnamefont {Cimatti}}, \bibinfo
  {author} {\bibfnamefont {R.}~\bibnamefont {Jimenez}}, \bibinfo {author}
  {\bibfnamefont {C.}~\bibnamefont {Maraston}}, \bibinfo {author}
  {\bibfnamefont {L.}~\bibnamefont {Verde}}, \bibinfo {author} {\bibfnamefont
  {D.}~\bibnamefont {Thomas}}, \bibinfo {author} {\bibfnamefont
  {A.}~\bibnamefont {Citro}}, \bibinfo {author} {\bibfnamefont
  {R.}~\bibnamefont {Tojeiro}}, \ and\ \bibinfo {author} {\bibfnamefont
  {D.}~\bibnamefont {Wilkinson}},\ }\href {\doibase
  10.1088/1475-7516/2016/05/014} {\bibfield  {journal} {\bibinfo  {journal}
  {JCAP}\ }\textbf {\bibinfo {volume} {05}},\ \bibinfo {pages} {014} (\bibinfo
  {year} {2016})},\ \Eprint {http://arxiv.org/abs/1601.01701} {arXiv:1601.01701
  [astro-ph.CO]} \BibitemShut {NoStop}%
\bibitem [{\citenamefont {Ratsimbazafy}\ \emph {et~al.}(2017)\citenamefont
  {Ratsimbazafy}, \citenamefont {Loubser}, \citenamefont {Crawford},
  \citenamefont {Cress}, \citenamefont {Bassett}, \citenamefont {Nichol},\ and\
  \citenamefont {V{\"a}is{\"a}nen}}]{Ratsimbazafy:2017vga}%
  \BibitemOpen
  \bibfield  {author} {\bibinfo {author} {\bibfnamefont {A.~L.}\ \bibnamefont
  {Ratsimbazafy}}, \bibinfo {author} {\bibfnamefont {S.~I.}\ \bibnamefont
  {Loubser}}, \bibinfo {author} {\bibfnamefont {S.~M.}\ \bibnamefont
  {Crawford}}, \bibinfo {author} {\bibfnamefont {C.~M.}\ \bibnamefont {Cress}},
  \bibinfo {author} {\bibfnamefont {B.~A.}\ \bibnamefont {Bassett}}, \bibinfo
  {author} {\bibfnamefont {R.~C.}\ \bibnamefont {Nichol}}, \ and\ \bibinfo
  {author} {\bibfnamefont {P.}~\bibnamefont {V{\"a}is{\"a}nen}},\ }\href
  {\doibase 10.1093/mnras/stx301} {\bibfield  {journal} {\bibinfo  {journal}
  {Mon. Not. Roy. Astron. Soc.}\ }\textbf {\bibinfo {volume} {467}},\ \bibinfo
  {pages} {3239} (\bibinfo {year} {2017})},\ \Eprint
  {http://arxiv.org/abs/1702.00418} {arXiv:1702.00418 [astro-ph.CO]}
  \BibitemShut {NoStop}%
\bibitem [{\citenamefont {Jiao}\ \emph {et~al.}(2023)\citenamefont {Jiao},
  \citenamefont {Borghi}, \citenamefont {Moresco},\ and\ \citenamefont
  {Zhang}}]{Jiao:2022aep}%
  \BibitemOpen
  \bibfield  {author} {\bibinfo {author} {\bibfnamefont {K.}~\bibnamefont
  {Jiao}}, \bibinfo {author} {\bibfnamefont {N.}~\bibnamefont {Borghi}},
  \bibinfo {author} {\bibfnamefont {M.}~\bibnamefont {Moresco}}, \ and\
  \bibinfo {author} {\bibfnamefont {T.-J.}\ \bibnamefont {Zhang}},\ }\href
  {\doibase 10.3847/1538-4365/acbc77} {\bibfield  {journal} {\bibinfo
  {journal} {Astrophys. J. Suppl.}\ }\textbf {\bibinfo {volume} {265}},\
  \bibinfo {pages} {48} (\bibinfo {year} {2023})},\ \Eprint
  {http://arxiv.org/abs/2205.05701} {arXiv:2205.05701 [astro-ph.CO]}
  \BibitemShut {NoStop}%
\bibitem [{\citenamefont {Moresco}(2023)}]{Moresco:2023zys}%
  \BibitemOpen
  \bibfield  {author} {\bibinfo {author} {\bibfnamefont {M.}~\bibnamefont
  {Moresco}},\ }\href@noop {} {\  (\bibinfo {year} {2023})},\ \Eprint
  {http://arxiv.org/abs/2307.09501} {arXiv:2307.09501 [astro-ph.CO]}
  \BibitemShut {NoStop}%
\bibitem [{\citenamefont {Jimenez}\ and\ \citenamefont
  {Loeb}(2002)}]{Jimenez:2001gg}%
  \BibitemOpen
  \bibfield  {author} {\bibinfo {author} {\bibfnamefont {R.}~\bibnamefont
  {Jimenez}}\ and\ \bibinfo {author} {\bibfnamefont {A.}~\bibnamefont {Loeb}},\
  }\href {\doibase 10.1086/340549} {\bibfield  {journal} {\bibinfo  {journal}
  {Astrophys. J.}\ }\textbf {\bibinfo {volume} {573}},\ \bibinfo {pages} {37}
  (\bibinfo {year} {2002})},\ \Eprint {http://arxiv.org/abs/astro-ph/0106145}
  {arXiv:astro-ph/0106145} \BibitemShut {NoStop}%
\bibitem [{\citenamefont {Moresco}\ \emph {et~al.}(2020)\citenamefont
  {Moresco}, \citenamefont {Jimenez}, \citenamefont {Verde}, \citenamefont
  {Cimatti},\ and\ \citenamefont {Pozzetti}}]{Moresco:2020fbm}%
  \BibitemOpen
  \bibfield  {author} {\bibinfo {author} {\bibfnamefont {M.}~\bibnamefont
  {Moresco}}, \bibinfo {author} {\bibfnamefont {R.}~\bibnamefont {Jimenez}},
  \bibinfo {author} {\bibfnamefont {L.}~\bibnamefont {Verde}}, \bibinfo
  {author} {\bibfnamefont {A.}~\bibnamefont {Cimatti}}, \ and\ \bibinfo
  {author} {\bibfnamefont {L.}~\bibnamefont {Pozzetti}},\ }\href {\doibase
  10.3847/1538-4357/ab9eb0} {\bibfield  {journal} {\bibinfo  {journal}
  {Astrophys. J.}\ }\textbf {\bibinfo {volume} {898}},\ \bibinfo {pages} {82}
  (\bibinfo {year} {2020})},\ \Eprint {http://arxiv.org/abs/2003.07362}
  {arXiv:2003.07362 [astro-ph.GA]} \BibitemShut {NoStop}%
\bibitem [{\citenamefont {Moresco}\ \emph {et~al.}(2022)\citenamefont {Moresco}
  \emph {et~al.}}]{Moresco:2022phi}%
  \BibitemOpen
  \bibfield  {author} {\bibinfo {author} {\bibfnamefont {M.}~\bibnamefont
  {Moresco}} \emph {et~al.},\ }\href {\doibase 10.1007/s41114-022-00040-z}
  {\bibfield  {journal} {\bibinfo  {journal} {Living Rev. Rel.}\ }\textbf
  {\bibinfo {volume} {25}},\ \bibinfo {pages} {6} (\bibinfo {year} {2022})},\
  \Eprint {http://arxiv.org/abs/2201.07241} {arXiv:2201.07241 [astro-ph.CO]}
  \BibitemShut {NoStop}%
\bibitem [{\citenamefont {Moresco}(2021)}]{mmorescoCCcovariance}%
  \BibitemOpen
  \bibfield  {author} {\bibinfo {author} {\bibfnamefont {M.}~\bibnamefont
  {Moresco}},\ }\href@noop {} {\enquote {\bibinfo {title} {Cc covariance
  components notebook},}\ }\bibinfo {howpublished}
  {\url{https://gitlab.com/mmoresco/CCcovariance/-/blob/master/examples/CC_covariance_components.ipynb}}
  (\bibinfo {year} {2021}),\ \bibinfo {note} {accessed on
  2025-03-20}\BibitemShut {NoStop}%
\bibitem [{\citenamefont {Foreman-Mackey}\ \emph {et~al.}(2013)\citenamefont
  {Foreman-Mackey}, \citenamefont {Hogg}, \citenamefont {Lang},\ and\
  \citenamefont {Goodman}}]{Foreman-Mackey:2012any}%
  \BibitemOpen
  \bibfield  {author} {\bibinfo {author} {\bibfnamefont {D.}~\bibnamefont
  {Foreman-Mackey}}, \bibinfo {author} {\bibfnamefont {D.~W.}\ \bibnamefont
  {Hogg}}, \bibinfo {author} {\bibfnamefont {D.}~\bibnamefont {Lang}}, \ and\
  \bibinfo {author} {\bibfnamefont {J.}~\bibnamefont {Goodman}},\ }\href
  {\doibase 10.1086/670067} {\bibfield  {journal} {\bibinfo  {journal} {Publ.
  Astron. Soc. Pac.}\ }\textbf {\bibinfo {volume} {125}},\ \bibinfo {pages}
  {306} (\bibinfo {year} {2013})},\ \Eprint {http://arxiv.org/abs/1202.3665}
  {arXiv:1202.3665 [astro-ph.IM]} \BibitemShut {NoStop}%
\bibitem [{\citenamefont {Lewis}(2025)}]{Lewis:2019xzd}%
  \BibitemOpen
  \bibfield  {author} {\bibinfo {author} {\bibfnamefont {A.}~\bibnamefont
  {Lewis}},\ }\href {\doibase 10.1088/1475-7516/2025/08/025} {\bibfield
  {journal} {\bibinfo  {journal} {JCAP}\ }\textbf {\bibinfo {volume} {08}},\
  \bibinfo {pages} {025} (\bibinfo {year} {2025})},\ \Eprint
  {http://arxiv.org/abs/1910.13970} {arXiv:1910.13970 [astro-ph.IM]}
  \BibitemShut {NoStop}%
\bibitem [{\citenamefont {Heavens}\ \emph {et~al.}(2017)\citenamefont
  {Heavens}, \citenamefont {Fantaye}, \citenamefont {Mootoovaloo},
  \citenamefont {Eggers}, \citenamefont {Hosenie}, \citenamefont {Kroon},\ and\
  \citenamefont {Sellentin}}]{Heavens:2017afc}%
  \BibitemOpen
  \bibfield  {author} {\bibinfo {author} {\bibfnamefont {A.}~\bibnamefont
  {Heavens}}, \bibinfo {author} {\bibfnamefont {Y.}~\bibnamefont {Fantaye}},
  \bibinfo {author} {\bibfnamefont {A.}~\bibnamefont {Mootoovaloo}}, \bibinfo
  {author} {\bibfnamefont {H.}~\bibnamefont {Eggers}}, \bibinfo {author}
  {\bibfnamefont {Z.}~\bibnamefont {Hosenie}}, \bibinfo {author} {\bibfnamefont
  {S.}~\bibnamefont {Kroon}}, \ and\ \bibinfo {author} {\bibfnamefont
  {E.}~\bibnamefont {Sellentin}},\ }\href@noop {} {\  (\bibinfo {year}
  {2017})},\ \Eprint {http://arxiv.org/abs/1704.03472} {arXiv:1704.03472
  [stat.CO]} \BibitemShut {NoStop}%
\bibitem [{\citenamefont {Li}\ \emph {et~al.}(2025{\natexlab{f}})\citenamefont
  {Li}, \citenamefont {Riess}, \citenamefont {Anand}, \citenamefont {Scolnic},
  \citenamefont {Murakami}, \citenamefont {Brout},\ and\ \citenamefont
  {Peterson}}]{Li:2025lfp}%
  \BibitemOpen
  \bibfield  {author} {\bibinfo {author} {\bibfnamefont {S.}~\bibnamefont
  {Li}}, \bibinfo {author} {\bibfnamefont {A.~G.}\ \bibnamefont {Riess}},
  \bibinfo {author} {\bibfnamefont {G.~S.}\ \bibnamefont {Anand}}, \bibinfo
  {author} {\bibfnamefont {D.}~\bibnamefont {Scolnic}}, \bibinfo {author}
  {\bibfnamefont {Y.~S.}\ \bibnamefont {Murakami}}, \bibinfo {author}
  {\bibfnamefont {D.}~\bibnamefont {Brout}}, \ and\ \bibinfo {author}
  {\bibfnamefont {E.~R.}\ \bibnamefont {Peterson}},\ }\href@noop {} {\
  (\bibinfo {year} {2025}{\natexlab{f}})},\ \Eprint
  {http://arxiv.org/abs/2504.08921} {arXiv:2504.08921 [astro-ph.CO]}
  \BibitemShut {NoStop}%
\bibitem [{\citenamefont {G{\'o}mez-Valent}(2022)}]{Gomez-Valent:2021hda}%
  \BibitemOpen
  \bibfield  {author} {\bibinfo {author} {\bibfnamefont {A.}~\bibnamefont
  {G{\'o}mez-Valent}},\ }\href {\doibase 10.1103/PhysRevD.105.043528}
  {\bibfield  {journal} {\bibinfo  {journal} {Phys. Rev. D}\ }\textbf {\bibinfo
  {volume} {105}},\ \bibinfo {pages} {043528} (\bibinfo {year} {2022})},\
  \Eprint {http://arxiv.org/abs/2111.15450} {arXiv:2111.15450 [astro-ph.CO]}
  \BibitemShut {NoStop}%
\bibitem [{\citenamefont {Notari}\ \emph {et~al.}(2024)\citenamefont {Notari},
  \citenamefont {Redi},\ and\ \citenamefont {Tesi}}]{Notari:2024rti}%
  \BibitemOpen
  \bibfield  {author} {\bibinfo {author} {\bibfnamefont {A.}~\bibnamefont
  {Notari}}, \bibinfo {author} {\bibfnamefont {M.}~\bibnamefont {Redi}}, \ and\
  \bibinfo {author} {\bibfnamefont {A.}~\bibnamefont {Tesi}},\ }\href {\doibase
  10.1088/1475-7516/2024/11/025} {\bibfield  {journal} {\bibinfo  {journal}
  {JCAP}\ }\textbf {\bibinfo {volume} {11}},\ \bibinfo {pages} {025} (\bibinfo
  {year} {2024})},\ \Eprint {http://arxiv.org/abs/2406.08459} {arXiv:2406.08459
  [astro-ph.CO]} \BibitemShut {NoStop}%
\bibitem [{\citenamefont {Trotta}(2008)}]{Trotta:2008qt}%
  \BibitemOpen
  \bibfield  {author} {\bibinfo {author} {\bibfnamefont {R.}~\bibnamefont
  {Trotta}},\ }\href {\doibase 10.1080/00107510802066753} {\bibfield  {journal}
  {\bibinfo  {journal} {Contemp. Phys.}\ }\textbf {\bibinfo {volume} {49}},\
  \bibinfo {pages} {71} (\bibinfo {year} {2008})},\ \Eprint
  {http://arxiv.org/abs/0803.4089} {arXiv:0803.4089 [astro-ph]} \BibitemShut
  {NoStop}%
\bibitem [{\citenamefont {Patel}\ \emph {et~al.}(2024)\citenamefont {Patel},
  \citenamefont {Chakraborty},\ and\ \citenamefont {Amendola}}]{Patel:2024odo}%
  \BibitemOpen
  \bibfield  {author} {\bibinfo {author} {\bibfnamefont {V.}~\bibnamefont
  {Patel}}, \bibinfo {author} {\bibfnamefont {A.}~\bibnamefont {Chakraborty}},
  \ and\ \bibinfo {author} {\bibfnamefont {L.}~\bibnamefont {Amendola}},\
  }\href@noop {} {\  (\bibinfo {year} {2024})},\ \Eprint
  {http://arxiv.org/abs/2407.06586} {arXiv:2407.06586 [astro-ph.CO]}
  \BibitemShut {NoStop}%
\bibitem [{\citenamefont {Sola~Peracaula}\ \emph {et~al.}(2019)\citenamefont
  {Sola~Peracaula}, \citenamefont {Gomez-Valent},\ and\ \citenamefont
  {de~Cruz~P{\'e}rez}}]{SolaPeracaula:2018wwm}%
  \BibitemOpen
  \bibfield  {author} {\bibinfo {author} {\bibfnamefont {J.}~\bibnamefont
  {Sola~Peracaula}}, \bibinfo {author} {\bibfnamefont {A.}~\bibnamefont
  {Gomez-Valent}}, \ and\ \bibinfo {author} {\bibfnamefont {J.}~\bibnamefont
  {de~Cruz~P{\'e}rez}},\ }\href {\doibase 10.1016/j.dark.2019.100311}
  {\bibfield  {journal} {\bibinfo  {journal} {Phys. Dark Univ.}\ }\textbf
  {\bibinfo {volume} {25}},\ \bibinfo {pages} {100311} (\bibinfo {year}
  {2019})},\ \Eprint {http://arxiv.org/abs/1811.03505} {arXiv:1811.03505
  [astro-ph.CO]} \BibitemShut {NoStop}%
\bibitem [{\citenamefont {Vincenzi}\ \emph {et~al.}(2025)\citenamefont
  {Vincenzi} \emph {et~al.}}]{DES:2025tir}%
  \BibitemOpen
  \bibfield  {author} {\bibinfo {author} {\bibfnamefont {M.}~\bibnamefont
  {Vincenzi}} \emph {et~al.} (\bibinfo {collaboration} {DES}),\ }\href
  {\doibase 10.1093/mnras/staf943} {\bibfield  {journal} {\bibinfo  {journal}
  {Mon. Not. Roy. Astron. Soc.}\ }\textbf {\bibinfo {volume} {541}},\ \bibinfo
  {pages} {2585} (\bibinfo {year} {2025})},\ \Eprint
  {http://arxiv.org/abs/2501.06664} {arXiv:2501.06664 [astro-ph.CO]}
  \BibitemShut {NoStop}%
\bibitem [{\citenamefont {Gialamas}\ \emph
  {et~al.}(2025{\natexlab{b}})\citenamefont {Gialamas}, \citenamefont
  {H{\"u}tsi}, \citenamefont {Kannike}, \citenamefont {Racioppi}, \citenamefont
  {Raidal}, \citenamefont {Vasar},\ and\ \citenamefont
  {Veerm{\"a}e}}]{Gialamas:2024lyw}%
  \BibitemOpen
  \bibfield  {author} {\bibinfo {author} {\bibfnamefont {I.~D.}\ \bibnamefont
  {Gialamas}}, \bibinfo {author} {\bibfnamefont {G.}~\bibnamefont {H{\"u}tsi}},
  \bibinfo {author} {\bibfnamefont {K.}~\bibnamefont {Kannike}}, \bibinfo
  {author} {\bibfnamefont {A.}~\bibnamefont {Racioppi}}, \bibinfo {author}
  {\bibfnamefont {M.}~\bibnamefont {Raidal}}, \bibinfo {author} {\bibfnamefont
  {M.}~\bibnamefont {Vasar}}, \ and\ \bibinfo {author} {\bibfnamefont
  {H.}~\bibnamefont {Veerm{\"a}e}},\ }\href {\doibase
  10.1103/PhysRevD.111.043540} {\bibfield  {journal} {\bibinfo  {journal}
  {Phys. Rev. D}\ }\textbf {\bibinfo {volume} {111}},\ \bibinfo {pages}
  {043540} (\bibinfo {year} {2025}{\natexlab{b}})},\ \Eprint
  {http://arxiv.org/abs/2406.07533} {arXiv:2406.07533 [astro-ph.CO]}
  \BibitemShut {NoStop}%
\bibitem [{\citenamefont {Zheng}\ \emph {et~al.}(2025)\citenamefont {Zheng},
  \citenamefont {Qiang},\ and\ \citenamefont {You}}]{Zheng:2024qzi}%
  \BibitemOpen
  \bibfield  {author} {\bibinfo {author} {\bibfnamefont {J.}~\bibnamefont
  {Zheng}}, \bibinfo {author} {\bibfnamefont {D.-C.}\ \bibnamefont {Qiang}}, \
  and\ \bibinfo {author} {\bibfnamefont {Z.-Q.}\ \bibnamefont {You}},\ }\href
  {\doibase 10.1088/1475-7516/2025/08/056} {\bibfield  {journal} {\bibinfo
  {journal} {JCAP}\ }\textbf {\bibinfo {volume} {08}},\ \bibinfo {pages} {056}
  (\bibinfo {year} {2025})},\ \Eprint {http://arxiv.org/abs/2412.04830}
  {arXiv:2412.04830 [astro-ph.CO]} \BibitemShut {NoStop}%
\bibitem [{\citenamefont {Mukherjee}\ and\ \citenamefont
  {Sen}(2025)}]{Mukherjee:2025fkf}%
  \BibitemOpen
  \bibfield  {author} {\bibinfo {author} {\bibfnamefont {P.}~\bibnamefont
  {Mukherjee}}\ and\ \bibinfo {author} {\bibfnamefont {A.~A.}\ \bibnamefont
  {Sen}},\ }\href@noop {} {\  (\bibinfo {year} {2025})},\ \Eprint
  {http://arxiv.org/abs/2503.02880} {arXiv:2503.02880 [astro-ph.CO]}
  \BibitemShut {NoStop}%
\bibitem [{\citenamefont {Liu}\ \emph {et~al.}(2024)\citenamefont {Liu},
  \citenamefont {Wang},\ and\ \citenamefont {Zhao}}]{Liu:2024gfy}%
  \BibitemOpen
  \bibfield  {author} {\bibinfo {author} {\bibfnamefont {G.}~\bibnamefont
  {Liu}}, \bibinfo {author} {\bibfnamefont {Y.}~\bibnamefont {Wang}}, \ and\
  \bibinfo {author} {\bibfnamefont {W.}~\bibnamefont {Zhao}},\ }\href@noop {}
  {\  (\bibinfo {year} {2024})},\ \Eprint {http://arxiv.org/abs/2407.04385}
  {arXiv:2407.04385 [astro-ph.CO]} \BibitemShut {NoStop}%
\bibitem [{\citenamefont {Caldwell}\ \emph {et~al.}(2003)\citenamefont
  {Caldwell}, \citenamefont {Kamionkowski},\ and\ \citenamefont
  {Weinberg}}]{Caldwell:2003vq}%
  \BibitemOpen
  \bibfield  {author} {\bibinfo {author} {\bibfnamefont {R.~R.}\ \bibnamefont
  {Caldwell}}, \bibinfo {author} {\bibfnamefont {M.}~\bibnamefont
  {Kamionkowski}}, \ and\ \bibinfo {author} {\bibfnamefont {N.~N.}\
  \bibnamefont {Weinberg}},\ }\href {\doibase 10.1103/PhysRevLett.91.071301}
  {\bibfield  {journal} {\bibinfo  {journal} {Phys. Rev. Lett.}\ }\textbf
  {\bibinfo {volume} {91}},\ \bibinfo {pages} {071301} (\bibinfo {year}
  {2003})},\ \Eprint {http://arxiv.org/abs/astro-ph/0302506}
  {arXiv:astro-ph/0302506} \BibitemShut {NoStop}%
\bibitem [{\citenamefont {Vikman}(2005)}]{Vikman:2004dc}%
  \BibitemOpen
  \bibfield  {author} {\bibinfo {author} {\bibfnamefont {A.}~\bibnamefont
  {Vikman}},\ }\href {\doibase 10.1103/PhysRevD.71.023515} {\bibfield
  {journal} {\bibinfo  {journal} {Phys. Rev. D}\ }\textbf {\bibinfo {volume}
  {71}},\ \bibinfo {pages} {023515} (\bibinfo {year} {2005})},\ \Eprint
  {http://arxiv.org/abs/astro-ph/0407107} {arXiv:astro-ph/0407107} \BibitemShut
  {NoStop}%
\bibitem [{\citenamefont {Ling}\ \emph {et~al.}()\citenamefont {Ling},
  \citenamefont {Du}, \citenamefont {Li}, \citenamefont {Zhang}, \citenamefont
  {Wang},\ and\ \citenamefont {Zhang}}]{Ling_ScienceDB_27772}%
  \BibitemOpen
  \bibfield  {author} {\bibinfo {author} {\bibfnamefont {J.-L.}\ \bibnamefont
  {Ling}}, \bibinfo {author} {\bibfnamefont {G.-H.}\ \bibnamefont {Du}},
  \bibinfo {author} {\bibfnamefont {T.-N.}\ \bibnamefont {Li}}, \bibinfo
  {author} {\bibfnamefont {J.-F.}\ \bibnamefont {Zhang}}, \bibinfo {author}
  {\bibfnamefont {S.-J.}\ \bibnamefont {Wang}}, \ and\ \bibinfo {author}
  {\bibfnamefont {X.}~\bibnamefont {Zhang}},\ }\href@noop {} {}\bibinfo
  {howpublished} {ScienceDB},\ \bibinfo {note}
  {\href{https://doi.org/10.57760/sciencedb.27772}{10.57760/sciencedb.27772}}\BibitemShut
  {NoStop}%
\end{thebibliography}%
\end{document}